\documentclass[pra,aps,superscriptaddress,nofootinbib,twocolumn]{revtex4-1}

\usepackage{amsmath,amssymb,amsthm,mathrsfs,amsfonts,dsfont}
\usepackage{silence}
\usepackage{epsfig}
\usepackage{tikz}
\usetikzlibrary{quantikz, math, fit}
\usepackage{pgfplots}
\usepgfplotslibrary{fillbetween, groupplots}
\pgfplotsset{compat=1.17,
             every tick label/.append style={font=\tiny},
             standard/.style={every axis plot post/.style={mark options={fill=white}}}}
\tikzmath{%
  function h1(\x, \lx) { return (9*\lx + 3*((\lx)^2) + ((\lx)^3)/3 + 9); };
  function h2(\x, \lx) { return (3*\lx - ((\lx)^3)/3 + 4); };
  function h3(\x, \lx) { return (9*\lx - 3*((\lx)^2) + ((\lx)^3)/3 + 7); };
  function skewnorm(\x, \l) {
    \x = (\l < 0) ? -\x : \x;
    \l = abs(\l);
    \e = exp(-(\x^2)/2);
    return (\l == 0) ? 1 / sqrt(2 * pi) * \e: (
      (\x < -3/\l) ? 0 : (
      (\x < -1/\l) ? \e / (8 * sqrt(2 * pi)) * h1(\x, \x*\l) : (
      (\x <  1/\l) ? \e / (4 * sqrt(2 * pi)) * h2(\x, \x*\l) : (
      (\x <  3/\l) ? \e / (8 * sqrt(2 * pi)) * h3(\x, \x*\l) : (
      sqrt(2/pi) * \e)))));
  };
}
\usepackage{braket}
\usepackage{bm}
\usepackage{enumerate}
\usepackage{color}
\usepackage{physics}
\usepackage{comment}
\usepackage{hyperref}
\hypersetup{
    colorlinks=true,
    linkcolor=blue,
    filecolor=magenta,      
    urlcolor=blue,
}

\newcommand{\qmaddress}{Quantum Motion, 9 Sterling Way, London N7 9HJ, United Kingdom}
\DeclareMathOperator{\sinc}{sinc}

\usepackage[algoruled,ruled,vlined,noline,linesnumbered]{algorithm2e}
\SetAlgoCaptionLayout{raggedright}
\newcommand{\codecomment}[1]{\textcolor{cyan}{// #1}}

\begin{document}

\title{Grid-based methods for chemistry simulations on a quantum computer}


\author{Hans Hon Sang Chan}
\email{hans.chan@materials.ox.ac.uk}
\affiliation{Department of Materials, University of Oxford, Oxford OX1 3PH, United Kingdom}

\author{Richard Meister}
\affiliation{Department of Materials, University of Oxford, Oxford OX1 3PH, United Kingdom}

\author{Tyson Jones}
\affiliation{Department of Materials, University of Oxford, Oxford OX1 3PH, United Kingdom}

\author{David P. Tew}
\affiliation{Department of Chemistry, University of Oxford, Oxford OX1 3TA, United Kingdom}
\affiliation{Duality Quantum Photonics, 6 Lower Park Row, Bristol BS1 5BJ, United Kingdom}

\author{Simon C. Benjamin}
\affiliation{Department of Materials, University of Oxford, Oxford OX1 3PH, United Kingdom}
\affiliation{\qmaddress}

\begin{abstract}
First quantized, grid-based methods for chemistry modelling are a natural and elegant fit for quantum computers. However, it is infeasible to use today's quantum prototypes to explore the power of this approach, because it requires a substantial number of near-perfect qubits. Here we employ exactly-emulated quantum computers with up to 36 qubits, to execute deep yet resource-frugal algorithms that model 2D and 3D atoms with single and paired particles. A range of tasks is explored, from ground state preparation and energy estimation to the dynamics of scattering and ionisation; we evaluate various methods within the split-operator QFT (SO-QFT) Hamiltonian simulation paradigm, including protocols previously-described in theoretical papers as well as our own techniques. While we identify certain restrictions and caveats, generally the grid-based method is found to perform very well; our results are consistent with the view that first-quantized paradigms will be dominant from the early fault-tolerant quantum computing era onward.

\smallskip
\noindent \textbf{TEASER}: Using emulated quantum computers, the impact that real quantum devices may have for chemistry dynamics simulation is studied.
\end{abstract}

\maketitle

\section{Introduction}

\color{black}
Quantum computers may prove to be transformative tools for exploration and prediction in chemistry. When conventional computers are used for first-principles quantum molecular dynamics simulation, which is an important technique for predicting reaction outcomes and various experimental observables, the required resources (i.e. the hardware and time duration) scale exponentially with the number of simulated particles. However these costs are expected to scale only polynomially for quantum computers, thus enabling simulations which are otherwise practically  impossible. Whether and when this promise will be realised can only be predicted with a comprehensive exploration of the quantum approach. 
It is relevant to note that recently, a study concluded that there is as-yet no evidence of fundamental `exponential quantum advantage' in the task of computing molecular ground state energies~\cite{GChan2022}. While that task is distinct from quantum dynamical simulation, the observation highlights a pressing need for clarity which will doubtless increase as more powerful quantum computers emerge (see e.g.~\cite{arute2019quantum}).

In this work, we investigate the prospects for accelerating chemical dynamics simulation on early fault-tolerant quantum computers using the \textit{first-quantized, real-space grid} approach~\cite{wiesner1996simulations, doi:10.1098/rspa.1998.0162, Kassal2008, Benenti2008, Cody_Jones_2012, somma2016quantum, Kivlichan_2017, Ollitrault2020, su2021fault, kosugi2021probabilistic, Childs2022, poirier2021fulldimensional, Ollitrault2022, Kosugi2022_geometry, Hirai2022}. By `early' we mean machines that have a limited number of error-corrected qubits, as we presently explain. This approach involves representing wavefunctions over a grid of points; thus the method explicitly encodes features such as particle symmetry (unlike the conventional second-quantized formulation). We select this approach as it is appealingly intuitive, but moreover first-quantized simulation is anticipated by many researchers to offer the optimal resource scaling for complex and interesting molecules~\cite{Kivlichan_2017, su2021fault, McClean2021}; indeed some have even suggested that first-quantized simulation can efficiently encode both nuclear and electronic degrees of freedom on an equal footing, potentially addressing the gap in simulating non-Born-Oppenheimer processes in modern chemical physics~\cite{Kassal2008}.

Real-space grid methods have been used with classical computers since at least the 80's~\cite{cerjan2013numerical, Tucker2000, Leforestier1991, PhysRevE.73.036708, harrison2016madness} even if in practice only simplified models with wavefunctions that are, so to speak, `heavily pixelated' can be stored and processed within conventional Random Access Memory. Even with quantum computers, first-quantized methods will require numerous qubits and deep circuits for meaningful realisations, making them impractical on the noise-burdened quantum computers of today. Most prior studies of such approaches have therefore focused on theoretical `pen and paper' analysis of the resource costs~\cite{Cody_Jones_2012, Kivlichan_2017, babbush2019quantum, Childs2022,su2021fault}. In this study, we take a different approach: we deploy very substantial classical computing resources to perform exact emulations of small but noise-free quantum computers; these emulated computers then simulate representative quantum molecular dynamics. Thus we are able to directly examine costs and performance measures.

The cost of emulation limits us to modest-sized quantum computers (we use at most 36 perfect qubits). However we find we can explore a number of informative scenarios within this restriction: 2D and 3D simulations of one- and two-electron systems. We select specific scenarios to elucidate two key areas of interest in chemistry. We now describe these and identify small-to-medium molecules that would be important targets for early fault-tolerant quantum computers; herein we leverage our results to estimate the quantum resources required.
\smallskip

\noindent \textit{Scenario I}: Simulation of dynamics in the presence of strong external fields. Our exploratory work here involves a suddenly-applied external field with resulting dipole oscillation and ionisation of a single bound electron. Ultimately efforts in this direction will encompass topics such as photochemistry and laser excitation. Some applications would require mature (rather than early) fault-tolerant quantum computers; for example, comprehensively modelling the dynamics of photosynthesis would be a profound accomplishment but would involve highly complex molecules e.g. the Fenna-Matthews-Olson complex~\cite{fenna1975chlorophyll}. A more near-term prospect is laser-driven dynamics: coherent quantum control of small molecules in this way has been considered one of the `holy grails' of chemical science~\cite{Wilson1995}.
A modest molecule well-worthy of study would be \textbf{ammonia (NH$_3$)}, investigated in the context of selective hydrogen atom removal~\cite{Tew2018simulating}. If quantum modelling of its dynamics under laser control were to reveal new synthesis options, the consequences could be profound since ammonia use lies at the heart of modern agriculture.
\smallskip

\noindent \textit{Scenario II}: Simulating the dynamics of particle scattering.  Our exploratory work here involves an incident electron scattering from a bound electron and potentially ionising it. In general, electron-molecule scattering is relevant in spectroscopy, astrochemistry, atmospheric chemistry, as well as manufacturing processes~\cite{RUHLE19961033}. While part of the computational challenge is scanning through possible initial energies of the incoming electron, predicting what happens upon collision and scattering is a highly quantum dynamical process difficult to model classically. Many cases involve reaction intermediates with fleeting lifetimes that are hard to observe, and occur under conditions experimentally challenging to access. An example currently beyond the reach of full-dimensional quantum dynamics simulation is \textbf{hexafluoro ethane (C$_2$F$_6$)}, a representative example of fluorocarbons~\cite{Hudson_2001} relevant in the chemistry of the ozone layer and in plasma etching.  
\smallskip

Our adopted approach is to perform wavepacket simulations with the split-operator quantum Fourier transform (SO-QFT) Hamiltonian simulation approach~\cite{Kassal2008, Ollitrault2020, kosugi2021probabilistic} based on the Lie-Suzuki-Trotter product formula. We model subatomic particles interacting directly via the Coulomb potential; a prerequisite for ultimately treating electrons and nuclei on a fully quantum basis.
The classical SO-FT algorithm has had decades of demonstrated success in nuclear wavepacket propagation~\cite{fleck1976time} on electronic potential energy surfaces, but as far as we know was never used with Coulomb potentials. Compared with other first-quantized real-space Hamiltonian simulation methods (see e.g. linear combination of unitaries~\cite{Kivlichan_2017} or qubitization~\cite{su2021fault, Childs2022}), SO-QFT can require the lowest number of qubits to implement time evolution~\cite{Kassal2008}.

Given that we use classically-emulated quantum computers to perform grid-based simulations, the reader might wonder if we are simply rehashing prior classical grid-based techniques under a new banner. We emphasise that this is not the case; our emulation is restricted to exactly the capabilities of real, albeit noise-free, quantum machines. This restriction is severe and manifests in multiple ways as we explore early fault-tolerant quantum computing techniques in the context of chemically-relevant quantum dynamics. 

Beyond the inherent value in executing previously-proposed quantum algorithms for the first time, and thus determining performance measures that hitherto could only be estimated, we make a number of contributions:

\begin{itemize}

    
    \item To perform scattering and ionisation modelling within the finite `simulation box' of the grid-based method, we create and explore a non-unitary wavepacket attenuation approach. It is inspired by complex absorbing potentials (CAPs) from classical simulation. The method uses measurement of a single entangled ancilla qubit to attenuate particles that are ejected from the finite simulation environment, preventing them from returning at the periodic boundary and interfering with the simulation. We note that sampling the outcome of the ancilla measurement doubles up as a means of tracking escape probabilities of wavepackets, and has potential for quantum computing reaction rates.
   We present visualisations of such dynamical events which a quantum computer would enable; the user of a real quantum processor would have access to analogous images for far more complex systems.
    
    \item Our use of the unmitigated Coulomb singularity creates a challenge in the spatial resolution, which we address by creating augmented split operator (ASO) approach. Here we use an additional small quantum circuit to correct the Trotter error incurred at every SO-QFT time evolution step when low spatial and temporal resolution is used.

    \item State preparation is a non-trivial challenge in quantum modelling. We assess prior methods and make our own contribution:
    \begin{itemize}
    \item We investigate an approach which uses the single-ancilla iterative phase estimation (IPE) measurement to project out excited states;
    \item we investigate an adaption of the single-ancilla probabilistic imaginary time evolution (PITE); method~\cite{kosugi2021probabilistic} for approximating small imaginary time evolution steps. 
    \item we build upon existing work~\cite{doi:10.1063/1.3115177} to create a method that explicitly generates the correct particle (anti)symmetry for first-quantized simulations.
    \end{itemize}
    
    \item Finally: In light of the above studies we estimate the quantum resource costs (time and hardware scale) for modelling the interesting molecules noted earlier, C$_2$F$_6$ and NH$_3$. We also indicate the hardware layout of a suitable quantum computer.
\end{itemize}
\smallskip

The paper is structured as follows. In Section~\ref{sec:results} we present a range of results from applying grid-based SO-QFT techniques to 2D and 3D systems with single and paired particles. We extrapolate from those results to estimate the quantum resources required for simulations beyond the reach of emulation, and we also present a suitable quantum hardware architecture. In Section~\ref{sec:summaryAndOutlook} we discuss implications and remark that the SO-QFT may be advantageous in applications well beyond molecular dynamics. In Section~\ref{sec:theory}, we describe the methods used in SO-QFT: Subsection~\ref{subsec:priorTheory} sets out ideas described in prior works but which are provided here for a self-contained explanation with consistent notation; expert readers may care to skip directly to subsection~\ref{sec:methods} where we set out the specific methods we employ.


\section{Results}\label{sec:results}
The numerical results in this section were obtained from exactly-emulated quantum processors, implemented using the open source tools QuEST~\cite{QuESTandHPC}, QuESTlink~\cite{QuESTlink} and  \texttt{pyQuEST}~\cite{pyquest}. Results are reported in Hartree atomic units, where the reduced Planck constant, electron mass, elementary charge and Bohr radius are treated to be unity $\hbar=m_e=e=a_0=1$. The particular techniques employed for each of the studies, are specified in the Methods sections and forward referenced from the present Section. Details of important configuration choices including the alignment between the grid of pixel functions and the nuclear potential, as well as the specific hardware employed, are given in the Supplementary Material.

\subsection{Commonly used Hamiltonian and initial states}
\label{sec:analtyicStates}
Here we frequently use the 2D hydrogenic system described by the Hamiltonian
\begin{equation}
    \hat{H}_\text{tot} = -\frac{\hbar^2}{2m_e}\nabla_\mathbf{r} - \frac{e^2}{4\pi\epsilon_0|\mathbf{R}-\mathbf{r}|},
    \label{eqn:2DHam}
\end{equation}
where we model the atomic nucleus as a classical discretised Coulomb potential, clamped with the origin between two pixels. Analytic solutions to Eqn.\,\ref{eqn:2DHam} have been reported in Refs.~\cite{Parfitt2002} and \cite{Yang1991}. We use an equation from the former,
\begin{align}
    \Psi_{n,m}(r, \theta) =& \sqrt{\frac{q_0^3(n-|m|)!}{\pi(n+|m|)!}} \times \left(2q_0r\right)^{|m|} \times e^{-q_0r}\nonumber\\
    & \times L^{2|m|}_{n-|m|}\left(2q_0r\right) \times e^{im\theta},
\end{align}
where $q_0 = \frac{1}{n+1/2}$ and $L$ are the generalised Laguerre polynomials.
The quantum numbers $n=0,1,2,\dots$, and there are $(2n+1)$ values of $m$.
The energy eigenvalues are
\begin{equation}
    E_n = -\frac{1}{2\left(n+\frac{1}{2}\right)^2}.
\end{equation}
We show two of the eigenstates used in this work in Fig.~\ref{fig:resolutionWF}. In a two-particle, 36 qubit simulation we also use states related to the well-known 3D hydrogenic eigenstates 
\begin{align}
    \Psi_{n,l,m}(r, \theta, \phi) =& \sqrt{\left(\frac{2Z}{n}\right)^3 \frac{(n-l-1)!}{2n(n+l)!}} \times \left(\frac{2Zr}{n}\right)^l  \nonumber \\
    & \times e^{-Zr/n} \times L^{2l+1}_{n-l-1}\left(\frac{2Zr}{n}\right) \nonumber \\
    & \times Y^m_l(\theta, \phi),
    \label{eqn:3Dhydrogen}
\end{align}
where $Y^m_l(\theta, \phi)$ is a spherical harmonic and $Z$ is the central nuclear charge. We also use Gaussian wavepackets of the form:
\begin{equation}
    \Psi(x) = e^{\Im(\gamma)}\left( \frac{2\Re(\alpha)}{\pi} \right)^{1/4} e^{-\alpha(x-x_c)^2 + ip_c(x-x_c) + i\gamma}
\end{equation}
where $x_c, p_c, \alpha$ and $\gamma$ are continuous parameters.

\subsection{Spatial and Temporal Resolution}
\label{sec:numerics}

A key topic to explore is the number of qubits, and the execution duration, required to achieve simulations of a given accuracy. In the grid-based method, the model's spatial resolution $\delta r^{-1}$ and the temporal resolution of the dynamics $\delta t^{-1}$ are crucial in determining accuracy; the qubit count per spatial dimension, $n_r$, is logarithmically related to the former (see Eqn.~\ref{eqn_resolution} in Methods).  

\begin{figure*}
    \centering
    \includegraphics[scale=1]{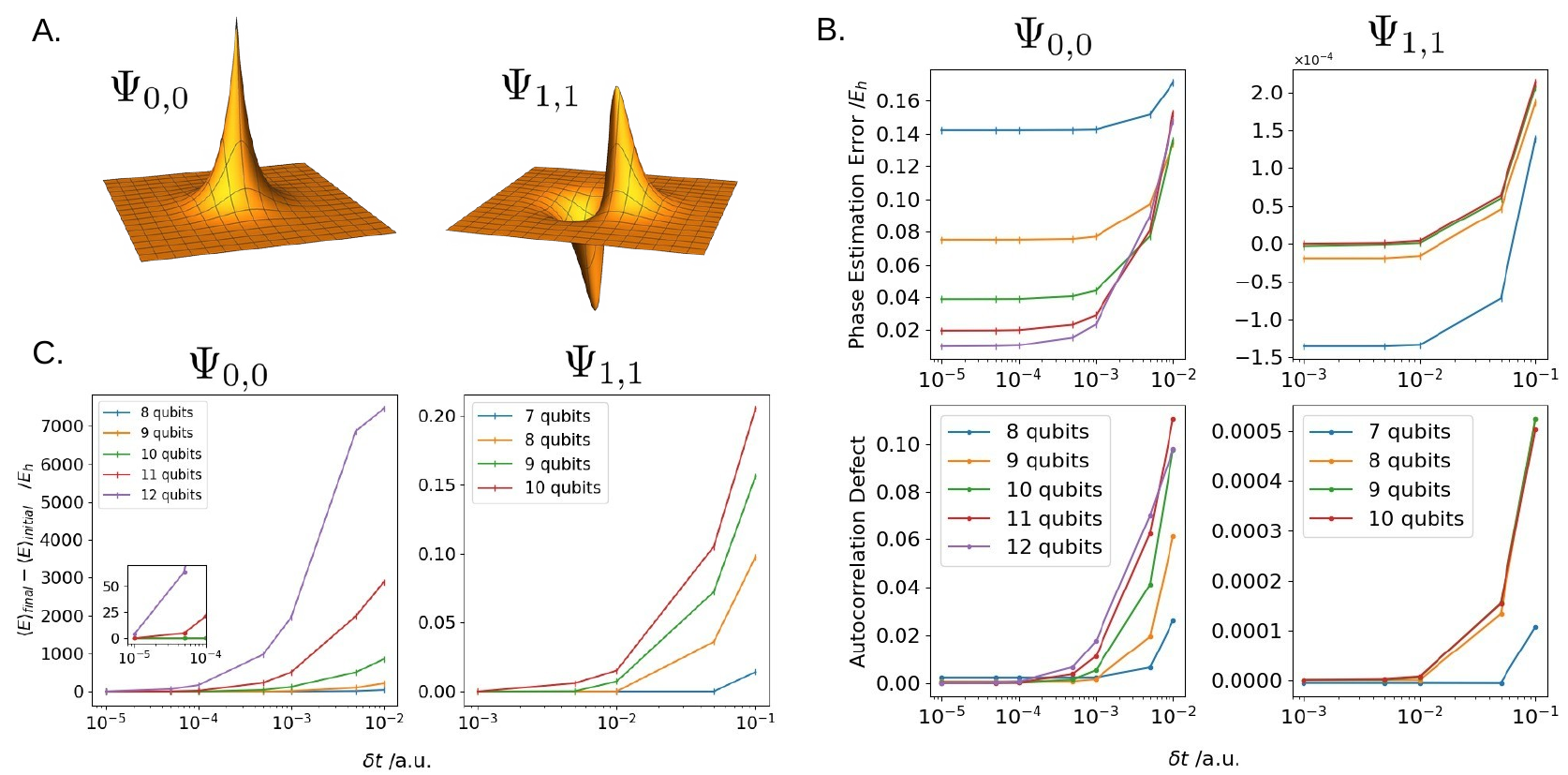}
    \caption{ \textbf{SO-QFT simulation of two-dimensional hydrogenic electron.} \textbf{(A)}~The ground $\Psi_{0,0}$ state (left) and a first excited $\Psi_{1,1}$ state (right) of two-dimensional hydrogen. Note that the plots here do not reflect the choice of simulation box size and are not to scale.
    \textbf{(B)}~Top panels represent difference between the energy from phase estimation and the analytic energy of 2D hydrogen. Bottom panels capture deviation of the simulation fidelity at the end of the propagation. In this series of experiments we initialised the ground state $\Psi_{0,0}$ centred in a simulation box with $L=10$~a.u., such that the origin of the Coulomb singularity lies half way between two central grid points. Each sub-register has a budget of $8 \leq n_r \leq 12$ qubits to store the wavefunction, corresponding to spatial resolutions of $0.039 \geq \delta r \geq 0.002$~a.u.. We also initialise the $\Psi_{1,1}$ excited state in a simulation box with sides of length 40~a.u., with budgets of $7 \leq n_r \leq 10$ qubits per sub-register and corresponding resolutions of $ 0.313\geq\delta r\geq 0.039 $~a.u.. These two states, represented at different spatial resolutions, are all time propagated using the first-order SO for 1.5 atomic time units. We used time steps between $0.00001 \leq \delta t \leq 0.01$ (150000 to 150 SO steps) for the $\Psi_{0,0}$ state, and between $0.001 \leq \delta t \leq 0.1$ (1500 to 15 SO steps) for the $\Psi_{1,1}$ state.
    \textbf{(C)}~The difference between the final and initial energy expectation value, measured by direct sampling of the state, of the ground state (left panels) and the first excited state (right panels), propagated at different spatial and time resolutions. The left inset plot zooms in on the energy error at high temporal resolutions for the ground state.
    }
    \label{fig:resolutionWF}
\end{figure*}

To explore these requirements, we first propagate eigenstates of the 2D hydrogen with different choices of $\delta r^{-1}$ and $\delta t^{-1}$. 
We start with loading the discretised analytic ground and excited states $\Psi_{0,0}$ and $\Psi_{1,1}$ into emulated qubit registers with different number of qubits per spatial degree of freedom $n_r$, then perform time-evolution experiments using the $1^\text{st}$-order SO-QFT. States were propagated for 1.5 atomic time units at different time step resolutions. Fig.~\ref{fig:resolutionWF} summarises these results.

As the initial states are eigenstates, ideally they would be static up to a global phase. Thus the final absolute value of the autocorrelation for each propagation sequence, specifically the deviation from unity, is a suitable fidelity metric. The lower pair of plots in Fig.~\ref{fig:resolutionWF}(B) displays this metric for $\Psi_{0,0}$ (left) and $\Psi_{1,1}$ (right). When a higher spatial resolution is used, correspondingly finer time steps are needed to conserve the fidelity (Section~\ref{sec:introResolution}). This is true of both cases but the relationship is more dramatic for $\Psi_{0,0}$, as expected given that its amplitude is peaked at the central Coulomb singularity. 

Single-ancilla Iterative Phase Estimation (IPE, see Section~\ref{sec:observablesAndPE}) is a simple means to extract an estimate of a system's energy from a simulation of its dynamics on a quantum computer. The upper panels in  Fig.~\ref{fig:resolutionWF}(B) plot the deviation of this estimate from the exact analytic result. One observes qualitatively the same behaviour as for the autocorrelation; there is a strong divergence when the $\delta t$ is insufficient, and that depends on both the $\delta r$ and the modelled state. 
For the ground $\Psi_{0,0}$ state, the error in the most accurate energy prediction (attained with the smallest time step propagation) halves when we increment $n_r$. While deviation of the energy inferred from phase estimation versus the exact value falls exponentially with the number of qubits $n_r$, chemical accuracy is not yet reached with $n_r=12$. 

We note that in additional simulations, not shown in the figure, we found that by quartering the size of the simulation box and increasing $n_r$ by 1 (effectively an additional 4 qubits) and propagating at $\delta t =10^{-5}$~a.u. for 4~a.u., we were able to achieve an error of 0.652~m$E_h$ from the estimated phase. In this case 400,000 SO-QFT cycles were used to cover only $\approx0.1$~femtoseconds of physical process. 
While these time requirements for accurate simulation of core-peaked states like $\Psi_{0,0}$ may seem daunting, it is worth reiterating that the approach
taken here is not optimised and we employed only the 1$^\text{st}$-order Trotter sequence. Moreover, as we show in Section~\ref{sec:ASOresults}, an Augmented Split-Operator approach can obtain accurate phase estimation with far fewer qubits and lower time resolution than the `brute force' method reported here. It is also noteworthy that accurate modelling of the $\Psi_{1,1}$ state is remarkably tolerant of low resolutions.

\color{black}

\subsection{Cautionary tale: A `bad' energy observable}
\label{sec:badObservable}
Here we examine a sampling-based method of estimating the system's energy which proves to be highly sensitive to inevitable imperfections in the model. Ultimately it will converge to give the correct expected energy once spatial and temporal resolution are sufficiently high. However it is profoundly inaccurate at resolutions where the IPE can already provide reliable, well-converged results. 

We are referring to the energy expectation as given by Eqn.\,\ref{eqn:expect}, viz. $
\langle E \rangle = \langle \hat{H}_\text{kin}\rangle + \langle \hat{H}_\text{pot}\rangle
$, and supposing that this would be obtained from our quantum computer as follows: Generate the desired state at time $t$, measure in the $k$-space representation, and repeat this many times in order to estimate the first term. Apply the same process but measuring in the real-space representation to estimate the second term. This method is inefficient in terms of the sampling cost, and is therefore already unattractive compared to phase estimation, but more problematically it is very sensitive to the resolution parameters. As shown in Fig.~\ref{fig:resolutionWF},
the error between the initial and final expected energy grows exponentially with the size of the elementary time step $\delta t$: for the ground state $\Psi_{0,0}$, the energy difference was more than 7000 $E_h$ accumulated over less than 40 attoseconds of simulated time in the worst offending case.
In the $\Psi_{1,1}$ simulation, the non-conservation of expected energy is also apparent when the temporal resolution does not match the spatial resolution, but is more contained relative to the ground state (in the worst case, it goes up to 0.20 $E_h$). It is evident that the core-peaked nature of the $\Psi_{0,0}$ state is a key issue.

It is the kinetic energy term $\langle \hat{H}_\text{kin}\rangle$ that exhibits this diverging behaviour. The explanation is as follows: Imbalance between the extreme potential and extreme kinetic energy near the Coulomb origin,
inevitable in our discrete grid representation, can allow a small amount of the amplitude to diffuse towards high frequency states in the plane wave representation.
The extent of this diffusion may be small relative to the initial state so that the fidelity of the state remains high (see Fig.~\ref{fig:resolutionWF}) and thus phase estimation methods can perform well. However, simply sampling $\ket{k}$ and squaring it to estimate $\langle \hat{H}_\text{kin}\rangle$ gives direct weight to this error that actually worsens as we improve the fineness of our resolution. Reducing $\delta r$, the separation of our spatial pixels, can reduce the leakage of amplitude (increasing the state's fidelity) but the maximum kinetic energy that the model can represent goes as $1/(\delta r)^2$. Amplitude leakage declines less rapidly than the rate at which the energy of the leaked-states increases: thus the problem worsens. One must use extremely high time resolution to ameliorate the effect.

A higher order Trotter formula would also presumably help, in the sense that a more modest time resolution could control the leakage. Nevertheless we anticipate that this approach to estimating energy will always be inferior to phase estimation.

\subsection{State Preparation}

\subsubsection{State Editing}
\label{subsec:phaseEstForPrep}
\begin{figure}[!tbp]
    \centering
    \includegraphics[clip,width=\columnwidth]{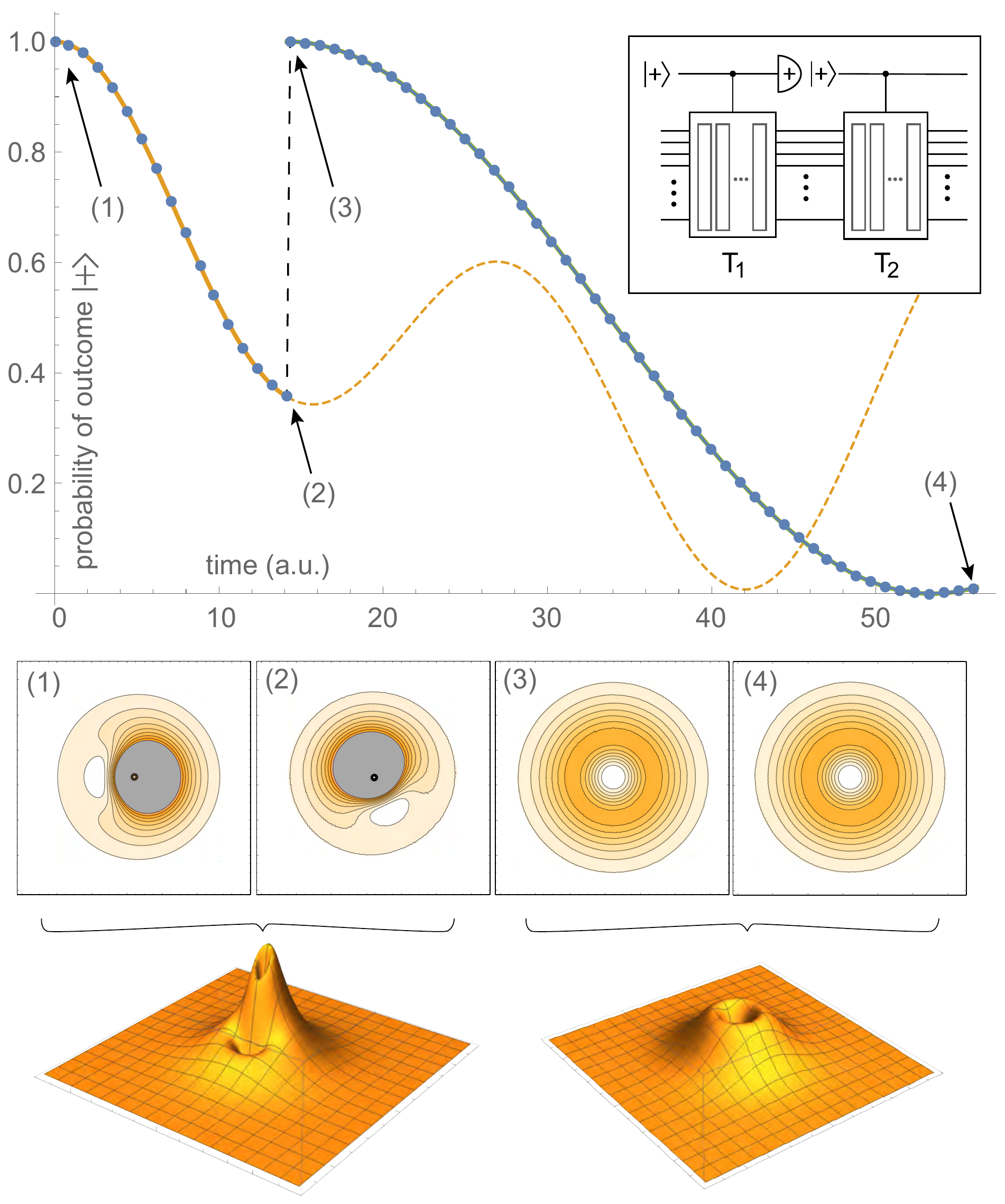}
\caption{\textbf{Demonstrating state editing by phase measurement.} A $1+2\times 8$-qubit quantum computer was emulated. The initial state is a superposition of non-degenerate eigenstates of 2D hydrogen. The representation uses $8$ qubits for each of the two dimensions, with a total simulation box width of $L=56$\,a.u. The state evolves, conditional on a controlling ancilla, for time $T_1$ chosen such that $T_1 E_1=\pi$; in this period the conditional evolution is a rotation of the state due to the accumulated phase difference between states (lower left plots). At $T_1$ the controlling ancilla is measured in the $\ket{+}$ as shown in the upper right inset. This projects the state into the $\Psi_{2,2}$ state.
}
\label{fig:statePreByMeasure}
\end{figure}

Earlier plots (Fig.~\ref{fig:resolutionWF}) have presented the results of IPE for energy estimation. Here we demonstrate preparation of an initial wavepacket state on a set of qubits using IPE (Section~\ref{sec:stateprepintro}). Fig.~\ref{fig:statePreByMeasure} shows the results of a simulation where the initial state (shown lower left and inset (i)) is a superposition of two eigenstates of 2D hydrogen: 
\[
\frac{1}{\sqrt{2}}\left(\Psi_{1,1}+\Psi_{2,2}\right).
\]
The state's amplitude is not symmetric about the nucleus, and when we apply our SO-QFT cycles we observe a rotation of the state due to the different rates at which the two superposed eigenstates acquire phase. At the time marked (ii), the total phase acquired by more tightly bound state $\Psi_{1,1}$ (having a more negative energy) has reached $\pi$; and thus, if this state were the sole one present, a control ancilla initially in state $\ket{+}$ would now certainly be in state $\ket{-}$. We indeed measure the ancilla at this point, but post-select on seeing the outcome $\ket{+}$. The probability of the desired outcome depends on the probability associated with the target $\Psi_{2,2}$ within the initial superposition (which was $0.5$) and the probability that this state, had it been prepared alone, would yield a $\ket{+}$ outcome at this point. The latter is $~0.713$ in the present case. In a scenario where the states to be distinguished are closer in energy, it may be optimal to simulate for several complete cycles of the undesired state before measuring.

In our numerical emulation, we indeed assume that the desired $\ket{+}$ state is obtained and we continue our time propagation. The new evolution of the ancilla state (green curve in the figure) is exactly that of the pure $\Psi_{2,2}$ state. The contour plots of the simulated state (insets (iii) and (iv)) confirm that we have prepared that pure state. Fidelity with respect to $\Psi_{2,2}$ was essentially identical to an initial state prepared directly in that state. 

This is a demonstration of the practically-useful capability to take an initial state that is not fully understood, and remove from it the components corresponding to states with energies that {\em are} understood. More generally, one could use Fourier analysis (see the Supplementary Material) to identify the components in the plot of Prob$(\ket{+})$ for the full state, and then use the post-selection method to stochastically isolate given components.

\subsubsection{Probabilistic Imaginary-Time Evolution}
\label{subsec:PITEForPrep}
We now compare with an alternative approach for preparing real-space ground states on a quantum computer, which approximates imaginary time evolution (Section~\ref{subsec:PITEForPrep}). As before, we model an attractive nucleus centred in a square simulation box. Instead of starting from an explicitly defined superposition of eigenstates as in the previous example, we initialise a Gaussian wavepacket centred about the origin of the Coulomb potential. The initial Gaussian wavepacket can obviously also be expanded in the eigenstate basis, which we assume has a large component corresponding to the ground state of the Hamiltonian.

\begin{figure}[!tbp]
    \centering
    \includegraphics[width=\columnwidth]{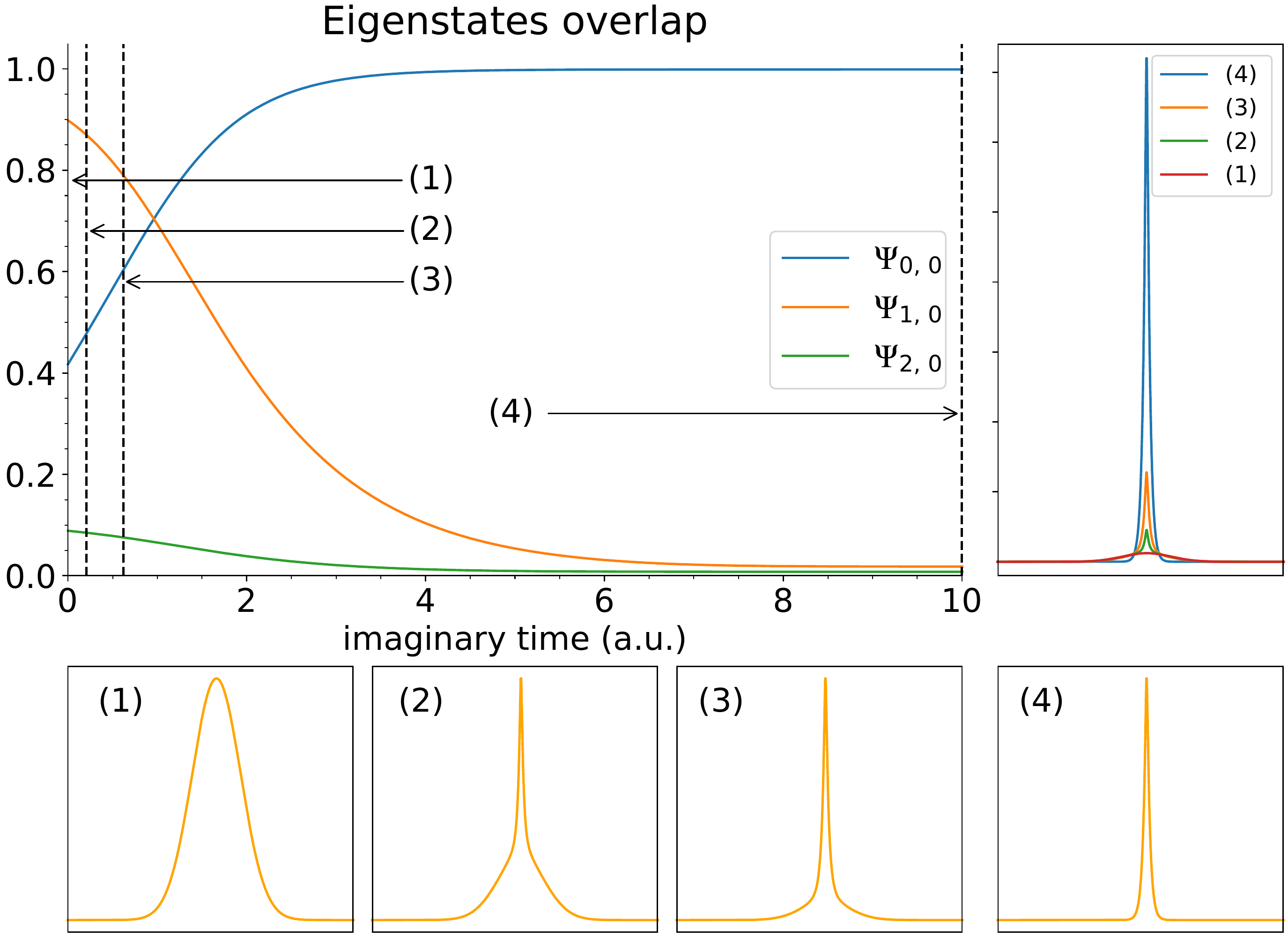}
    \caption{\textbf{Preparing the ground state of 2D hydrogen using the PITE technique.} The method was emulated on a $1+2\times10$-qubit quantum computer. Post-select the successful outcome at all times. The main plot shows the overlap of the propagated state with analytic eigenstates. Bottom panels are scaled cross-sections of the electron probability density sampled at labelled points during the imaginary time evolution. Top right panel shows the same probability densities plotted on the same scale.
}
\label{fig:statePreByPITE}
\end{figure}

We then propagate the state using PITE for 300,000 steps, using $m_0=0.9$ and $\delta t=2\times 10^{-5}$ (note the actual imaginary-time $\delta\tau$ rescales $\delta t$). Fig.~\ref{fig:statePreByPITE} shows the state evolving under the approximate imaginary-time evolution. The main plot shows the overlap of the state with the analytic eigenstates of Section~\ref{sec:analtyicStates}. Only eigenstates peaked at the origin (the equivalent of $s$-states in 3D) have large overlap with the state throughout the propagation, with the $n=1$ excited state contributing more to the initial Gaussian wavepacket than the ground state. In the long time limit, the ground state overlap approaches unity, whereas its overlap with higher energy states decays. The overlap signal does not go exactly to 1, and nor do the contributions of higher energy states go exactly to 0; this is because the state prepared is the ground state of the pixelated model Hamiltonian, which nonetheless has a high overlap with the true analytic ground state $\Psi_{0,0}$ digitised to the same spatial resolution. This disparity should vanish with higher spatial resolution.

The probability distribution of the state, plotted at the bottom, also shows that the broad initial wavepacket contracting to a sharp state peaked at the origin. Visually it would appear that at a short evolved time, the state already resembles the ground state with a singular central peak. However, superimposing the sampled distributions (top right panel), the subsequent long time evolution appears necessary to render increased sharpness.

We further assess the state prepared from PITE by subjecting it to real-time SO-QFT propagation for 4~a.u., and estimate its energy via IPE. The fidelity does not drop below $4\times10^{-8}$, and the estimated phase agrees with the converged energy at this spatial resolution reported in the previous section, further confirming that the PITE converges to the ground state of the model at this resolution.

This scenario however demonstrates clearly the main drawback of PITE: At every measurement, there is a substantial probability of measuring the undesired outcome; this is then a failure of the procedure. In the case described here that probability is about 0.33; this means the cumulative success probability falls before $10^{-4}$ after only about 23 measurement steps. 
The demonstration here, with 300,000 steps, would therefore have (essentially) zero success probability on a real quantum computer. However the method may be useful in `quantum-inspired' classical algorithms given its attractive feature of not requiring {\it a priori} knowledge of the states. Moreover the authors of Ref.~\cite{kosugi2021probabilistic} suggest that amplitude amplification methods might address the issue of vanishing success probability.
\color{black}

\subsection{Quantum Dynamics Demonstrations}
\label{sec:efieldAndScatterResults}
\begin{figure*}[!tbp]
    \centering
\includegraphics[clip,width=1.9\columnwidth]{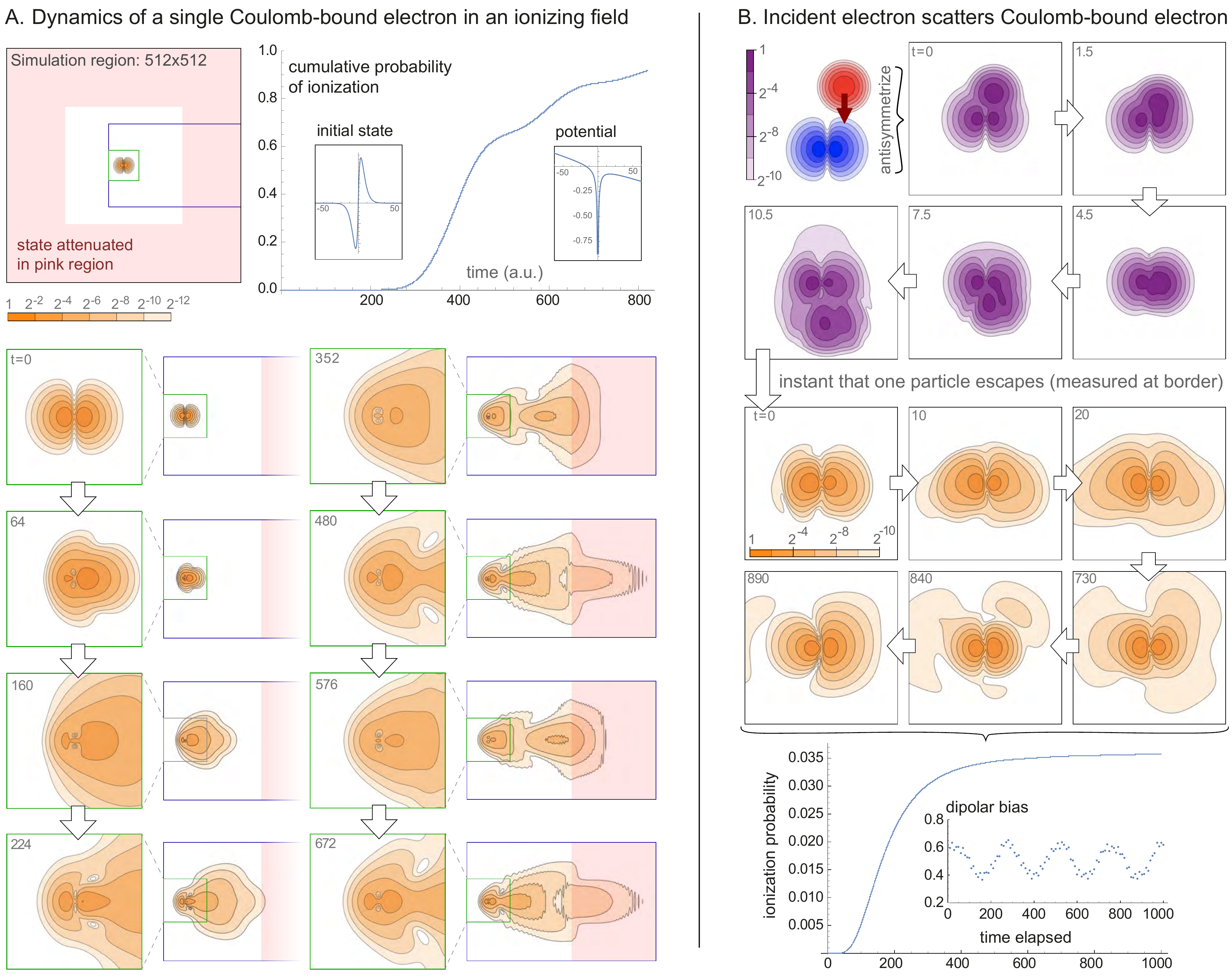}
\caption{{\bf Single- and two-electron dynamics.}
\textbf{(A)}~Figures showing a 2D simulation of a single electron ionised by a strong electric field, executed on a $1+2\times9$-qubit quantum computer. The initial state is a low-lying bound state of 2D hydrogen ($\Psi_{1,1}+\Psi_{1,-1}$). The panel in the upper left shows the initial state as a contour plot; it occupies a small region in the centre of the entire simulation box (black boundary). The regions marked with green and blue borders are the zones into which we zoom in the lower panels, which are a series of contour plots with the time index shown in the corner of each plot. The pink region corresponds to attenuation (a complex potential). The total probability of the particle having been found in the outer, pink region is shown by the upper right plot.
\textbf{(B)}~2D simulation of two-particle scattering using a $25$-qubit emulated quantum computer. The initial state is the antisymmetrised version of the following: an electron (blue) in a low-lying bound state of 2D Hydrogen while a second electron (red) is in a Gaussian state and moving in the negative $y$-direction (downward in the plots). The simulation reveals the scattering event (purple plots) until we deem one particle to have exited. A second phase of simulation shows the evolution of the state of the remaining electron (orange) which has a probability of about $3.5\%$ of ultimately ionising (escaping the nuclear potential). Interestingly, in the event that the remaining electron does not ionise, it nevertheless remains in an state where it oscillates back and forth; the inset plot shows the quantity $\langle x_L \rangle>$ as defined in the text. 
}
\label{fig:FieldAndScattering}
\end{figure*}

We describe two studies which are proof-of-concept real-space grid simulations relevant to the two scenarios that we described in the Introduction: ionisation by strong external field, and electron-electron scattering. The corresponding data are shown in Fig.\,\ref{fig:FieldAndScattering}. In both these studies, we employ a method of amplitude attenuation via weak measurements, which we developed as an analogue of the complex absorbing potentials used in modelling with non-quantum computers. Our method allows one to track the rate at which particle(s) exit the simulation box and prevents them becoming incident due to the periodic boundary conditions; it is explained in Sec. \ref{sec:AttenAndScatIntro}.

\subsubsection{Scenario I: Electric field ionisation}

In panel (A) of Fig.\,\ref{fig:FieldAndScattering} we show the performance of a $19$ qubit emulated quantum computer, modelling a single 2D particle ($9+9$ qubits represent the state, one qubit is used for the weak measurements). The modelled system at $t=0$ is a state within the first excited manifold of 2D hydrogen, specifically 
\[
\frac{1}{\sqrt{2}}(\Psi_{1,1}+\Psi_{1,-1}).
\]
An additional $E_x$ term in the Hamiltonian corresponds to a strong, static electric field applied in the horizontal direction; combination of the Coulomb potential and the electric field is shown the Figure inset. Because of the electric component, the initial state is no longer an eigenstate, and the simulation can determine whether the electron will indeed be removed from the nucleus. 

The initial state occupies only a small central region in the simulation box. The contour plots in the main part of the figure show the evolution, both in the centre part of the simulation box (the box bordered in green) and a larger region encompassing the centre and the region to the right (the box bordered in blue). The pink region, constituting the outermost $50\%$ of the simulation box in both the $x$ and $y$ directions, is the region that is monitored by weak measurements. 

The figure also shows the cumulative probability that the particle will indeed have `escaped' i.e. that it will have been measured to be in the pink region. The curve ultimately approaches unity, indicating that the particle will eventually ionise with certainty. Interestingly, we observe oscillations in this curve, which we can account for by examining the contour plots of particle density shown below. 

Note that the contour plots show the particle's probability density post-selected on it not having yet escaped; therefore the normalisation is unity in each case. Focusing on the green boxes, those zoomed in close to the nucleus, we observe that the part of the state that remains close to the nuclear core actually oscillates in a dipole-like fashion. Note that the sequence of green panels, labelled $352$ to $672$, exhibit this: panel $352$ is similar to $576$ while $480$ is similar to $672$. Moreover, examining the corresponding blue regions we observe waves of density propagating away from the nucleus, synchronised with the dipole oscillation; whenever the oscillation favours the `down stream' right side, there is an enhanced probability that the particle will escape; in due course this leads to a fluctuation in the probability of observing the particle in the pink, attenuating region. It is interesting to note that this fluctuation probability is reminiscent of the bond-breaking of sodium iodide observed with femto-second pulsed lasers~\cite{Zewail1990}, an experiment recognised by the Nobel Prize in Chemistry in 1999. 

On a real quantum processor, the contour plots of Fig.\,\ref{fig:FieldAndScattering} can be obtained by repeated sampling, simply by measuring the state of the particle's register at a given time $t$. Obtaining these outputs through `brute force' sampling would obviously represent a multiplicative cost depending on the accuracy with which we require the plots. 
The plot of the cumulative probability can also be obtained by repeatedly executing the simulation; however, it only requires measurement on a single ancilla, and directly produces particle location data that can be used to determine, for example, rates in a chemical reaction dynamics simulation. We argue therefore that this approach is more useful for studying real chemistry problems on early fault-tolerant quantum computers than direct state sampling.

\subsubsection{Scenario II: Two-Particle Scattering}

 Panel (B) of Fig.\,\ref{fig:FieldAndScattering} shows the results of a two-particle scattering simulation. There are two simulation phases. In the initial phase, we have the both interacting particles present: an electron initially in a bound state of 2D hydrogen corresponding to $(\Psi_{1,1}+\Psi_{1,-1})/\sqrt{2}$, and an incident electron in a Gaussian state (but with the total state properly antisymmetrised). This simulation runs until one of the particles is measured to have `escaped' our simulation box. We then proceed to a second phase of simulation where we study the dynamics of the surviving particle. We find that it has a small ($\sim 3.5\%$) probability of ionising due to the perturbation of the prior `impact' with the incident particle.

In the first phase we use a $25=1+4\times 6$ qubit emulated quantum processor, where the $x$ and $y$ coordinates of each particle are represented with $2^6=64$ states. As in the case of the electric field ionisation described, we monitor with weak measurement a set of spatial pixels near the boundary of the simulation box. In the initial phase of this two-particle simulation the width of that region is $25\%$ rather than the $50\%$ width used in the electric field case. The contour plot panels in the figure show the central, non-attenuated region. The state is antisymmetrised so that the probability density plots do not distinguish one particle from the other, but informally we can say that the incident particle interacts with the bound particle before passing on, away from the nucleus. In the case shown in the figure, we deem that a particle has been detected in the attenuation region exactly at the point when the cumulative probability of detection reaches $50\%$ (this is an arbitrary choice; in with a real quantum processor the user would of course be unable to specify this). This event occurs shortly after the last of the panels in the upper part of the figure. 

The simulation to that point would not teach us much about the nature of the scattering event. We could in principle measure, e.g, any deflection in the trajectory of the incident particle, and we can confirm (from longer simulations) that there is near-$100\%$ probability of a particle exiting the simulation region; the incident particle does not become bound. However, it is more interesting to study the subsequent behaviour of the remaining particle. Because this particle is represented by a register in the quantum processor that has not `collapsed', we can simply continue to simulate its evolution; the component $U_{V}$ in our SO-QFT cycle, corresponding to particle-particle interaction, will no longer be applied. We can now choose to vary other parameters such as $\delta t$, anticipating that further dynamics are on a slower timescale since the high energy particle has exited the simulation. Moreover, we can reallocate some of the qubits that were previously used to model the now-exited particle, re-purposing them to model a larger simulation box. Given that the outer regions of such a box have zero amplitude associated with them at the moment the first particle exits, there is no difficulty in simply introducing those qubits. (In the case that we employ two's complement, so that the coordinate $(0,0)$ is in the centre of the simulation box, we simply append the new qubits to the high-order end of each sub-register and perform a CNOT gate on each new qubit controlled by the prior highest-order qubit.) This was indeed performed in the simulation shown in the right panel of Fig.\,\ref{fig:FieldAndScattering}, and the second phase of the simulation uses $1+2\times 8$ qubits to provide a much larger attenuation region.

In this second phase we observe that the remaining particle has been perturbed by the passage of the incident particle: its distribution at $t=0$ (now measuring time relative to the exit of the other particle) is noticeably more irregular than the simple symmetric initial form. The probability distribution is lopsided, favouring the left side. Over the remaining period of simulation, the particle exhibits mild dipolar oscillation between a left-favoured and a right-favoured distribution. Defining $P_\text{left}(t)$ as the probability that the particle would be found left-of-centre, we observe the oscillation shown in the inset to the time-series plot in the right panel of Fig.\,\ref{fig:FieldAndScattering}. Moreover, as the particle oscillates it sheds probability -- i.e. there is a finite probability that the particle will escape the simulation region. In contrast to the electric field simulation, this probability is shed symmetrically (both left and right) and it converges to a small cumulative probability of about $3.5\%$ (see main plot).

\subsection{Augmented Split-Operator}
\label{sec:ASOresults}
The concept of the ASO is described in Section\,\ref{sec:introASO}. The intent is to optimise the fidelity of simulation without resorting to very high spatial or temporal resolutions, by introducing additional elements to the basic SO-QFT cycle. We assess the method by using it to simulate the dynamics of states peaked at the singularity, i.e. the most challenging cases. 

\begin{figure}[!tbp]
    \centering
\includegraphics[clip,width=\columnwidth]{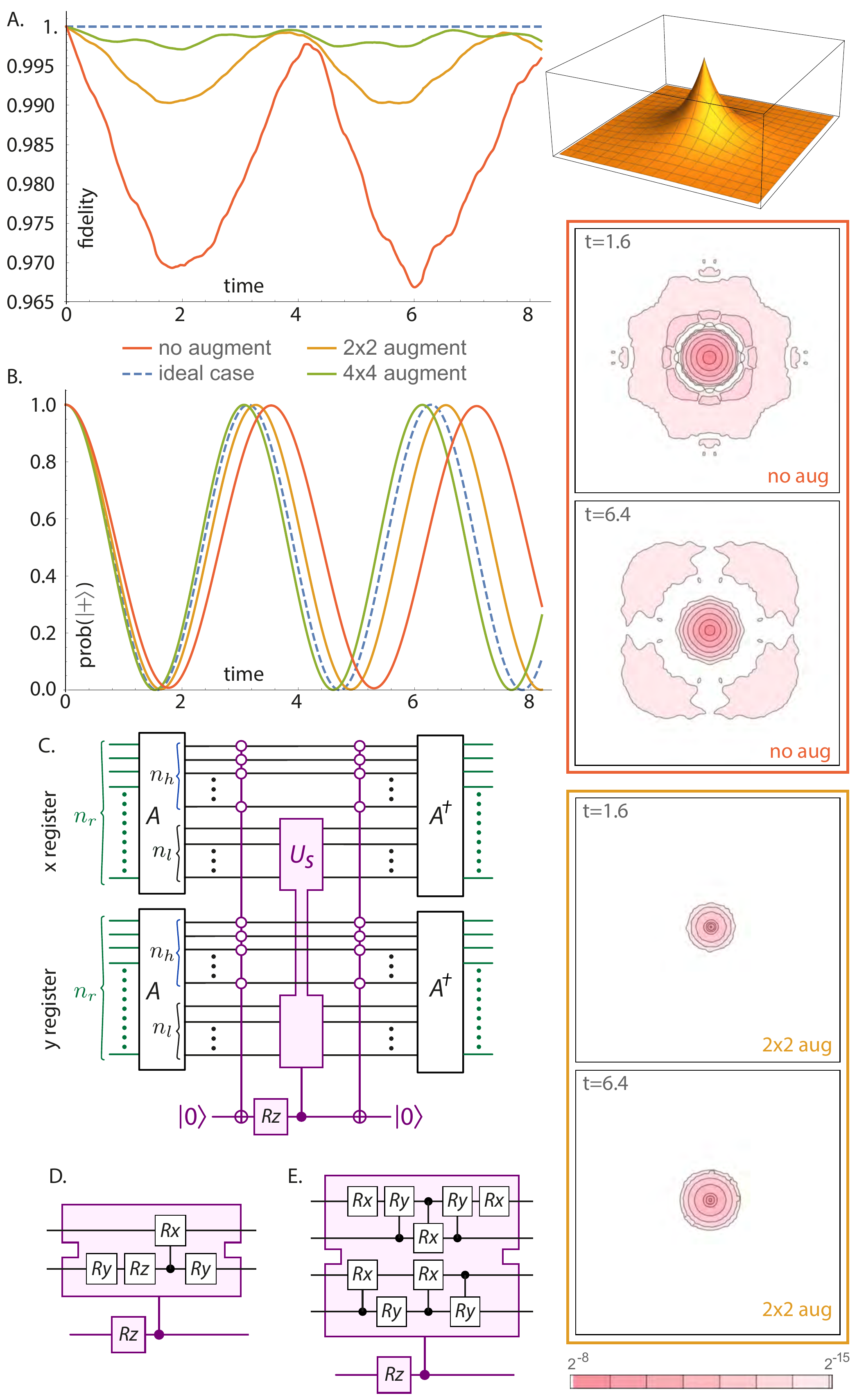}
\caption{{\bf The performance of the ASO technique.} The emulated quantum computer has $13=1+6\times 6$ qubits. \textbf{(A)} The 3D inset depicts eigenstate $\Psi_{0,0}$ of 2D hydrogen, within its simulation box. The graph shows the autocorrelation of the state at time $t$ with respect to the initial state, which ideally remains at unity (blue dashed line). The red, orange and green lines are respectively the cases of no augmentation, a $2\times 2$ augmentation and a $4\times 4$ augmentation.  \textbf{(B)} The result of phase estimation for the same three cases, with the ideal again shown with a blue dashed line. The contour plots on the right show the absolute difference in probability density, with respect to the initial state, for the case of no augmentation (red) and the $2\times 2$ augmentation (orange).
\textbf{(C)} Shows generic circuit employed, while \textbf{(D)} and \textbf{(E)} specify the particular circuits used for our $2\times 2$ and $4\times 4$ augmentations, respectively. The $A$ operators simply increment the indexing of the spatial `pixels' so that the lower right pixel is (0,0). In the case of the $2\times 2$ augmentation, the increment used is $G=1$ so that $A$ maps the indices of the pixels of interest, i.e. $(-1,-1),(-1,0),(0,-1),(0,0)$, to $(0,0),(0,1),(1,0),(1,1)$ respectively. These indices are now exactly those states for which the most-important $n_r-1$ qubits are zero, facilitating the application of the small augmentation circuit to {\it only} the four target states. 
}
\label{fig:coreStabFigs}
\end{figure}

In Fig.\,\ref{fig:coreStabFigs} we present numerical results from our study of the ASO method. We consider the ground state of 2D hydrogen $\Psi_{0,0}$ which is peaked (with discontinuous gradient) at the origin of the classical Coulomb potential. We use a relatively modest resolution corresponding to $n_r=6$ qubits per register (i.e. a $64\times 64$ grid of spatial pixels) to represent the state, and set the simulation box to be optimal (see upper right graphic in the Figure). 

Step~\ref{ASO:UcoreStep} in Section\,\ref{sec:introASO} states that, having decided that our core patch will involve a subspace of $Q$ pixels, we `derive a small $Q\times Q$ unitary $U_\text{core}$ that closely matches $U_\text{repair}$ in that subspace'. The operator $U_\text{repair}$ is simply the matrix that maps from the SO cycle as actually applied, to the ideal time increment operator. We therefore begin by calculating these matrices, each an $2^{2n_r}\times2^{2n_r}$ object, explicitly using Mathematica. Having thus obtained $U_\text{repair}$ we can proceed.

In simulations reported in Fig.\,\ref{fig:coreStabFigs} we consider two cases: core patches of size $Q=2\times2$ and $Q=4\times4$ pixels. For each case, we write down the $Q\times Q$ matrix $M_\text{core}$ which is composed of the elements of $U_\text{repair}$ that lie in the $Q$-pixel subspace. $M_\text{core}$ will not be unitary and therefore cannot be implemented deterministically on our emulated quantum computer; we need a unitary matrix $U_\text{core}$ that is close to $M_\text{core}$. This was obtained by performing a standard singular value decomposition of $M_\text{core}$ and using the components to construct $U_\text{core}$:
\[
M_\text{core}=u\,\Sigma\,v^\dagger\ \ \ \ \text{then}\ \ \ \ U_\text{core}\equiv u\,v^\dagger.
\]
Note that $\Sigma$ matrix would be the identity if  $M_\text{core}$ were unitary; it is not, but by simply omitting $\Sigma$ we generate our unitary approximation. The final step~\ref{ASO:circSynthesis} of finding a circuit to implement the stabilisation was performed using the circuit synthesis tool described in Ref.~\cite{Meister2022}; for the sizes used here, a trivial task.

We perform a series of standard and ASO simulations, in all cases fixing the time resolution at $\delta t=0.004$, and monitor the absolute value of the autocorrelation. Graph (A) in Fig.\,\ref{fig:coreStabFigs} shows the result for three cases: The simple SO-QFT protocol (red), and protocols with a small (orange) and a medium (green) scale core stabilization augmentation. The small augmentation involves a circuit that modifies only the amplitudes associated with the $2\times 2$ spatial pixels that are closest to the singularity (note that we align the spatial pixel lattice such that the singularity is mid-way between the four central pixels). The medium augmentation involves a larger set of $4\times 4$ spatial pixels. We observe that there is a dramatic improvement in the autocorrelation, by an order of magnitude between the red and green plots. 

The plots in graph (B) of Fig.\,\ref{fig:coreStabFigs} show the result of the single-qubit phase-estimation method described in Section~\ref{sec:observablesAndPE}. Ideally, the autocorrelation plot would match the dashed blue curve, which corresponds to phase acquisition  according to the analytically derived energy. We observe that the red, orange and green lines are again progressively closer to the ideal.  We should emphasise that the ASO method is agnostic to the state simulated; therefore, while it would be trivial to `cheat' and apply an exactly compensating phase to obtain the blue dashed curve, in fact the circuits we have employed are derived without foreknowledge of the specific simulation task (i.e. $\Psi_\text{0,0}$ in no way enters the derivation of $U_\text{core}$).

Finally, a third lens on the simulation fidelity is provided by evaluating the difference between the $t=0$ probability density (over the grid of $64^2$ spatial pixels) and the density at a later time. Ideally, of course, this difference would be zero. In the contour plots on the right of Fig.\,\ref{fig:coreStabFigs}, we contrast the case with no augmentation with the small augmentation scenario. While there is still a discrepancy, it is far more localised and stable (the un-augmented simulation involves wider, more dramatic fluctuations in the probability distribution).  

The circuits that implement both the small and the medium augmentations are shown in the bottom left of Fig.\,\ref{fig:coreStabFigs}. They were derived through the process explained in Section\,\ref{sec:introASO}. Note the preparatory step of applying a unitary $A$ that simply adds an integer to all states in the spatial representation, i.e.
\begin{equation}
A\ket{n}=\ket{n+G}
\nonumber
\end{equation}
where the addition is understood to be modulo $2^{n_r}$. The main figure shows the implementation of this for a general `size' of the augmentation patch $n_l$, where in our numerical studies $n_l=1$ for the small case and $n_l=2$ for the medium case. The figure caption defines the shift for the case of the $2\times 2$ augmentation, where we will use $G=1$. The $A$ operators can be avoided entirely if we simply define the origin of our spatial coordinates to a bounding corner of our core stabilisation region. The implementation of the spatial part of the SO would need to be corrected for such shift, but that may prove to have lower total cost. Regardless, we include the $A$ operators in the figure so as to make the it directly consistent with the other circuits and expressions in the present paper. 

We note that the multi-controlled \small{NOT} operation appearing in the circuits of Fig.\ref{fig:coreStabFigs} can be compactly realised by recently discovered circuits~\cite{gidney2021CCP} that involve $4n-6$ T-gates (single-qubit phase gates) and a comparable number of control-\small{NOT}s, together with an ancilla that is measured during the process.

We anticipate that the time cost of moving from the simple SO-QFT approach to the ASO method should be far less than the cost of the `brute force' increase to the temporal resolution needed to assure proper behaviour of core-peaked states (Section~\ref{sec:introResolution}). For the present demonstration, the performance of the $n_r=6$ qubit registers was able to approximately match that of the $n_r=11$ qubit registers under a `brute force' approach (Fig.\,\ref{fig:resolutionWF}, upper left panel) and moreover this was achieved with a time step $\delta t$ an order of magnitude greater, so facilitating more rapid runtimes. The ASO method is therefore highly relevant when one operates with the strict Coulomb interaction, as in all the numerical studies in the present paper. It remains to be seen if it would also be useful in scenarios where the Coulomb singularity is approximated by some of the other means listed in Section~\ref{sec:introResolution}. 
The ASO method is distinct from, and compatible with, the use of higher order Trotter sequences. Throughout the paper we restrict to the lowest order Trotter pattern, but extending this is an interesting direction for further work.

\subsection{3D Helium simulation}
\label{sec:3Dhelium}
Finally, we extend the low-dimensional demonstrations to simulating the dynamics of a helium atom: two electrons interacting via a repulsive Coulomb interaction, both bound by a central attractive Coulomb potential, in three spatial dimensions. As the true electron eigenstates of the helium atom cannot be solved exactly, we approximate the 2-electron initial state by combining two single-electron solutions of the 3D hydrogen-like Schr\"odinger equation (Eqn.\,\ref{eqn:3Dhydrogen}), with a central nuclear charge of $Z=2$. Note that this would be an exact eigenstate if there were no electron-electron interaction. The two sets of quantum numbers $(n, l, m)$ we used were $(2, 1, 0)$ and $(2, 1, -1)$, of which the former is the $2p_z$-orbital and the latter is the atomic orbital $2p_{-1} = (2p_x - i\,2p_y)/\sqrt{2}$. The complete initial (triplet) state is then the antisymmetric wave function
\begin{multline}
 \Psi_\text{init}(\vec{r}_1, \vec{r}_2)=\\
 \Psi_{2,1,0}(\vec{r}_1)\,\Psi_{2,1,-1}(\vec{r}_2) - \Psi_{2,1,-1}(\vec{r}_1)\,\Psi_{2,1,0}(\vec{r}_2).   
\end{multline}
As our simulation box we use a cube with sides lengths of 25\,a.u. and discretise the initial function $\Psi_\text{init}$ as per Eqn\,\ref{eqn:3DrealSpacePparticles} using $n_r=6$ qubits per register providing 64 divisions per axis and per particle. We then propagate the full $36$ qubit state forwards in time for 500 steps, where each step is of length 0.05\,a.u., thus a total evolution of 25\,a.u. It is worth noting that this relatively simple initial state has symmetries that could be exploited for a more compact representation; but we wished to test the full 3D, two-particle grid representation and therefore we did not exploit any such properties. 

\begin{figure}[!tbp]
    \centering
\includegraphics[clip,width=0.95\columnwidth]{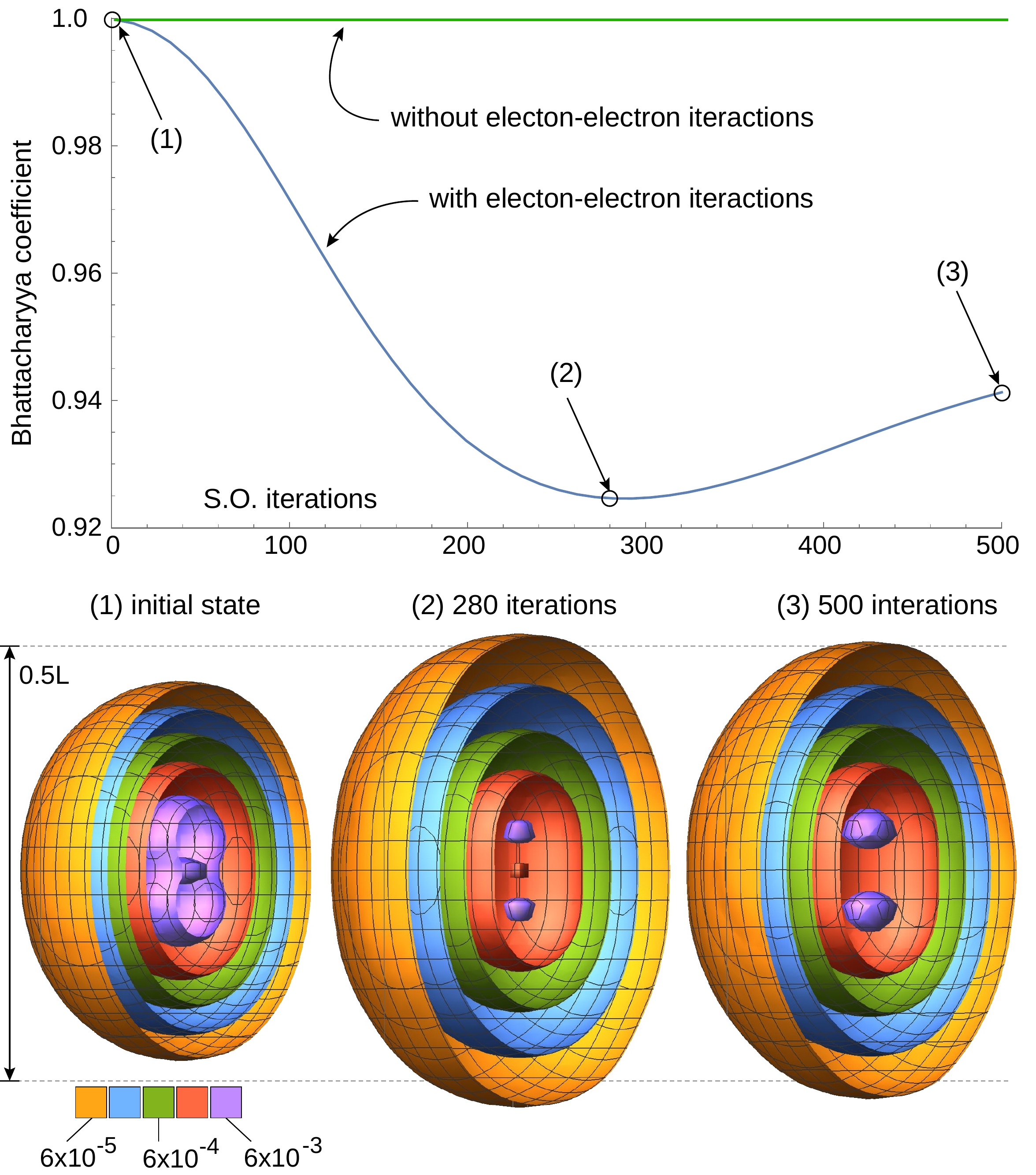}
\caption{\textbf{Simulation of a Helium atom in real space.} The top plot is the Bhattacharyya coefficient (with and without enabled electron-electron interactions), and the bottom is the real-space electron density distributions during the time evolution of the Helium atom simulation. Coloured shells are electron probability density isosurfaces, within a simulation box with $L=25$\,a.u. Distributions are shown for the initial state, at the time where it is maximally spread out, and at the end of the simulation. The 500 SO cycles correspond to propagation of 25~a.u. ($\approx0.6$~fs).
}
\label{fig:3dHe}
\end{figure}

The single-ancilla phase estimation method was used in previous sections (see e.g. Fig.\,\ref{fig:statePreByMeasure}) to track the evolution of the simulated system's state. However this requires our emulator to use double the memory that would be needed simply to represent and propagate the state. Since the resource costs for the present emulation are already considerable, we opted instead to compute and record, at every SO-QFT time step, the probability density of one of the electrons (since the two electrons are indistinguishable, the probability density is equal). Specifically, we record the probability associated with each computational basis state of the three registers corresponding to one of the particles. Using these probability distributions we can compute the Bhattacharyya coefficient~\cite{bhattacharyya1943measure} of the distribution at time $t$ with respect to the distribution of the initial state $\Psi_\text{init}$. This quantity is 
\[
\sum_i \sqrt{p_i}\sqrt{q_i}
\]
for our two discrete probability distributions $p$ and $q$. It can be thought of as a classical analogue of the usual inner product fidelity. It is plotted in Fig.\,\ref{fig:3dHe}.

As the initial state is not an eigenstate, we expect the distribution to vary over time. As shown in Fig.~\ref{fig:3dHe}, the electron density is initially distributed with rotational symmetry around the vertical $z$-axis, with charge accumulations in the positive and negative $z$-directions. Because the electrons partly shield each other from the core, the chosen central charge of the initial state $Z=2$ is too large, and thus the electron orbitals are too close to the core. The time evolution shows that the charge initially spreads out away from the core in every direction, then returns slightly, but not to its original distribution. To confirm that the interaction between the electrons is the cause for this behaviour, we also performed an identical calculation, but with the $e$-$e$-interaction disabled. Fig.~\ref{fig:3dHe} shows that in this case the probability distribution simply stays constant. We thus confirm that our simulation directly shows the effect of electron shielding in this hypothetical configuration. 

One could repeat the experiment to find the value for the effective nuclear charge that allows our analytic initial state to most closely approximate a true helium eigenstate; moreover, one could use the methods of Section\,\ref{sec:stateprepintro} to actually prepare the eigenstate from such an initial approximation. These are interesting tasks for further study.

\subsection{Antisymmetrisation of the initial state}
\label{sec:AntiSymmResults}

In Section\,\ref{sec:stateprepintro} we discuss the preparation of an initial state of our grid-based simulator with proper antisymmetrisation of the electrons. We assume that there is some set of $P$ single-particle basis states $\psi_i$, each of which we know how to prepare on a register (possibly via a repeat-until-success probabilistic method). We note that it would be convenient to simply prepare a product state over our $P$ particle registers, and subsequently antisymmetrise it. 

We observe that one means of doing so involves first finding  a Hamiltonian $H_\text{synth}$ with the property that 
\begin{equation} \label{eqn:Hsynth}
H_\text{synth}\ket{\psi_i}=E_i\ket{\psi_i},\ \ \ \ \ \ \ E_{i+1}>E_i.    
\end{equation}
Here $H_\text{synth}$ need not correspond to any physically legitimate scenario. One means of obtaining $H_\text{synth}$ would be to start from a description of the chemical system of interest, and then introduce modifications to conveniently localise Hamiltonian terms and to break any degeneracies in the single-particle solutions. In this way, the available basis states $\ket{\psi_i}$ will be close to the canonical choice of basis states that might be made in, e.g., chemistry modelling with a conventional computer. While this is an interesting topic to consider further, we will not do so in the present paper but rather we will simply assume that $H_\text{synth}$ can be found.

Given that $H_\text{synth}$ is synthetic, it can of course be scaled and shifted arbitrarily. We will align the energies conveniently with respect to the binary states that can be represented by a register of $t$ bits. For example we can set $E_0=0$ and $E_{P-1}=2^t-1$ where $t$ is sufficiently large that the smallest energy gap $\min({E_{i+1}-E_i})$ is at least unity. 

We can proceed in at least two distinct ways, and we will describe the first in detail. It requires a modest number of additional qubits: For each of the $P$ particle registers, introduce an `tag' ancilla register of $t$ qubits.
\begin{enumerate}
\item Prepare the particle registers as $\ket{\psi_0}\ket{\psi_1}...\ket{\psi_{P-1}}$.
\item Set the $i^\text{th}$ tag register to the the integer closest to $E_i$, which we write as $E^\prime_i$.
\item Permute the entire object (the particle registers and their tag registers) into antisymmetric form by any means that can permute an initially-ordered list of integers; for example, the inverse sorting network method of Ref.~\cite{berry2018improved}. We simply apply this method to the tags while `dragging along' the particle registers `for the ride'.
The result is the state 
\begin{equation}
\frac{1}{\sqrt{P!}}\sum_i^{P!} \text{perm}_i\big((\ket{E_0^\prime}\ket{\psi_0})...(\ket{E_{P-1}^\prime}\ket{\psi_{P-1}})\big)
\label{eqn:permedWithTags}
\end{equation}
with the notation $(\ket{\text{tag}}\ket{\text{particle state}})$.
\item Erase the tag registers through phase estimation: apply a QFT to every tag register, and then apply operations of the form controlled-$\exp(i H_\text{sym}C\pi)$ from each tag register qubit to its main register, where $C$ is an appropriate power of two. 
\item \label{finalStepAS} Measure out the tag register qubits in the $x$ basis; with high probability, all should be in the $\ket{+}$ state and this outcome is our success criterion.
\end{enumerate}

If all energies $E_i$ can be exactly represented with $t$ bits, then $E_i^\prime=E_i$ for all $i$ and the method will succeed with certainty. Moreover, even in the in the event that the energies cannot be perfectly represented in this binary form (as will be the case, for example, if two energies differ by an irrational number) then the only consequence is that success criterion in Step\,\ref{finalStepAS} will have a reduced probability of occurring. Given that it does occur, the final state remains ideal. 

We explored this using classical emulation, on inputs of up to $P=5$ particles. In this case, $30$ qubits were used: $t=3$ tag qubits for each particle, and a further $3$ qubits to represent each corresponding $\psi$ state. We considered several choices of spectrum $E_i$ and proceeded in each case as follows. We prepared the state given in Eqn.\,\ref{eqn:permedWithTags} directly in the emulator's RAM: it is readily verifiable by inspection that the first three steps indeed lead to this state regardless of the choice of $E_i$, whereas the additional qubits required to permute via the sorting network would have greatly increased the emulation cost. We then performed the final steps explicitly and noted the performance. 

For $P=5$ particles, with the ideal case that all energies correspond to integers, we confirmed the expected success probability of unity. For cases where the all energies differed from an integer, we found success probabilities of $0.990$, $0.960$, and $0.850$ for deviations of $0.025$, $0.05$ and $0.1$ respectively, showing an anticipated quadratic behaviour. However, when only one of the energies differs from an integer by some specified discrepancy, then we find that the success probability is essentially constant regardless of $P$. Thus the performance for a large number of particles will depend on how many of the energies differ from an integer, and to what extent (with respect to $2^{-t}$). We note that in each emulated scenario, we confirmed that given the `all-$\ket{+}$' success criterion, the resulting state is ideal (up to a meaningless global phase).

In the Supplementary Material we outline the alternative method, which is similar but allows one to create and destroy the tag register on the fly, thus reducing the qubit count while increasing the computation time.  

\color{black}


\subsection{Quantum computer resources and architectures}
\label{sec:postClassicalSims}
In the following two subsections we assess on the resource requirements for undertaking modelling beyond the reach of classical algorithms, and the related question of suitable quantum architectures. 

\subsubsection{Resources for post-classical modelling}
In this section we reflect upon the implications of our numerical results for the resource demands of post-classical chemistry modelling. We will not undertake a formal resource scaling analysis, noting instead that asymptotic analyses have been made in the last few years for real-space grid first-quantized Hamiltonian simulation~\cite{Cody_Jones_2012, Kivlichan_2017, su2021fault} and specifically the SO-QFT method~\cite{Childs2022}.
Complementing these analyses, our present work implemented grid-based simulations using emulated quantum computers that have proven large enough to elucidate practical issues, such as the number of basis functions required for given levels of accuracy in specific observables. Data of this kind will be of use in estimating the constants that must appear in any resource scaling analysis. 

The lowest possible qubit count (for a given simulation  accuracy) results from selecting  the  smallest adequate ‘simulation box’ length $L$, and setting the spatial resolution $\delta r$ (Eqn.\,\ref{eqn_resolution}) to be just sufficient to capture the most curved elements of the wavefunction (see Section~\ref{sec:introResolution}). The number of qubits per particle, per spatial dimension, is then $n_r=\log_2(L/\delta r)$. The quantity $L$ simply specifies the region of space outside of which our multi-particle wavefunction only has negligible amplitude `clipped' by the boundary. Moreover as we explain in the attenuation Section~\ref{sec:AttenAndScatIntro} we can study processes that would, over the simulation time, go beyond the simulation box due to scattering or ionisation. It is reasonable to assume, as in Ref.~\cite{kosugi2021probabilistic} for example, that the Volume $L^3$ goes linearly with the number of particles $P$; a molecule with twice the number of particles requires (of order) twice the volume. Meanwhile, the severity of the wavefunction's curvature should scale directly with $Z_\text{max}$, the maximum nuclear charge of any of the nuclei in the system (Section~\ref{sec:introResolution}). Thus, increasing the molecule's size without increasing $Z_\text{max}$ should not require any adjustment in $\delta r$, but the simulation box may have to be larger to accommodate more particles. Importantly, the accuracy with which we model the particle is essentially unaffected by this change.

In view of the above observations we can expect that, for suitable constants $C_i$,
\begin{eqnarray}
n_r=\log_2\left(\frac{L}{\delta r}\right)&\approx&\log_2
\left(\frac{C_1P^\frac{1}{3}}{C_2 Z^{-1}_\text{max}}\right)\nonumber\\
&=&C_3+\log_2(Z_\text{max})+\frac{1}{3}\log_2(P).\ \ \ \ 
\label{eqn:scaling}
\end{eqnarray}
The total qubit count will be $3Pn_r$ for $P$ particles in 3D. Note that this simple expression for $n_r$ does not account for the potentially helpful fact that atomic radii have a highly non-linear (and sub-linear) dependence on the number of electrons~\cite{covalentRadii}. In the Supplementary Material, we make an estimate of the root term $C_3$ from our numerical results; we argue that $C_3 \sim 10$ is optimistic but not unreasonable.

Using some further assumptions we can make an estimate of the number of qubits necessary to model the important scenarios that we described in the Introduction: electron scattering of hexafluoro ethane (C$_2$F$_6$) and quantum coherent control of the ammonia molecule (NH$_3$).
In the Supplementary Material, we suggest that grid-based modelling of C$_2$F$_6$ may require about 2250 computational qubits, even with frugal use of ancilla qubits. Fortunately our estimate for the coherent control of NH$_3$, being one of smallest relevant cases, is far lower at $450$ qubits (with relatively optimistic assumptions). 

The overall time cost for simulation depends of course on the hardware realisation, but is certainly interesting to discuss.
As the methods we propose will almost certainly require a fault-tolerant quantum computer, the most relevant metric for time complexity on such a machine may be the T-gate count. This is a measure of how many steps in the algorithm correspond to the costly non-Clifford operations that cannot be directly performed in stabiliser codes, and efforts to minimise the count for standard subroutines are ongoing~\cite{beverland2020}. For example, a recent note has reduced the number of such gates needed for multi-controlled rotations to a remarkably frugal level\,\cite{gidney2021CCP}; such rotations are key in both our attenuation and ASO techniques. More generally, the trade-off between time and qubit count is a research topic in its own right and recent papers have shown how dramatically this balance can be adjusted~\cite{gidneyAndFowler2019,Litinski2019}.

The overall `wall clock' duration of a simulation is determined by the number of T-gates needed for each complete SO-QFT cycle, the duration each of these steps represents (the time resolution $\delta t$), and the total duration of the dynamical process under investigation. In the Supplementary Material we use various observations and assumptions to estimate the gate depth for simulating interesting chemical physics, noting that dynamical processes where quantum effects are meaningful can occur from sub-femtosecond to picosecond timescales. For simulations of rapid events such as ionisation, we estimate an algorithmic gate depth of $O(10^8)$, while for a more challenging simulation of physics over longer timescales we suggest that a depth of $O(10^{11})$ may be required.

Translating gate count to execution time will vary dramatically depending on the native physical gate error rate (typically assumed to be $10^{-3}$ or $10^{-4}$), the speed of a stabiliser cycle and the option to trade computation speed for higher qubit overheads~\cite{Litinski2019}. For surface code based implementations with solid state platforms, a credible stabiliser speed~\cite{PhysRevA.86.032324} is 1$\mu$s (with faster speeds being conceivable).
Assuming the state preparation procedure only requires a polynomial-scaling overhead, and is thus not the dominant cost, this gives a clock time of the order of minutes for the more simple simulations, while the challenging long-timescale process would require on the order of a day.

\color{black}
We must also note that generating certain interesting plots (equivalent to those shown in this paper) will require many repeated executions of the simulation. Fortunately such repetitions can be perfectly parallelised over independent quantum processors, which need not have quantum interlinks or even be co-located. 

The same full-dimensional real-space grid simulation of the reaction on classical machines of course is not possible, simply because the sheer number of pixels in the simulation is beyond memory available on most high-performance computer clusters. An equivalent classical procedure of selecting reduced reaction coordinates where the dynamics is expected to be relevant, computing the corresponding electronic potential energy surfaces, and subsequent dynamics propagation, can require months of effort; finalising the electronic structure itself can often be the main bottleneck. We therefore conclude that quantum Hamiltonian simulation approach presented can be more efficient than an equivalent classical method, based on the fact that a more complete picture is used and the effort of computing potential energy surfaces is circumvented.

\color{black}

\subsubsection{Quantum computer architectures}\label{sec:architectures}
The preceding Section obtained back-of-the-envelope estimates of qubit counts ranging from several hundred to several thousand. While this sounds encouraging, we note the very likely need for fault tolerance and the resulting multiplicative increase in physical qubit count. Presently, the most well-understood codes can require many hundreds of physical qubits per logical qubit, assuming relatively deep algorithms and physical error rates comparable to today's best QC prototypes. Even if one makes the very optimistic assumption that some form of error mitigation can suffice in place of full fault tolerance, at least for small molecular simulations, the more powerful forms of mitigation can require a multiplicative increase in the number of physical qubits~\cite{Koczor2021, PhysRevX.11.041036}. Thus the number of physical qubits required for the modelling considered in the preceding section could easily reach the high thousands or millions. This raises the question of what kind of architecture would be needed. 

\begin{figure}[!tbp]
    \centering
\includegraphics[clip,width=1.0\columnwidth]{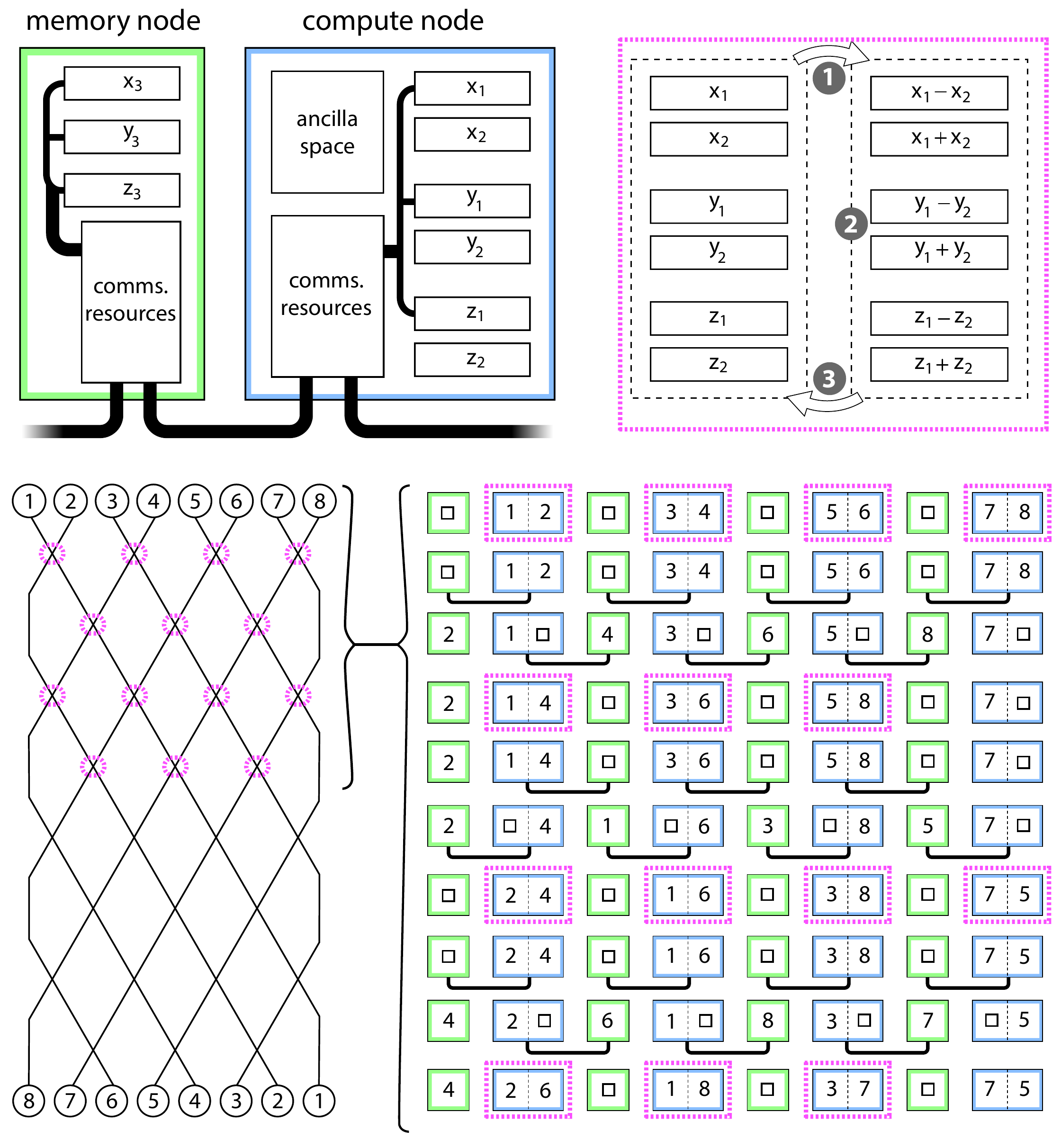}
\caption{\textbf{A possible `multicore' architecture.} Based on distributing the $P$-particle state over $P/2$ memory nodes ($3n_r$ qubits) and  $P/2$ compute nodes ($6n_r$ qubits) which are interlinked either on-chip \cite{jnane2022multicore} or at the macro-scale. The upper-right panel indicates that three steps occur in a compute node. In step 1, quantum addition and subtraction are used to move to the relative and total coordinates (and step 3 will reverse this). In the middle step 2, the relative coordinate is used to apply phase shifts required by the SO cycle, and optionally we apply an augmentation step.  The lower panels show how parallelised phases alternatingly process and permute the data. The numbers within the circles and squares are the labels of particles; they correspond to the subscripts of the $x$, $y$ and $z$ symbols in the upper-right panel.
}
\label{fig:architecture}
\end{figure}

In particular, we are motivated to explore whether some form of network architecture might be compatible with the split-operator (or ASO) approach. Such a network might involve quantum computers interlinked within a building, analogously to a conventional HPC facility, and relevant methods of linking processors have been experimentally realised.
Alternatively, for suitably compact platforms the network might correspond to linked QC processors on a single chip, analogous to today's multicore CPUs -- indeed multicore quantum computing has recently been explored~\cite{jnane2022multicore}. In either case, it is realistic to assume that in a network of processor nodes the inter-node operations are slower than the intra-node ones.

Fortunately, the SO-QFT method is quite compatible with a network paradigm; there is a natural partitioning of the problem into nodes that each contain the registers (or sub-registers) associated with a given simulated particle. While data exchange between nodes is obviously required, it need not be a dominant component of the overall resource costing even if the physical links are slow. 

In Fig.\,\ref{fig:architecture} we show one possible partitioning -- it is not the most granular, since one could assign individual sub-registers to cores, but it does strike a good balance between the intra- and inter-core operations. Note that the required connectivity between cores is merely linear and nearest-neighbour. We suppose that there are two forms of core: the compute nodes are responsible for all the processing that is associated with the SO-QFT method, while the simpler memory nodes only store registers transiently. 

Each node includes communication resources which facilitate transfer of quantum information between nodes, for example through the use of teleportation enabled by high-quality shared Bell pairs. The `comms resources' would thus correspond to Bell pair distribution, purification, and buffering. Such processes can occur independently of the main computation and simultaneously with it, and need not involve a large number of qubits; see for example the analysis in Refs.\,\cite{jnane2022multicore,Nigmatullin_2016}. 

In the scheme illustrated in Fig.\,\ref{fig:architecture} the transfer of a register from one node to another involves writing into an `empty' register where all qubits are in state $\ket{0}$. If the individual qubits are in fact encoded logical states formed of many physical qubits, then this would introduce a multiplicative factor in the Bell pair count, but it would remain linear in the register size. Ideally the `comms' hardware would be capable of generating the required Bell pairs within the time that the compute node requires for a full implementation of the particle-particle interaction component for the current pair (and any augmentation, as per Section~\ref{sec:introASO}). Given that this computation will require a gate depth of at least $n_r^2$, there is scope for a factor of $10-100$ in the relative speeds of the intra-core computations and the inter-code Bell preparation before the latter would become a bottleneck.

\section{Discussion}\label{sec:summaryAndOutlook}
In this work, we explore the SO-QFT approach on exactly-emulated qubits, and test resource-frugal techniques that facilitate augmentation and monitoring of first quantized, real-space quantum chemistry simulations.
We test known quantum techniques and others that we introduce, covering all key aspects of quantum simulation: state preparation, Hamiltonian simulation, and the extraction of physical observables. Thus we characterise the resources needed to realise a `digital experiment' of quantum molecular dynamics~\cite{McClean2021} on early fault-tolerant quantum computers. 

\smallskip
The methodologies we presented can become part of a learning/prediction cycle which augments physical experiments, providing accurate data sets for machine-learned emulators that can accelerate chemical discovery. We believe the SO-QFT method, in tandem with the resource-frugal approaches presented here, may prove itself superior to classical quantum molecular dynamics simulations relatively early in the era of fault-tolerant machines. We have already noted that the technique itself leads to robust methods for measuring observables such as phase estimation. We have also noted that we are free to check certain properties quite cheaply at any time, analogous to stabilisers of a code, and their cost of evaluation can be low even if preparing the state was not. In the PITE and attenuation cases, we have also used frequent measurement to modify the evolution of the system in a non-unitary manner.
In light of this, it is interesting to ask whether the grid method can be inherently robust to errors. It may be possible to craft a version of the algorithm where the majority of harmful errors will cause the state to fail a validation check, while less damaging errors are mitigated by a non-unitary component in the dynamics. 

While combining SO-QFT with other Hamiltonian simulation methods may lead to hybrid quantum-classical approaches for real-space simulations, the high qubit count for encoding the first-quantized grid representation and generally deep circuits will likely prevent its applicability in the NISQ era. Nevertheless in the early fault-tolerant regime small executions of these methods might offer synergies with real-space electronic structure approaches such as density functional theory (DFT); one can imagine using small exact calculations enabled by real-space quantum simulations to improve DFT functionals, or using particle densities provided by an initial DFT precomputation to inform the single particle functions that are loaded into particle registers.

A natural next step is to explore the use of multi-resolution grids as employed in e.g. MADNESS~\cite{harrison2016madness, PhysRevA.85.033403}, the highly successful classical computing program for real-space grid simulations. We are well-motivated to incorporate such ideas given that the present paper has revealed remarkable variation in resolution requirements -- even across different states of a single system. Moreover, particles within many-body systems can often be considered as localized, presenting the opportunity for frugal representations based on that locality.
However while multi-resolution grids might reduce the qubit count considerably, it would likely be at the cost of more sophisticated time propagators and basis transformations.
Thus it is important to establish whether the methods in e.g.~\cite{harrison2016madness, PhysRevA.85.033403} can indeed be translated successfully to the quantum context.

\smallskip
It is obvious that the SO-QFT simulation techniques presented here can be generalised to modelling systems beyond quantum chemistry. The solution to any Cauchy type initial value problem 
\begin{equation}
    \frac{\partial}{\partial t}\Psi(x, t) = \hat{\mathcal{D}}(t)\Psi(x, t)
\end{equation}
with a time-dependent differential operator that may be separated into operators that are respectively diagonal in position- and momentum-space
\begin{equation}
    \hat{\mathcal{D}}(t) = \hat{\mathcal{D}}_1(t) + \hat{\mathcal{D}}_2(t) 
\end{equation}
can be approximated with the SO-QFT approach~\cite{BAUKE20112454}. Many problems of interest can be modelled with Cauchy type partial differential equations.
For example, Dirac and Klein-Gordon equations, which are of Cauchy form, reconcile quantum mechanics with special relativity and may be used for modelling of high energy particles~\cite{RUF20099092}.
A very different application is financial engineering.
Quantum advantage is often promised for the modelling of how the prices of assets, such as options and derivatives, evolve over time \cite{Stamatopoulos2020, Chakrabarti2021}. This is key to executing purchases and sales that maximise the eventual payoff from trading such assets. These assets often have complex underlying dependence on random variables, which have in practice been modelled using computationally expensive stochastic Monte Carlo methods.
In the same manner as pixelating real-space wavefunctions and storing them in the computational basis states of a quantum computer, probability distributions corresponding to asset prices can be discretised and loaded into quantum registers. A very relevant model for time-propagation of these distributions is the Black-Scholes-Merton equation \cite{BlackScholes1973, Merton1973}, which is a Cauchy type partial differential equation.
Beyond these use cases, it remains to be seen how non-unitary operations achieved through ancilla measurements can extend the applications of the SO-QFT model.


\section{Methods and Materials}
\label{sec:theory}
Our Methods section is divided into two parts. The first introduces the theoretical framework, i.e. the essential physics and notation, as well as the core concepts for the grid paradigm. The second describes the specific methods explored and evaluated in Section~\ref{sec:results}, including techniques developed for this study.

\subsection{Theoretical Framework}
\label{subsec:priorTheory}
The non-relativistic time evolution of a quantum state is governed by the time-dependent Schr\"odinger equation
\begin{equation}\label{eqn:grandH}
    \frac{\partial}{\partial t}\Psi(\mathbf{r},\mathbf{s}, t) = -i\hat{H}\Psi(\mathbf{r},\mathbf{s}, t) ,
\end{equation}
where $\Psi$ is the normalised, complex-valued, many-body wavefunction defined by the spatial $\mathbf{r}$ and spin $\mathbf{s}$ coordinates of the constituent particles  (note that throughout this work we are using atomic units where $\hbar=1$). For the systems of interest here, the Hamiltonian $\hat{H}$ is:
\begin{equation}
\hat{H}_\text{tot} = \hat{H}_\text{kin}+\hat{H}_\text{int},
\end{equation}
where
\begin{equation}
    \label{eqn:kinH}
    \hat{H}_\text{kin}=-\sum_{p=1}^P \frac{1}{2m_p}\nabla_p^2
\end{equation}
represents the kinetic energy of each of the $P$ particles present, and $\hat{H}_\text{int}$ encompasses all interactions. In many cases it is convenient to further resolve according to
\begin{equation}
\hat{H}_\text{int}= \hat{H}_\text{U}+\hat{H}_\text{V} ,
\label{eqn:H_int}
\end{equation}
where $\hat{H}_\text{U}$ represents single-particle interactions with e.g. classical fields, while $\hat{H}_\text{V}$ represents particle-particle interactions. For interacting charged particles (electron-electron, nucleus-electron, nucleus-nucleus), we would write
\begin{eqnarray}
\hat{H}_\text{V}=\sum_{p,q=1; p\neq q}^P \frac{q_{p,q}}{|{\mathbf r}_p-{\mathbf r}_q |}
\label{eqn:H_V}
\end{eqnarray}
for suitable constants $q$. In the case of atomic and molecular systems, we could opt to model each nucleus as a full quantum particle in its own right, in which case we may set $\hat{H}_U=0$ unless there are external e.g. electric or magnetic fields. In their seminal paper, Kassal {\it et al.} argue that this is the natural choice given the relatively modest additional resources needed~\cite{Kassal2008}.

If the nuclei are not treated explicitly within the model, we then opt to model only the electrons and employ classical fields to represent the $M$ nuclear potentials originating at fixed locations $\mathbf R_m$ according to 
\begin{eqnarray}
\hat{H}_\text{U}&=&\sum_{m=1}^M\, \sum_{p=1}^P \frac{Z_{p,m}}{|{\mathbf r}_p-{\mathbf R}_m |}\nonumber \\
&=&\sum_{p=1}^P \frac{Z_p}{|{\mathbf r}_p|}\ \ \text{for }M=1\text{ nucleus at }{\mathbf R}_0={\mathbf 0}
\end{eqnarray}
for suitable values of nuclear charges $Z_p$ or $Z_{p,m}$. In the 2D and 3D atomic simulations we report, we consider $M=1$ nucleus as in the equation above. However the techniques generalise naturally to $M>1$. External and possibly time-dependent potentials can also be included in the Hamiltonian. Presently we will consider the case of uniform electric field $\mathbf{E}$, by including within $\hat{H}_U$ a term \[
\hat{H}_\text{E}=\sum_{p=1}^P Q_p{\mathbf r}_p\cdot\mathbf{E}.
\]


Formally, the solution to Eqn.\,\ref{eqn:grandH} is
\begin{equation}
    \Psi(\mathbf{r},\mathbf{s}, t) = e^{-i\hat{H}t}\Psi(\mathbf{r},\mathbf{s}, 0) .
\end{equation}
For the multi-electron Hamiltonian with more than two particles, it is not possible to analytically evaluate the action of the time evolution operator $e^{-i\hat{H}t}$, and one has to resort to numerical techniques to solve the initial value problem. A practical time propagation method therefore involves selecting a representation of the state and then applying (some approximation of) the time evolution operator. 

We begin by providing a brief summary of the real- and momentum-space ($k$-space) grid representations of a many-body wavefunction, suitably encoded on a quantum computer.
We then discuss the SO-QFT method for simulating Hamiltonian dynamics.
We refer the reader to the Supplementary Materials for a detailed presentation of these topics.
We highlight that this method of exploiting quantum computers, explored by earlier authors in \cite{Kassal2008, Benenti2008, somma2016quantum, Ollitrault2020, Cody_Jones_2012, kosugi2021probabilistic, su2021fault, Hirai2022, Kosugi2022_geometry}, is an adaption of the classical computing methods developed in~\cite{fleck1976time, FEIT1982, FEIT1983, KOSLOFF1986363}, which we also review in the Supplementary Materials.

\subsubsection{Representations in real- and momentum-space} \label{sec:representations}

\begin{figure*}
\includegraphics[width=0.95\textwidth]{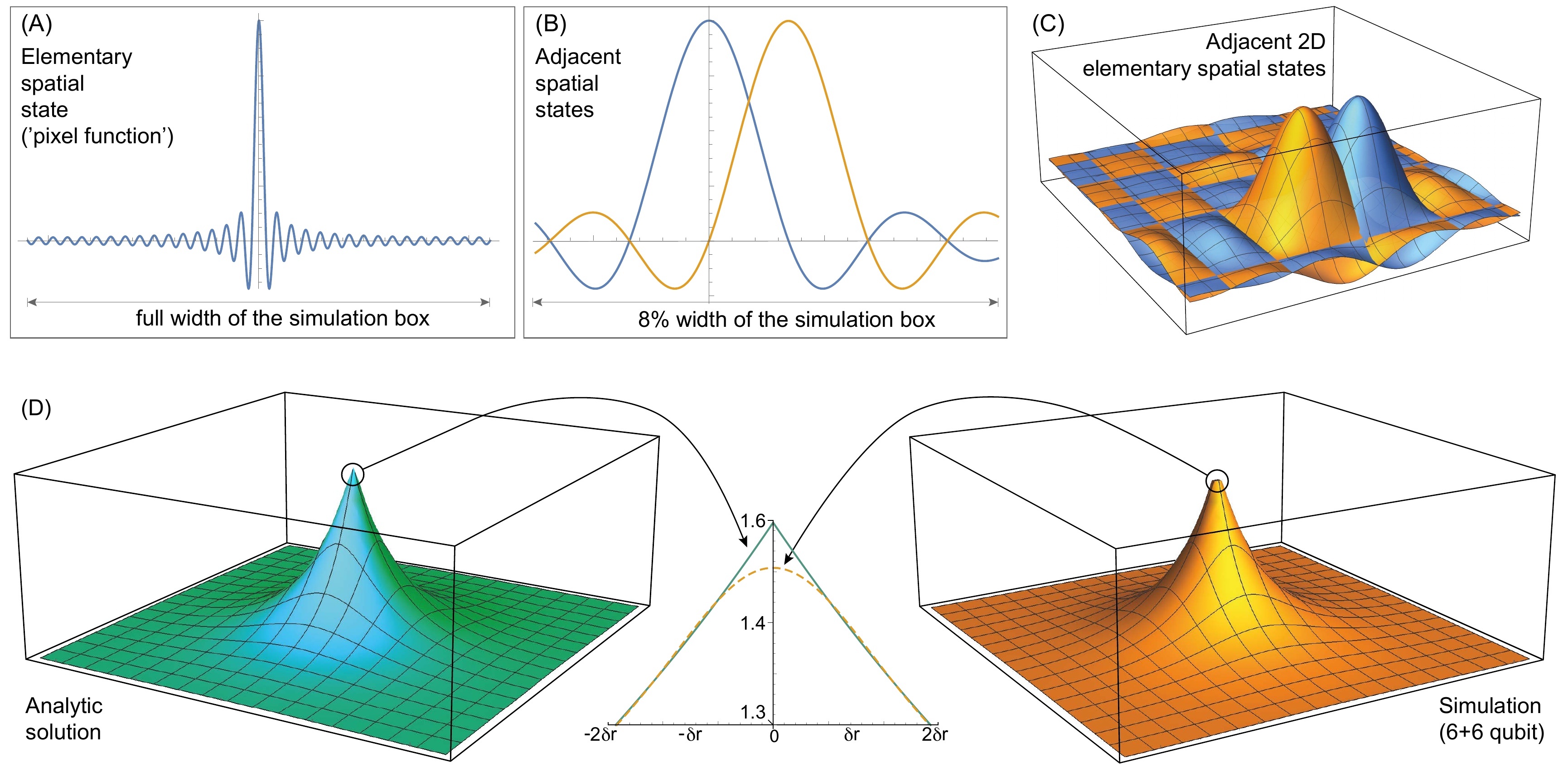}
\caption{{\bf Basis of spatial DVR wavefunctions modelled by a quantum computer.} \textbf{(A)}~A `pixel function' $P_{n_r}^n(x)$ with $n_r=6$ qubits, and $n=0$ (neglecting its complex phase, see Eqn.\,\ref{eqn:Pfunction}). \textbf{(B)}~Adjacent pixel functions $n=0$ and $n=1$; where a function has its primary peak, all others are zero. \textbf{(C)}~2D pixel functions formed from products of the 1D cases.
\textbf{(D)}~Representing a continuous state in this DVR. The left (blue-green) figure shows the exact ground state $\Psi_{0,0}$ of 2D hydrogen.
The right (orange) plot is the same state approximately modelled by a $6+6=12$ qubit quantum computer. The fidelity of the modelled state with respect to the analytic state (integrating over all space) is $0.99946$. If instead we project the analytic state into the box and renormalize, then the infidelity between the model and the analytic solution falls below $10^{-4}$. The discrepancy is a $\approx6\%$ variation localised at the Coulomb singularity, where the analytic expression has a gradient discontinuity, as highlighted by the central plot. The sharpness of the approximation is characterised by $\delta r^{-1}$ and so increases exponentially with $n_r$.}
\label{fig:kSpaceAndRealSpace}
\end{figure*}

Consider a system of $P$ quantum particles in $d$ spatial dimensions, well-localised within a region of volume $L^d$ (has negligible amplitudes beyond throughout the simulation), which we refer to as the `simulation box'. In this work, we use the approach where this system is represented on a quantum computer by partitioning the qubits into $P$ registers, and each register is further divided into $d$ spatial sub-registers each with $n_r$ qubits. Each particle is thus discretised into an evenly-spaced grid with $2^{dn_r}$ basis functions; either in a spectal, finite basis representation (FBR) or its dual pseudo-spectral basis, also called the discrete variable representation (DVR). The coefficients of the wavefunction expansion in this grid representation map directly onto the amplitudes of the computational basis~\cite{wiesner1996simulations, doi:10.1098/rspa.1998.0162, Kassal2008}. The number of qubits in the register therefore scales linearly as $O(dPn_r)$ and logarithmically with the number of grid basis functions. The favourable asymptotic scaling is one of the main advantages of first-quantized real-space grid based encoding; in the second-quantized representation, the required number of qubits scales linearly with the number of basis functions (sites or orbitals)~\cite{Kassal2008, su2021fault}.

We choose a FBR where the plane wave basis state of the modelled system is represented by a state of the computer's sub-register as follows
\begin{equation}
\phi_k(x)= L^{-\frac{1}{2}}\,\exp(\frac{i 2\pi k x}{L})\ \ \leftrightarrow\ \ 
 \ket{k}.
\label{eqn:k-element-map}
\end{equation}
Here $k$ is an integer and $\ket{k}$ refers to the computational basis state which, regarded as a binary string, corresponds to $k$.
Defining $\rho=2^{n_r-1}$ and noting that we have $2\rho$ basis states in our computer's sub-register, 
a natural choice for the allowed $k$ is to run from $-\rho$ through zero to $\rho-1$. Therefore a 1D single-particle state $\Psi$ would be represented by our sub-register $\ket{\psi}_{x}$ according to
\begin{equation}
\Psi=L^{-\frac{1}{2}}\sum_{k=-\rho}^{\rho-1} a_k e^{i 2\pi k x/L}\ \ \leftrightarrow\ \ 
\ket{\psi}_{x}^\text{KS}=\sum_{k=-\rho}^{\rho-1} a_k \ket{k}.\ \ \ 
\label{kSpaceMapping}
\end{equation}
The superscript KS denotes the $k$-space representation.

We generate the dual DVR on a quantum computer by applying, to each sub-register, a quantum Fourier transform (QFT) denoted by $U_\text{QFT}$ (see the Supplementary Material for its quantum circuit).
The sub-register as a whole will be transformed as
\begin{equation}
\ket{\psi}_{x}^\text{RS}=U_\text{QFT}\ket{\psi}_{x}^\text{KS}=\sum_{n=-\rho}^{\rho-1} b_n \ket{n},
\label{transformingToR}
\end{equation}
where 
\begin{equation}
b_n=\frac{1}{\sqrt{2\rho}}\sum_{k=-\rho}^{\rho-1} 
\exp(i \frac{n \pi}{\rho}k)\,a_k.
\end{equation}
The superscript RS indicates the real-space representation. When we wish to return to the original, $k$-space representation we employ the inverse QFT: 
\begin{equation}
\ket{\psi}_{x}^\text{KS}=U_\text{QFT}^{\dagger}\ket{\psi}_{x}^\text{RS}=\sum_{k=-\rho}^{\rho-1} a_k \ket{k},
\label{transformingToK}
\end{equation}
where of course
\begin{equation}
a_k=\frac{1}{\sqrt{2\rho}}\sum_{n=-\rho}^{\rho-1} 
\exp(-i \frac{k \pi}{\rho}n)\,b_n.
\label{ak_from_bn}
\end{equation}
We find that the basis functions represented by each computational basis $\ket{n}$ appearing in Equation~\ref{transformingToR} are peaked at (but not strictly localised around) the spatial point $x_n=\frac{n}{\rho}\frac{L}{2}$. The upper left panel of Fig.~\ref{fig:kSpaceAndRealSpace} shows a plot for the case that $n_r=6$, $n=0$. Specifically, the inferred mapping is
\begin{equation}
\phi_n(x)=P_{n_r}^n(x)
\ \ \ \leftrightarrow\ \ \
\ket{n}
\label{eqn:1Dpixel}
\end{equation}
with
\begin{equation}
P^n_{n_r}(x)=
\exp({-\frac{i\pi x^\prime}{L}})\sqrt{\frac{2}{\rho L}}
\ \sum_{j=1}^{\rho}\cos\frac{\pi(2j-1)x^\prime}{L}
\label{eqn:Pfunction}
\end{equation}
where $x^\prime=x-x_n$ with $x_n= \frac{nL}{2\rho}$, and $\rho=2^{n_r}-1$.
The function $P^n_{n_r}(x)$, which we informally refer to as a ``pixel function'', serves the role of an approximation or ``smear" of the Dirac delta function $\Delta(x-x_n)$, the sharpness increasing as $2^{n_r}$ tends to infinity.
The separation between the peaks of adjacent pixel functions, e.g. $x_2-x_1$, is 
$\delta r=L/2^{n_r}$, and
we now define the model's {\it spatial resolution} as the reciprocal of this quantity, i.e. as the number of pixel functions per unit distance:
\begin{equation}
\delta r^{-1}=2^{n_r}/L.
\label{eqn_resolution}
\end{equation}
A suitable decomposition to represent any 1D wavefunction $\Psi(x)$ in our quantum register is therefore very intuitive:
\begin{equation}
\Psi(x)\approx
\frac{C}{\sqrt{\delta r}}\sum_{n=-\rho}^{\rho-1} \Psi(x_n)P^n_{n_r}(x)
 \leftrightarrow\ 
\frac{C}{\sqrt{\delta r}}
\sum_{n=-\rho}^{\rho-1}\Psi(x_n)\ket{n}\ 
\label{eqn:1DpixelatedWF}
\end{equation}
i.e. the required amplitude of $\ket{n}$, the state representing the wavefunction peaked at point $x_n$, is found simply by sampling the target wavefunction at that point. Here $C$ is a normalisation constant that will be close to unity providing that (a) the target wavefunction has negligible amplitude outside of the simulation box and (b) the target wavefunction varies slowly with respect to $\delta r$. Intuitively, one can think of these spatial basis functions $P^n_{n_r}(x)$ analogously to pixels as used in conventional digital photographs: the greater the number of spatial pixels or grid points, the more features of the wavefunction are adequately captured.

The duality between the momentum- and real-space representations under the $U_\text{QFT}$ has been explored in multiple grid-based quantum simulation studies, including Refs.~\cite{Kassal2008, Cody_Jones_2012, Ollitrault2020, kosugi2021probabilistic}.
Whereas the $k$-space representation maps qubit basis states to wavefunctions with sharp values of $k$, in the dual representation the mapping is to wavefunctions that are not perfectly sharp around points in real-space. Nonetheless, because of their Dirac-like nature, one can analyse techniques and protocols {\it as if} they were Dirac functions, secure in the knowledge that in the high-resolution limit this becomes exact. This gives the visually intuitive picture that a particle \textit{is} a (pixelated) wavefunction in real-space, supported by a basis of sharp, evenly-spaced functions; this insight is employed in most prior works~\cite{Kassal2008, Ollitrault2020, kosugi2021probabilistic, Childs2022}. The appealing conceptual simplicity over second quantization can be regarded as another merit of first-quantized grid representation. 

The generalisation to 2D or 3D is the natural one: the sub-registers tensor together to form the complete representation of a given particle. The 3D analogue of Eqn.\,(\ref{eqn:1DpixelatedWF}) is
\begin{eqnarray}
\Psi(\mathbf{r}) &\approx& C\delta r^{-\frac{3}{2}}\sum_{\mathclap{n,m,l=-\rho}}^{\mathclap{\rho-1}} \Psi(x_n,y_m,z_l) P^n_{n_r}(x)P^m_{n_r}(y)P^l_{n_r}(z) \nonumber \\
&\leftrightarrow& \ 
C\delta r^{-\frac{3}{2}}\sum_{\mathclap{n,m,l=-\rho}}^{\mathclap{\rho-1}} \Psi(x_n,y_m,z_l) \ket{n} \ket{m}\ket{l}.\label{eqn:3DrealSpace}
\end{eqnarray}
When we generalise to represent a $P$-particle wavefunction $\Psi(\mathbf{r}_1,...,\mathbf{r}_P)$ we need only extend in the natural fashion,
\begin{equation}
C\delta r^{-\frac{3P}{2}}\sum_{\mathclap{\{n_1..l_P\}=-\rho}}^{\rho-1} \Psi(x_{n_1},y_{m_1},...,z_{l_P}) \ket{n_1} \ket{m_1}...\ket{l_P}.\label{eqn:3DrealSpacePparticles}
\end{equation}

\subsubsection{Split-Operator Propagation}\label{sec:SOprop}

\begin{figure*}
    \centering
    \includegraphics[scale=0.9]{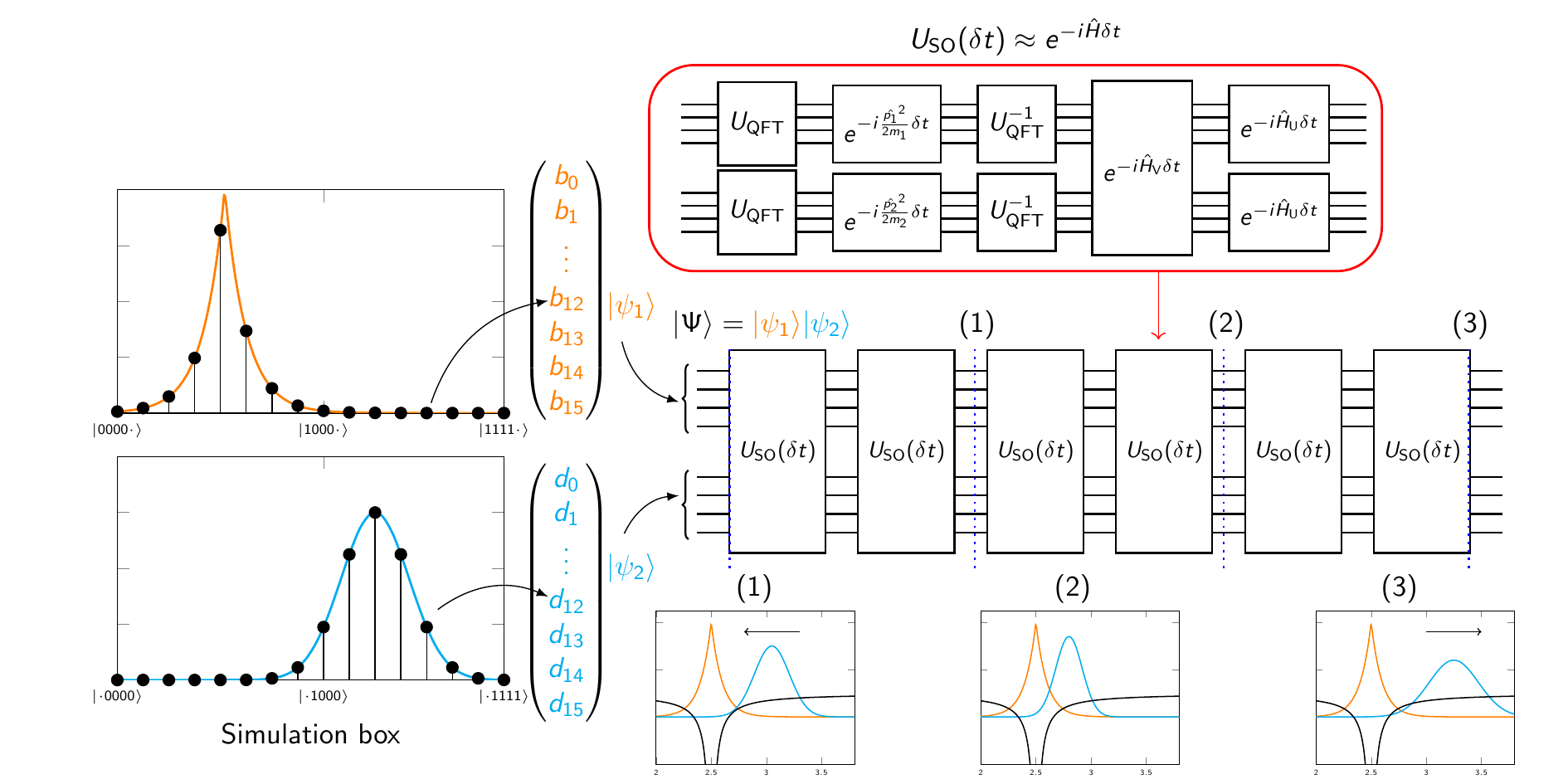}
    \caption{\textbf{Real-space grid Hamiltonian simulation using the SO-QFT approach.}
    In this example, the dynamics of two electrons in a simulation box is simulated: One is modelled as a Gaussian wavepacket, and the other as the bound ground state hydrogen solution. The two interact via a repulsive Coulomb potential, and the atomic nucleus is modelled as a classical attractive Coulomb potential.
    The particles are digitised into regularly spaced pixels, encoded into the state of two qubit subregisters of the same size. The starting state is a tensor of these two states (a Hartree product), and upon antisymmetrisation of the two particles (which would generate a Slater determinant), the initial state can then be propagated in time by repeated application of the SO-QFT small-time evolution operator $U_\text{SO}(\delta t)$ (depicted by the top circuit). A possible sequence of the subsequent dynamics, where the unbound electron scatters with the bound electron, is plotted at the bottom.}
    \label{fig:SO-QFT}
\end{figure*}

The SO-QFT exploits these representation choices, and the low computational cost incurred to transform between them, to approximate the time evolution operator $e^{-iH\delta t}$. Evolution by a total time $t$ is discretised into short time intervals $\delta t$ such that
\begin{equation}
    e^{-i\hat{H}t} = \underbrace{e^{-i\hat{H}\delta t}e^{-i\hat{H}\delta t}\dots e^{-i\hat{H} \delta t}}_{N \text{ times}} = U^N(\delta t) \quad,
\end{equation}
and the SO-QFT approximates the unitary short time propagator using Lie-Trotter-Suzuki product formula (or Trotterisation), splitting the Hamiltonian into its kinetic and interacting parts
\begin{eqnarray}
    \label{eqn:splitoperator}
    e^{-i\hat{H}\delta t}&=&e^{-i(\hat{H}_\text{kin} + \hat{H}_\text{int}) \delta t} \nonumber\\
     &=& e^{-i\hat{H}_\text{kin}\delta t}e^{-i\hat{H}_\text{int}\delta t} + O(\delta t^2) \nonumber\\
     &=& U_\text{SO}(\delta t) + O(\delta t^2)\nonumber.
\end{eqnarray}
Higher order splitting schemes and their numerical properties are well-documented~\cite{doi:10.1063/1.5092611, doi:10.1063/1.5094046, Kosloff1988, suzuki1991, childsTrotter2021}. For simplicity, and in order to compare different techniques on an equal basis, in this work we focus on the 1$^\text{st}$ order SO-QFT, summarised in Fig.~\ref{fig:SO-QFT}.
The methods we employ are equally compatible with any Trotter sequence; though where $[\hat{H}_\text{kin}, \hat{H}_\text{int}]\neq 0$, the dynamics will be imperfectly modelled and gives rise to the Trotter error terms as in Eqn.\,(\ref{eqn:splitoperator}); we discuss further in Section~\ref{sec:introResolution}.

The real- and momentum-space grid representations (detailed in Section \ref{sec:representations}) are natural options for state representation when computing the approximate time evolution operator of Eqn.\,(\ref{eqn:splitoperator}): In the $k$-space representation, the kinetic part of the Hamiltonian $\hat{H}_\text{kin}$ is separable and exactly local (diagonal); in the real-space representation, the interaction part of the Hamiltonian $\hat{H}_\text{int}$ is approximately diagonal, and would be exact if the basis of pixel functions were Dirac delta functions. Because the gate complexity for the $U_\text{QFT}$ only scales quadratically with the number of qubits per particle sub-register $n_r$ (see Supplementary Material for the QFT circuit), the SO-QFT can very efficiently compute the two phases of the short-time propagator by periodically transforming each sub-register independently into their preferred, diagonal basis:
\begin{equation}
\ket{\psi(t+\delta t)}^\text{RS}=U_\text{SO}(\delta t)\,\ket{\psi(t)}^\text{RS},
\nonumber
\end{equation}
where
\begin{equation}
U_\text{SO}(\delta t)=e^{-iD_\text{int}\delta t}\left(U_\text{QFT}^\dagger\ e^{-iD_\text{kin}\delta t}\  U_\text{QFT}\right).
\label{eqn:usingD}
\end{equation}
Here $D_\text{kin}$ and $D_\text{int}$ are diagonal real matrices, and $U_\text{QFT}$ is the quantum Fourier transform applied to all sub-registers.

We discuss in detail the evaluation of these operators on quantum computers in the Supplementary Material, but provide a summary here. From Eqn.~\ref{eqn:kinH}, we observe that propagation under the kinetic Hamiltonian $\hat{H}_\text{kin}$ separates exactly into a product of operators acting independently on each particle and in each spatial sub-register because the components commute:
\[
U_\text{kin}(\delta t)=\prod_{p=1}^P\prod_{q\in \{x,y,z\}}\exp(-i\frac{\delta t}{2m_p} k_q^2).
\]
where $k$ again refers to the momentum state $k$ from Eqn.~\ref{eqn:k-element-map}. The required quadratic phases are introduced onto our computational basis states in the momentum-space representation according to
\begin{equation}
\ket{k}\,\Rightarrow\,e^{-i C\delta t\, k^2 }\ket{k},
\label{eqn:littleKineticTime}
\end{equation}
and can be achieved using a sequence of single- and two-qubit phase gates (see for example \cite{Ollitrault2020, Cody_Jones_2012}) which scales as $O(n_r^2)$.

Propagation under the interaction potentials $\hat{H}_\text{U}$ and $\hat{H}_\text{V}$ is only modestly more complex than the former kinetic propagation. For interaction of particles with classical fields $\hat{H}_\text{U}$, we are primarily interested in an attractive Coulomb potential representing a nucleus (although we discuss a variation including a static electric field presently). For a single nucleus with charge $Z$, we write a time evolution operator 
\[
U_\text{U}(\delta t)=\prod_{p=1}^P\exp(-i\frac{Z\delta t}{\sqrt{x^2+y^2+z^2}}).
\]
The operations are independent between the registers corresponding to different particles, but not independent between sub-registers assigned to a given particle.
The phases changes we apply to our quantum registers are
\begin{eqnarray}
\ket{n}\ket{m}\ket{l}\,&\Rightarrow&\,
\exp(\frac{-iZ\,\delta t}{\sqrt{x_n^2+y_m^2+z_l^2}} )\ket{n}\ket{m}\ket{l}\nonumber\\
&=&\, \exp(\frac{-iZ\,\delta t}{\delta r\sqrt{n^2+m^2+l^2}} )\ket{n}\ket{m}\ket{l}\ \ 
\label{eqn:littleNuclearTime}
\end{eqnarray}
Efficient evaluation of functions such as the inverse-square-root on quantum computers is an active area of development; we highlight Refs.~\cite{Kassal2008, Cody_Jones_2012, haner2018optimizing, poirier2021fulldimensional} in these directions. Here it suffices to note that the number of gates required can scale quadratically with the number of qubits $n_r$~\cite{haner2018optimizing, Cody_Jones_2012}, and we discuss this further in the Supplementary Material.

For interaction between quantum particles $\hat{H}_\text{V}$, the propagation is of the form 
\begin{equation}
U_\text{V}(\delta t)=\exp(-i\delta t\sum_{p,q=1; p\neq q}^P \frac{q_{p,q}}{|{\mathbf r}_p-{\mathbf r}_q |}).
\end{equation}
To compute this propagation, we consider two particle registers, each composed of three sub-registers, which represent a given pair of particles $p=1$ and $p=2$. The basis states will be updated as 
\begin{equation}
\ket{n_1\,m_1\,l_1}\ket{n_2\,m_2\,l_2}
\,\Rightarrow\,
e^{-i \Theta}\ket{n_1\,m_1\,l_1}\ket{n_2\,m_2\,l_2}
\nonumber
\end{equation}
where
\begin{eqnarray}
\Theta&=& \frac{Z_{p,q}\,\delta t}{\sqrt{(x_{n_1}-x_{n_2})^2+(y_{m_1}-y_{m_2})^2+(z_{l_1}-z_{l_2})^2}} \nonumber\\
&=& \frac{Z_{p,q}\,\delta t}{\delta r\sqrt{(n_1-n_2)^2+(m_1-m_2)^2+(l_1-l_2)^2}}
\label{eqn:interactingForm}
\end{eqnarray}
We discuss how this is computed in Section\,\ref{sec:architectures} when we consider appropriate computer architectures, and in further detail in the Supplementary Material.

We again emphasise that, when propagating $H_\text{U}$ and $H_\text{V}$ in the real-space representation, the operators are approximated to be diagonal. This would be exact in the limit of infinite spatial resolution and the spatial states were Dirac delta functions. Under this approximation, the two terms commute, and we expect that for sufficiently small $\delta t$ and a smoothly varying potential,an adequate spatial resolution $\delta r^{-1}$ will result in dynamics that converge to the exact behaviour. However the Coulomb potential is singular at $\mathbf{r}=\mathbf{0}$ and consequently we will need to investigate the behaviour in this region carefully.

\subsection{Techniques for SO-QFT modelling}
\label{sec:methods}
We now present techniques and important considerations specific to SO-QFT modelling. We outline the methods known in the literature but tested for the first time here in a quantum chemistry setting, while highlighting those methods that (to our knowledge) have not been previously described.

\subsubsection{Energy Observable}
\label{sec:observablesAndPE}
In this first section we review energy measurement for grid-based quantum simulations; the experienced reader may care to skip to Section~\ref{sec:stateprepintro}.
\smallskip

\noindent \textbf{Energy Expectation} --
The energy expectation is the most ubiquitous observable in quantum chemistry.
In the grid-based approach it is possible to split the Hamiltonian and calculate the kinetic energy expectation in $k$-space, followed by the potential energy expectation in real-space
\begin{equation}
    \label{eqn:expect}
    \langle E \rangle = \bra{\Psi}U_\text{QFT}^\dagger \hat{H}_\text{kin}U_\text{QFT}\ket{\Psi} + \bra{\Psi}\hat{H}_\text{int}\ket{\Psi}
\end{equation}
where here $\ket{\Psi}$ is understood to be the spatial representation.
Aside from requiring inefficient repeated sampling of the quantum state, as we demonstrate in Section~\ref{sec:badObservable}, the aforementioned Trotter error gives rise to non-conservation of this energy expectation, and renders the method unsuitable for extracting the correct physics from simulation. 
\smallskip 

\noindent \textbf{Iterative Phase Estimation} --
The energy of a state can also be extracted by tracking the global phase it acquires during time propagation. This is given by the phase of the autocorrelation function, which is a discrete-time series of the inner product between a propagated state and the initial state. 
Suppose we time propagate an eigenstate $\Psi_n$ of $\hat{H}_\text{tot}$ with energy $E_n$. The corresponding autocorrelation signal is
\begin{equation}
    \bra{\Psi_n(t)}\ket{\Psi_n(t=0)} = e^{-iE_nt},
\end{equation}
where the absolute value $|\bra{\Psi(t=0)}\ket{\Psi(t)}|$ should be close to unity and in this work we use it as one measure of a simulation's veracity.

On a quanutm computer, this otherwise-unobservable global phase is efficiently extracted using phase estimation. Phase estimation through the use of ancilla qubits~\cite{kitaev1995quantum} is one of the fundamental techniques employed in diverse applications of quantum computing, and its utility in the context of SO-QFT and first quantized simulation is well recognised (see e.g.~\cite{Kassal2008}).

\begin{figure*}
    \centering
    \includegraphics[scale=0.85]{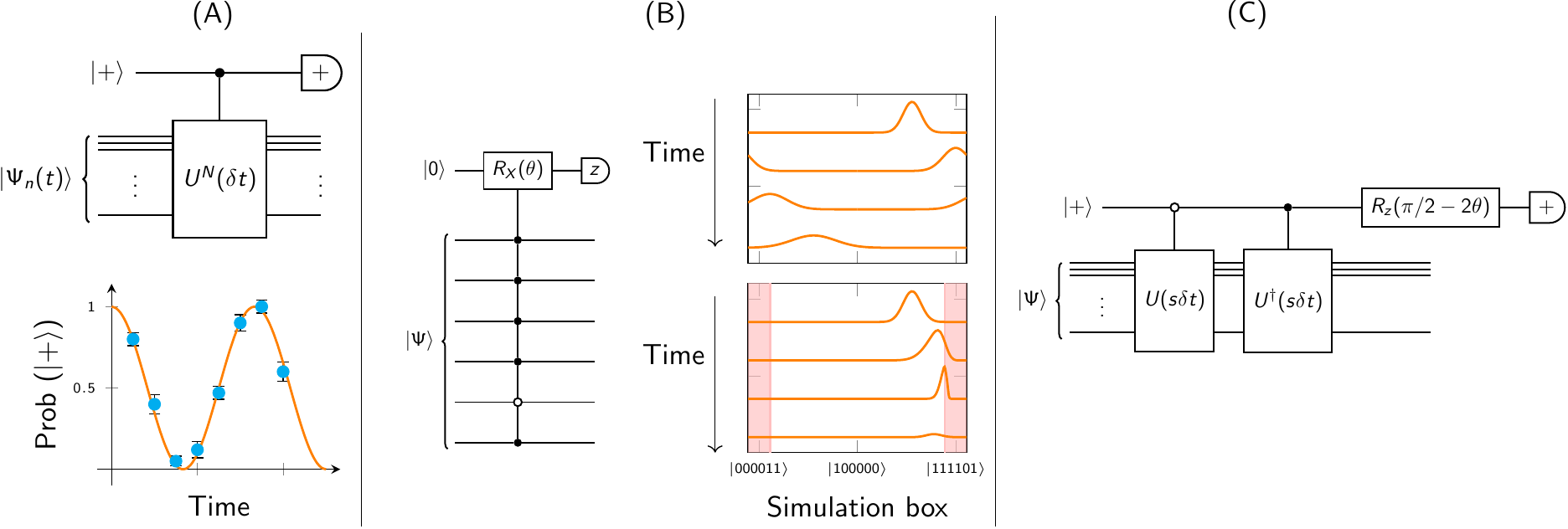}
    \caption{\textbf{Three early fault-tolerant quantum circuit techniques for real-space chemistry explored in this work.}
    \textbf{(A)}~The single-ancilla iterative phase estimation method emulated in this work. The global phase is encoded in the probability of measuring an entangled ancilla qubit, which controls the application of $N$ split-operator cycles, in the $\ket{+}$ state. To obtain this probability one must repeat the propagation and measurement at each point in time where one wishes to sample the signal.
    \textbf{(B)}~Attenuation of a wavepacket on a qubit register. The top right panel depicts the dispersion of a Gaussian wavepacket across the periodic boundary. Addition of complex absorbing region in pink (bottom right panel) attenuates the scattered wavepacket by reducing its norm. In the illustrated case, the process is not quite perfect: There is some reflection caused by the attenuation being too severe.
    The left panel is a circuit that performs probabilistic wavepacket attenuation at a select pixel. The pixel $\ket{111101}$ is selected, corresponding to the attenuating region on the right hand side of the simulation box in the figure.
    \textbf{(C)}~Preparing ground states using the PITE circuit. The filled circles indicate `control by $\ket{1}$' and the open circles `control by $\ket{0}$' Post-selecting on the $\ket{+}$ outcome yields a state with, to 1$^\text{st}$ order, an imaginary time evolution step applied. Given choice of an appropriate parameter $m_0$, the time rescaling factor $s=m_0/\sqrt{1-m_0^2}$ and the rotation $\theta=\kappa\arccos{[m_0+\sqrt{1-m_0^2}/\sqrt{2}]}$, where $\kappa=\text{sgn}(m_0-1/\sqrt{2})$ (see Ref.~\cite{kosugi2021probabilistic}).
    }
    \label{fig:circuit_techniques}
\end{figure*}

In the Iterative Phase Estimation (IPE) approach, even a single ancilla~\cite{PhysRevA.76.030306, somma2016quantum} is sufficient to learn this phase; a resource cost saving that will be welcome in the early fault-tolerant regime.
The method is summarised in the left panel of Figure~\ref{fig:circuit_techniques}. We conditionally apply $N$ SO-QFT steps $U^N(\delta t)$ controlled by an ancillary qubit in the $\ket{+}$ state. At the $N^\text{th}$ step, we measure the ancillary qubit in the $\ket{+}$ basis, at which point the state is discarded. We see that for an eigenstate, the global phase information is encoded in the relative phase between the $\ket{0}$ and $\ket{1}$ state of the ancillary qubit
\begin{eqnarray}
    \ket{+}\ket{\Psi_n(0)} &\Rightarrow& \frac{1}{\sqrt{2}}\Big(\ket{0}\ket{\Psi_n(0)}
    + \ket{1}\ket{\Psi_n(t)} \Big)\ \ \ \ \ \nonumber\\
    &=& \frac{1}{\sqrt{2}}\Big(\ket{0} + e^{-iE_nt}\ket{1} \Big)\ket{\Psi_n(0)}.
\end{eqnarray}

The probability of finding the ancillary qubit in state $\ket{+}$ fluctuates as the phase. We use this to extract a periodic time signal $a(t)$ where the frequency is proportional to the energy of the simulated wavefunction
\begin{equation}
    a(t) = \cos^2\left(\frac{E_n}{2} t\right).
\end{equation}
Because the number of qubits we can classically emulate is limited, using the single-ancilla IPE for our demonstrations herein is a natural choice. We report the exact evolution of $a(t)$ plotted at regular time points; this is straightforward since we use classically-emulated quantum processors. On a real device, because the single-ancilla projection probability is statistical in nature, the time propagation and measurement will have to be repeated multiple times. 

If more ancilla qubits are available this naturally extends to the standard Fourier phase estimation; for completeness we include this in the Supplementary Material, where we also note the use of classical Fourier analysis to extract features if indeed the hardware is limited to a single ancilla.

\subsubsection{State Preparation}
\label{sec:stateprepintro}
Preparing an appropriate starting state is a crucial and non-trivial component of dynamics simulation. To be consistent with the performance of the subsequent time-evolution, we should ideally perform state preparation in polynomial time. Furthermore, since we are working in the first quantized picture, we must explicitly realise the correct (anti-)symmetry therein. The canonical quantum algorithmic approach is to start by initialising an easily prepared reference state, and then drive it towards e.g. the ground state; recent studies~\cite{GChan2022} have highlighted ongoing challenges and uncertainties associated with this task. In this Section, we briefly review them, and introduce the state preparation methods assessed in this work; a quantum algorithm for preparing antisymmetric states, state preparation with IPE described in Section \ref{sec:observablesAndPE}, and the probabilistic imaginary-time evolution (PITE)~\cite{kosugi2021probabilistic} for preparing ground states. We provide a brief review of the state preparation literature in the Supplementary Materials.
\smallskip

\noindent \textbf{Initial state loading and antisymmetrisation} --
For many-body systems in which the majority of particles are indistinguishable, the challenge of preparing an initial state with proper exchange symmetry is non-trivial. Here we will assume that we wish to prepare antisymmetric states, but the methods to prepare bosonic symmetric states are near-identical. The choice of method to accomplish antisymmetrisation determines the options for actually loading the initial single-particle basis states, as we now explain. 

Protocols for creating an antisymmetric superposition on a quantum computer have been explored in several publications dating back at least as far as Abrams and Lloyd in 1997~\cite{PhysRevLett.79.2586}. The relatively-recent 2018 paper of Berry \textit{et al.}~\cite{berry2018improved} describes a deterministic, poly-logarithmic algorithm to create a state of the form, 
\begin{equation}
\ket{\text{prm}}\equiv\frac{1}{\sqrt{P!}}\sum_i^{P!} \text{perm}_i\large(\ket{0}\ket{1}...\ket{P-1}\large).
\end{equation}
Here we have $P$ sets of qubits, each of size $s=\lceil \ln P \rceil$, which represent the binary numbers $0$ to $P-1$. The notation $\text{perm}_i()$ indicates the $i^\text{th}$ permutation, and is understood to include the sign appropriate to antisymmetry (that is, $(-1)^k$ if there are $k$ pair-wise swaps needed to permute from the canonical ascending order). 

The state $\ket{\text{prm}}$ is a superposition of binary labels, rather than a superposition of states in the grid-based representation as we require. In principle we can obtain the latter once we have the former. Specifically, we prepare $\ket{\text{prm}}$, and distribute each of the $P$ sets of qubits as follows: Each will occupy the first $s$ qubits of one of our $P$ particle-representing registers, while the remaining qubits within each particle register are in state zero, $\ket{\underline 0}$. We now apply to each register an operator that maps these integer numbers to the desired grid-based eigenstates as follows:
$
M_\text{unpack}\left(\ket{j}\ket{\underline{0}}\right)=\ket{\psi_j}
$
where $\psi_j$ with $j=0...(P-1)$ are the single-particle basis states from which we are building the initial state. While this is conceptually simple, the operator $M_\text{unpack}$ may be challenging to implement since it must encode knowledge of every $\psi_j$ and moreover it should be (near-)deterministic if it is to succeed over all $P$ particle registers.

Instead one might wish to begin by preparing a simple product state across our particle-registers, i.e.
\begin{equation}
\ket{\psi_0}\ket{\psi_1}...\ket{\psi_{P-1}},
\label{eqn:simpleLoad}
\end{equation}
and then find some means to antisymmetrise this, thus generating a Slater determinant~\cite{szabo1996modern}. The preparation of each state $\ket{\psi_j}$ would be relatively easy, performed independently for each particle register and optionally involving non-deterministic approaches with a high success probability per-attempt.
In the Supplementary Material we highlight recent advances in preparing known (analytic) single-particle states $\ket{\psi}$ on a quantum register. This route is clearly attractive, but relies upon an efficient means of subsequently antisymmetrising the product state in Eqn.\,\ref{eqn:simpleLoad}.

In~\cite{doi:10.1063/1.3115177} the authors propose performing an antisymmetrisation on a product state using an adaption of the methods in Ref.~\cite{PhysRevLett.79.2586} (i.e. permuting a product state by co-sorting a state of the form $\ket{\text{prm}}$), however they do not explicitly specify how the resulting swap flags are to be uncomputed (moving from Eqn.(4) to Eqn.(5) in Ref.~\cite{doi:10.1063/1.3115177} appears non-trivial). In Section~\ref{sec:AntiSymmResults} we introduce, to our knowledge for the first time, two simple solutions to this issue.  We will rely on the following assumption (which may itself be non-trivial): we must specify a Hamiltonian $H_\text{synth}$ with the property that 
\begin{equation*}
H_\text{synth}\ket{\psi_i}=E_i\ket{\psi}_i\ \ \ \ \ \ \ E_{i+1}>E_i
\end{equation*}
(reiterating Eqn.~\ref{eqn:Hsynth}). This Hamiltonian is synthetic in the sense that it need not correspond to any physically legitimate scenario. The condition implies that the states $\ket{\psi_i}$ must be orthogonal; moreover we desire well-separated energies.  Section~\ref{sec:AntiSymmResults} specifies the key steps and also gives indicative performance numbers from numerical emulation of antisymmetrisation for $P=5$ particles. 
\smallskip

\noindent \textbf{Refining the initial state: Ground state preparation} -- The remarks above concern the initialisation of our grid-based machine into a state that we understand and can describe analytically. Of course, we may then wish to drive this state into a more accurate representation of e.g. the ground state of the real system prior to exploring its dynamics, assuming non-zero overlap between the initial reference and the target state. While many quantum algorithms have been proposed for this (see Supplementary Material), in this work we demonstrate two techniques which exploit the non-unitary nature of single-ancilla measurements.
\smallskip

\begin{itemize}
    \item \textit{State Editing} Use of ancilla measurements in phase estimation is a well-established technique for ground state preparation~\cite{kitaev1995quantum, PhysRevLett.102.130503, berry2018improved}.
    
    Here we consider the single-ancilla variant which uses IPE described in Section \ref{sec:observablesAndPE}, where we regard the ancilla measurement as a mid-point rather than an end-goal. We emulate an approach which uses this measurement to remove a known state from superposition to 2D hydrogen which we report in Section~\ref{subsec:phaseEstForPrep}.

    Suppose we can prepare an initial state which contains a superposition of eigenstates including our desired state. Let us assume that we know the energy $E_\kappa$ of the state we wish to remove, which can be revealed using the method of Section~\ref{sec:observablesAndPE} and classical Fourier analysis. We prepare our ancilla in the $\ket{+}$ state and then perform conditional evolution previously described. To remove state(s) $\Psi_\kappa$ with eigenvalue $E_\kappa$, we require measurement at $T_\kappa$, such that $e^{-iE_\kappa T_\kappa}=-1$, thus
    \begin{equation}
        T_\kappa = N_\kappa\delta t = \frac{\pi}{E_\kappa}
    \end{equation}
    Now the undesired component has zero probability of yielding a $\ket{+}$ from the ancilla measurement, so post-selecting on that outcome will entirely eliminate it's contribution from the register. The more conventional approach for state preparation with phase estimation, which seeks to amplify a target state with known energy, is reviewed in the Supplementary Material.
    
    \item \textit{Probabilistic Imaginary-Time Evolution (PITE)} The second state preparation method we explore is a recently introduced approach for performing imaginary-time evolution~\cite{kosugi2021probabilistic, liu2021probabilistic} on quantum computers. The method does not require \textit{a priori} knowledge of Hamiltonian eigenvalues however unfortunately it does suffer an exponentially vanishing success probability, as the authors note and we presently explore.

    Suppose again that we have an initial state which contains a superposition of eigenstates $\ket{\Psi} = \sum_n c_n\ket{\Psi_n}$. Imaginary-time evolution applies the non-unitary evolution operator $e^{-\hat{H}\tau}$ to the initial state 
    \begin{equation}
        \ket{\Psi(\tau)} = e^{-\hat{H}\tau}\ket{\Psi} = \sum_n c_n e^{-E_n\tau}\ket{\Psi_n}
    \end{equation}
    and therefore approaches the lowest energy state in the initial superposition exponentially fast at the long $\tau$ limit. The PITE approximates, to 1$^\text{st}$ order, the Hermitian small imaginary-time evolution operator $m_0e^{-\hat{H}\delta\tau}$ (where $0<m_0<1$ and $m_0\neq 1/\sqrt{2}$) applied to the state on the main register, by controlling both forward and reverse real-time evolution operators $U(\delta t)$ and $U^\dagger(\delta t)$ with a single ancillary qubit prepared in superposition. Post-selecting on the ancilla measurement outcome of $\ket{0}$ applies the desired evolution on the register. We prepare the same $\ket{+}$ ancilla state as per the IPE, and start from the circuit in Figure~\ref{fig:circuit_techniques}, noting that it achieves the exact same non-unitary operation proposed in Ref.~\cite{kosugi2021probabilistic} but with fewer gates. In Section~\ref{subsec:PITEForPrep} we emulate such a circuit, and test it numerically for the first time in preparing the ground state of a 2D hydrogen atom.
    
    Like state-editing by phase estimation, this approach is also probabilistic and preparing an initial state with large overlap with the ground state is key to maximising the rate of success. In this case however, the probability that every sequential ancilla measurement yields the desired outcome decays exponentially with the number of measurements. We refer the reader to Ref.~\cite{kosugi2021probabilistic}, which introduced the technique and applied it to wavepackets in parabolic potentials, for details of this drawback and possible ways to address it.  We highlight that, while the current method may be impractical on a real quantum computer, it is nonetheless an effective quantum-inspired ground state preparation method for classical emulations.
\end{itemize}

\subsubsection{Attenuation and scattering}\label{sec:AttenAndScatIntro}

In this work, we developed a quantum circuit analogue of complex absorbing potentials (CAPs), a well-established technique from classical simulation~\cite{KOSLOFF1986363}. Here we present this method and show that it can also be used to track the probabilities of wavepackets being found at specific pixels, important to the determination of reaction rates in molecular processes. We test these techniques in the ionisation of a single bound electron by a strong applied electric field and two-particle scattering scenarios in Section~\ref{sec:efieldAndScatterResults}.

The use of the $k$-space mapping imposes a periodic boundary condition in real space. 
For non-periodic systems, this introduces artificial interactions and state interference through the boundary and one of the requirements on the simulation  box is that it is sufficiently large that these have negligible impact on the dynamics. Given that, for fixed spatial resolution, the width can be increased with only a logarithmic cost both in the number of additional qubits and the `wall clock' execution time (see Section~\ref{sec:postClassicalSims}), this does not usually pose a problem. However, there are simulation tasks where simply enlarging the box may not be an ideal solution: for example, a study of a scattering process where the quantum states of scattered particles have contributions that travel at speed towards the edges of the simulation box. 

One alternative is to attenuate wavefunction amplitude near the boundary. Any such attenuation would be a non-unitary process involving measurement and implying some probabilistic element; however the cost associated with any need for repetition may be preferable to the increase in the number of qubits required in the former approach. Furthermore, we show that from the rate of failure of this probabilistic process, one can infer the probability of a particle being measured at a given pixel. 

The interaction potential part of the Hamiltonian $\hat{H}_\text{pot}$ is modified to contain both a real and (negative) imaginary part
\begin{equation}
    \hat{H}_\text{int} = \hat{H}_\text{U} + \hat{H}_\text{V} - 
    i\hat{V}
\end{equation}
The time evolution operator is thus
\begin{equation}
    \Psi(\mathbf{r}, t) = e^{-i\hat{H}t}e^{-\hat{V}t}\Psi(\mathbf{r},0)
\end{equation}
The absorbing potential $\hat{V} = V(\mathbf{r})$ should be balanced between sharp, abrupt attenuation which reduces the size of the attenuation region while suppressing transmission, and smooth, gradual attenuation which suppresses reflection. For brevity, we will only assert that it is possible to construct complex absorbing potentials which are completely reflection and transmission free \cite{Riss1995}, and refer the reader to the rich body of classical quantum dynamics literature on complex potentials \cite{JOLICARD1985106, Neuhauser1989, Riss_1993, Balint-Kurti1993, Scheit2006, Gatti2017numericalmethods}.

On a quantum computer, the objective is to achieve the following non-unitary transformation in (though not limited to) the real space representation at every SO-QFT step
\begin{equation}
    \sum a_n\ket{n} \Rightarrow \sum a_n e^{-V(x_n)\delta t} \ket{n},
\label{eqn:attenuate}
\end{equation}
followed of course by the necessary renormalization. 
Here $x_n$ is the location of the peak of the pixel function $P_{n_r}^n(x)$ as per Eqn.\,\ref{eqn:1Dpixel}. Of course $\delta t$ is merely a constant, corresponding to the temporal granularity of the simulation. The analogous mapping should occur for $y$ and $z$, and for each particle in the system -- ideally this should be done in a fashion that perfectly preserves any exchange symmetry relevant to the system.

To achieve this, we could proceed as follows for each $x_n$ that lies within our attenuating region (i.e. where $V(x_n)\neq0$). At the end of each spatial part of our SO-QFT cycle we prepare an ancilla qubit in state $\ket{0}$ and we perform a multi-controlled rotation $e^{i \sigma_x \theta}$ on that ancilla. The rotation occurs if and only if the sub-register is exactly in state $\ket{n}$ which corresponds to the pixel function peaked at $x_n$. We then measure the ancilla. If we obtain outcome $\ket{1}$ we say that the particle in question has been measured to be that point, and we may either halt the simulation or proceed with the remaining particles -- both scenarios are examined in the numerical emulations presented. However, if we obtain outcome $\ket{0}$ then the simulation simply proceeds, except that the amplitude of state $\ket{n}$ has been attenuated from $a_n$ to $a_n\cos(\theta)$ (and the state has been renormalised). By repeating this process for all $x_n$ within the attenuating region, selecting the proper $\theta=\arccos(e^{-V(x_n)\delta t})$ in each case, we implement Eqn.~\ref{eqn:attenuate} (see centre panel of Figure~\ref{fig:circuit_techniques}). In parallel the same process can be applied for other sub-registers and other particles. 

If the attenuation region is narrow, corresponding to only a few values of $x_n$, then the above protocol can be relatively efficient. Remarkably frugal constructions for multi-controlled Paulis do exist~\cite{gidney2021CCP}, and of course a controlled Pauli can implement the general $\theta$ rotation described above with the use of additional single-qubit rotations. However if the attenuation region is a main portion of the simulation box then the method would be exponentially inefficient as it requires action for each element of the spatial superposition. Fortunately however, one can simply set a uniform attenuation strength $V$ over a range of $x_n$ values and implement it in one step using an appropriate subset of register qubits in the control process. The simplest example occurs when the width of the attenuating region is $2^{-m} L$ for some integer $m$ ($L$ being the width of the simulation box). In this case, only the most significant $m$ qubits from the sub-register control the rotation. This is the approach taken in the numerical emulations presented in this work.

By repeating the entire simulation multiple times and keeping track of the probability that the ancilla qubit yields $\ket{1}$, we are also recording the time-dependent probability that a particle has reached a given pixel in the simulation box. One could in principle obtain this information by generating the probability distribution of particles at different times through direct sampling of particle subregisters; indeed, we generate plots that would be obtained this way. However, on a real quantum computer, selecting specific pixels and entangling them with a single qubit that would then be measured is a much more cost efficient statistical approach.
One might envisage the determination of reaction rates by selecting key pixels, much like the selection of a reaction coordinate in traditional reaction dynamics, and using an equivalent approach to track the probability that a particle wavepacket crosses said pixels.  
We show a proof-of-concept here by directly measuring particle scattering and ionisation probabilities, in combination with CAPs.

\subsubsection{The Coulomb potential: Demands on spatial and temporal resolution}
\label{sec:introResolution}
Here we review considerations regarding the granularity of simulating the Coulomb potential using either the classical or quantum split-operator; 
a more detailed discussion can be found in the Supplementary Material.

The split-operator uses finite spatial basis set and discrete time approximations, which become exact in the limit of infinite resolution. To optimise computation cost, it is sensible to use the minimum finite resolution necessary for obtaining physically accurate results. The spatial discretisation determines the fidelity of a model's state and (potential) operators to reality, and bounds the maximum attainable accuracy of measurable properties.
The time resolution is chosen such that the Trotter error during time propagation is sufficiently low; but because the error term of any Trotter sequence depends on the commutator between $\hat{H}_\text{kin}$ and $\hat{H}_\text{int}$, which in turn depends on the spatial derivatives of the model potential, using a fine spatial resolution will therefore require commensurately fine time-stepping to suppress Trotter error during propagation~\cite{childsTrotter2021}.

For propagation under smooth (e.g. harmonic) potentials, low spatial and time resolutions are generally sufficient. On the other hand, because of the sharp singularity in the Coulomb potential, accurately describing wavefunction propagation near the singularity is more involved, even if we are to effectively model a capped potential by never collocating the origin at a grid point. We expect that a high spatial resolution is necessary to attain accurate operator physics for states with large amplitudes at the singularity, which will demand a high time resolution to control propagation errors. In Section~\ref{sec:numerics}, we use numerical studies to estimate the minimum spatial and temporal resolution required for sufficiently accurate time-propagation of chemically-relevant systems, using the first-order SO-QFT.

\color{black}

\subsubsection{Augmented Split-Operator}\label{sec:introASO}

We now describe an alternative means to tackle simulation of states that may have large amplitudes at the Coulomb singularity, instead of using a `brute force' increase in spatial and temporal resolution. We call the approach the augmented split-operator (ASO) as it involves introducing an additional step in the SO-QFT cycle.
The results of applying the protocol below with an emulated quantum computer are presented in Section~\ref{sec:ASOresults}.

Following the implementation of the exponential operators in a SO-QFT time step, and while still in the real-space representation, we apply a {\em core stabilization} process. This is intended to `fix' the most severe deviations of the applied unitary from the ideal unitary $e^{-i \hat{H}_\text{tot}\delta t}$. We emphasise that the fix is at the operator level, and is agnostic to the specific state that is being modelled. However, the motivation is to make the dynamics of core-peaked states more accurate. We aim to reduce the overall time cost of a simulation with a given accuracy, without increasing the qubit count.

We describe the process for a simulation of a single particle and comment on the generalisation to $N$ particle systems further below. The following is performed on a classical computer, as a one-off preparation for our simulation. We make a fixed choice of $\delta t$, and we must repeat the analysis for each value of $\delta t$ we wish to use. 

\begin{enumerate}
\item 
\label{ASOstep1} We compute the ideal iterate $U_\text{ideal}=e^{-i \hat{H}_\text{tot}\delta t}$ and the actual iterate defined by Eqn.\,\ref{eqn:usingD} viz.
\[
U_\text{SO}(\delta t)=e^{-iD_\text{int}\delta t}\left(U_\text{QFT}^\dagger\ e^{-iD_\text{kin}\delta t}\  U_\text{QFT}\right).
\]
These objects are computed in the full basis of pixel functions for the particle (and are therefore far from diagonal).

\item\label{ASO:UcoreStep} We examine $U_\text{repair}$ which is defined by 
\[
U_\text{ideal}=U_\text{repair}U_\text{SO}(\delta t)
\]
and we will find that $U_\text{repair}$ is close to the identity except for elements corresponding to pixel functions that are near to the Coulomb singularity. Therefore we select a set of $Q$ such pixels, and derive a small $Q\times Q$ unitary $U_\text{core}$ that closely matches $U_\text{repair}$ in that subspace.
\item We define augmentation $U_\text{aug}$ as the identity operator except for the $Q$-state subspace, where it corresponds to $U_\text{core}$.
\item\label{ASO:circSynthesis} We find a circuit $C_\text{aug}$ that can implement $U_\text{aug}$ to a good approximation. This circuit is not merely a set of phase operations since  $U_\text{aug}$ is not diagonal in the $Q$-state subspace; $C_\text{aug}$ will cause a flow between the amplitudes of those states.
\end{enumerate}

This classical analysis is non-trivial but tractable since the unitaries involved are `only' of size $2^{n_rD}$ for a $D$-dimensional problem where $n_r$ is the number of qubits per sub-register. Moreover there are helpful symmetries in the analysis, and notwithstanding the description in Step~\ref{ASOstep1} we need not compute $U_\text{ideal}$ and $U_\text{SO}$ entirely but only in the subspace where we are likely to identify the $Q$ core pixels. 
With these steps completed, we are in a position to perform simulations with the augmented SO iterate: 
\begin{eqnarray}
U_\text{ASO}(\delta t)&\equiv&U_\text{aug}\,U_\text{SO}(\delta t)\label{eqn:ASOdef}\\
&=&U_\text{aug}\,e^{-iD_\text{int}\delta t}\left(U_\text{QFT}^\dagger\ e^{-iD_\text{kin}\delta t}\  U_\text{QFT}\right).\nonumber
\end{eqnarray}

In the Results section we report on the performance of this method for two cases, $2\times 2$ and $4\times 4$ pixels. As demonstrated by those examples, even with such small corrections the method is quite effective, and the cost of the augmentation can be modest: The circuit $C_\text{aug}$ is compact and need only target a small subset of qubits, with the others acting as controls. 

The generalisation of the ASO method to the $P$-particle case is straightforward except for a caveat. The main time cost of the SO-QFT method lies in the implementation of the $e^{-i \hat{H}_\text{int}\delta t}$ part, because of the $P(P-1)/2$ particle pairings involved. The ASO method can be efficiently implemented by applying an $U_\text{aug}$ augmentation after each such pairing, subsequent to implementing the $e^{-i C\delta t/|\mathbf{r}|}$ and while the registers are still representing relative coordinates $x_i-x_j$, $y_i-y_j$ etc. Because $U_\text{aug}$ is not diagonal, ensuring the proper evolution requires implementing the full orthogonal transformation to relative and central coordinates:
\[
\{x_i,x_j\}\Rightarrow\{x_i-x_j,x_i+x_j\}\ \text{rather than}\ \ \{x_i-x_j,x_i\}
\]
and similarly for $y$ and $z$.
Fortunately the additional arithmetic step that this implies, is trivial.


\newpage
\bibliography{library.bib}
\bibliographystyle{rsc}

\smallskip

\section*{Acknowledgements}
H.H.S.C. would like to thank J. Wu, P. Ollitrault, A. Lieberherr, A. Rattew, H. Jnane, J. P. Malhado, T. Carrington, J. Richardson and D. Manolopolous for helpful discussions.
The authors also thank D. Marti-Dafcik, B. Koczor and S. McArdle for their expert advice.
\textbf{Funding}~H.H.S.C. is funded by the Croucher Foundation, Hong Kong. S.C.B. acknowledges support from the EPSRC QCS Hub EP/T001062/1, from U.S. Army Research Office Grant No. W911NF-16-1-0070 (LOGIQ), and from EU H2020-FETFLAG-03-2018 under the grant agreement No 820495 (AQTION). The authors would like to acknowledge the use of the University of Oxford Advanced Research Computing (ARC) facility\,\cite{oxfordARC} in carrying out this work. \textbf{Author Contributions}~H.H.S.C., S.C.B. and D.P.T. designed the research. H.H.S.C., R.M., T.J. and S.C.B. contributed to the software used in this work. H.H.S.C., R.M., S.C.B. and D.P.T. contributed to the analysis. All authors contributed to the manuscript. \textbf{Conflict of Interest}~All authors declare that they have no competing interests.

\section*{Data and Materials Availability}
All data needed to evaluate the conclusions are present in the paper or the Supplementary Materials. The code for performing real space simulation can be found at \href{https://zenodo.org/record/7540464}{https://zenodo.org/record/7540464}.

\begin{figure*}
    \section*{Supplementary Materials}
\end{figure*}
\newpage

\renewcommand{\thesection}{S\arabic{section}}  
\renewcommand{\thetable}{S\arabic{table}}  
\renewcommand{\thefigure}{S\arabic{figure}}
\renewcommand{\theequation}{S\arabic{equation}}
\setcounter{section}{0}
\setcounter{table}{0}
\setcounter{figure}{0}
\setcounter{equation}{0}

\section{Background}\label{classicalBackground}
In seeking to use a controlled quantum machine as our simulator, one may adapt the rich set of numerical methods for solving the time dependent Schr\"odinger equation on `classical' (pre-quantum) computing platforms. Though many state representation and time propagation schemes exist \cite{Kosloff1988, Kosloff1994, Leforestier1991, Tucker2000, PhysRevE.73.036708, tannor2007introduction, Gatti2017numericalmethods}, the one of particular interest here is the dynamic Fourier or split-operator Fourier transform (SO-FT) method, which dates back at least as far as the work of Feit and Fleck in 1976 \cite{fleck1976time, FEIT1982, FEIT1983}.
The scheme is popular and often employed in small-molecule applications, predominantly in propagating nuclear wavepackets on pre-calculated static or time-dependent electronic potential energy surfaces. It has found uses in reactive scattering \cite{Sadeghi1993, Zhang1995} and excited-state chemistry \cite{Serov2001, doi:10.1063/1.2356477, doi:10.1063/1.465362, Greene2017, Curchod2018}, as well as modelling quantum processes like tunnelling~\cite{doi:10.1063/1.1560636, doi:10.1063/1.1766298} and superfluid turbulence~\cite{trofimov2009comparison}, and indeed quantum control for designing quantum devices~\cite{DION2014407, PhysRevB.102.125406}. Application to the relativistic Klein-Gordon equation~\cite{RUF20099092} and non-linear Schr\"odinger equation~\cite{doi:10.1063/1.36847, 10.2307/2157521, faou_gauckler_lubich_2014} extends the method's scope. 
Its success is partly attributed to the simplicity of its construction, as it only involves updating the vector describing the state, and partly to the low computational price paid for obtaining efficiently converging results~\cite{Kosloff1993}.

Building on the works of Zalka \cite{doi:10.1098/rspa.1998.0162} and Wiesner \cite{wiesner1996simulations} from the mid 1990's, it was observed just over a decade ago by Kassal \textit{et al.} \cite{Kassal2008} that it will be advantageous to store high-dimensional spatial grid representations of molecular wavefunctions on digital quantum computers using qubits that scale only linearly with the degrees of freedom. They also showed that, by exploiting the remarkably efficient quantum Fourier transform (QFT), a quantum version of the SO method (the SO-QFT)  will be able to simulate the dynamics of a chemical Hamiltonian in polynomial time. Indeed, the authors challenged the \textit{status quo} of limiting the split-operator method to propagating nuclei on electronic potential energy surfaces.

Although the SO-QFT method is oriented towards fault-tolerant quantum computing, it is nonetheless gaining increasing interest. The method has been investigated in the context of propagating one-dimensional Gaussian wavepackets \cite{Benenti2008}, quantum harmonic oscillator \cite{somma2016quantum}, non-adiabatic dynamics across coupled, pre-computed potential energy surfaces \cite{Ollitrault2020}, imaginary-time evolution for molecular geometry optimisation \cite{Hirai2022} as well as ground state, Gibbs state and partition function determination \cite{kosugi2021probabilistic}.
We highlight works geared towards understanding the scaling of the SO-QFT at mid-to-large scale, specifically that of Jones~\textit{et al.} which looked at the resource necessary for a fault-tolerant implementation \cite{Cody_Jones_2012}, the extensive investigation into first quantized basis-set simulations for chemistry of Su \textit{et al.}, which also included block-encoding the SO-QFT Hamiltonian for simulation using qubitization and interaction picture techniques~\cite{su2021fault}, and analysis by Childs \textit{et al.} which provided concrete upper bounds for the gate complexity of SO-QFT to arbitrary order~\cite{Childs2022}. Other attempts at direct quantum simulations in real-space grids include using finite difference stencils to approximate the kinetic operator in second quantization \cite{PhysRevX.8.011044}, and in first quantization with propagation based on the truncated Taylor series algorithm~\cite{Kivlichan_2017}, as well as a Cartesian component-separated approach~\cite{poirier2021fulldimensional}.

\section{Theoretical Framework}
\subsection{Representations in momentum- and real-space} \label{appendix:representations}
We wish to model the dynamics of a multi-particle problem with Hamiltonian $\hat{H}_\text{tot}$. We divide the qubits of our quantum computer into registers, each associated with one of our particles, then further divide each register into sub-registers corresponding to the dimensions of the model. In general, the registers corresponding to different particles need not be the same size, nor do the sub-registers need to match in size. Indeed, many scenarios would naturally suggest variations, e.g. a nuclear particle may be adequately modelled within a smaller `box' than its bound electrons, or we might model e.g. a solid-state interface where the $z$-direction is restricted versus the $x$-$y$ plane. In the present work we confine ourselves to considering cases where all sub-registers are of equal size, so that the computer has size of order $O(d\,P\,n_r)$ (neglecting ancillas used in temporary roles) where $d$ is the spatial dimensionality of the problem (2 or 3 for us), $P$ is the number of particles (1 or 2 in our simulations), and $n_r$ is the number of qubits in each sub-register.

We now elaborate on the finite basis (plane-wave) and discrete variable (real-space) representations described in the introduction. Suppose that we wish to model a 1D system which is well-localised within a region $-\frac{L}{2}<x<\frac{L}{2}$. We refer to this region as the `simulation box' and thus $L$ is the box width. For now we will assume that the real system's state has negligible amplitude outside this box, both initially and throughout the anticipated simulation. This condition is relaxed presently when we consider scattering and ionisation. We choose a $k$-space (or `spectral') representation in which a plane wave basis state of the modelled system is represented by a state of the computer's sub-register as
\begin{equation}
\phi_k(x)= L^{-\frac{1}{2}}\,\exp(\frac{i 2\pi k x}{L})\ \ \leftrightarrow\ \ 
 \ket{k}.
\label{appeqn:k-element-map}
\end{equation}
Note that a negative value of $k$ implies the two's complement binary representation. The same meaning is intended whenever we write a Latin letter in a ket.

Defining $\rho=2^{n_r-1}$ and noting that we have $2\rho$ basis states in our computer's sub-register, 
a natural choice for the allowed $k$ is to run from $-\rho$ through zero to $\rho-1$. With this choice subsequent expressions have a compact form; but the mapping from plane wave to computational basis state can employ a shift so that $k$ runs from (say) $0$ to $2\rho-1$. Indeed, in the simulations performed presently, underlying code using both conventions is tested. The eventual choice will be a matter of optimising circuit depths in the real quantum processor. As a further aside, note that an interesting alternative to Eqn.\,\ref{appeqn:k-element-map} is to use  $2\pi(k+1/2)$ rather than $2\pi k$. Then any state of the model must satisfy $\Psi(x=L/2)=-\Psi(x=-L/2)$. Therefore state $\Psi$ of a 1D modelled system would be represented by our sub-register $\ket{\psi}_{x}$ according to
\begin{equation}
\Psi=L^{-\frac{1}{2}}\sum_{k=-\rho}^{\rho-1} a_k e^{i 2\pi k x/L}\ \ \leftrightarrow\ \ 
\ket{\psi}_{x}^\text{KS}=\sum_{k=-\rho}^{\rho-1} a_k \ket{k}.\ \ \ 
\label{appeqn:kSpaceMapping}
\end{equation}
Formally this is a bijective encoding of the Fourier components of the modelled state as computational basis states. Throughout this paper we use the double-headed arrow $\leftrightarrow$ to indicate such a mapping: encoding of Fourier components (modelled system) on the left, to the quantum computer's state on the right. Practically, the meaning of $\leftrightarrow$ is that when we apply operations to the qubit register, we will do so with our mapping to the modelled system in mind; this will be clear presently. We use the superscript KS for $k$-space. For simplicity we do not put `$p=1$' in the subscript, but we understand this is for a given particle.

Our choice of mapping in Eqn.\,(\ref{appeqn:kSpaceMapping}) together with the equivalent representations for other particles and dimensions allows our quantum computer to represent an arbitrary $k$-space wavefunction of the multi-particle system, up to the limitation that plane wave components of spatial frequency higher than $2^{n_r-1}$ are omitted. 
Note that the represented state is periodic in space, since $\ket{\psi}_{x+1}=\ket{\psi}_{x}$ and similarly for other sub-registers. Throughout this paper we restrict our interest to the range $-L/2<x<L/2$ and refer to this as the `simulation box' with $L$ therefore the simulation box width. In the present section we have, and will, set $L=1$ in order to keep the expressions in a clean form. The periodicity is not exploited but we do need to be mindful of it in, e.g., the scattering/ionisation modelling described. 

It would be possible to exclusively use the $k$-space representation in grid-based modelling, and indeed there may be advantages to doing so \cite{babbush2019quantum, su2021fault}. However, in the present study we employ the approach in which the registers are periodically transformed into a `dual' representation \cite{PhysRevX.8.011044}.
Specifically, we apply to each sub-register a quantum Fourier transform (QFT) denoted by $U_\text{QFT}$ and defined by the circuit shown in Fig.~\ref{fig:qft_circuit_diagram}:
\begin{figure}[!htbp]
    \centering
    \includegraphics[width=0.95\columnwidth]{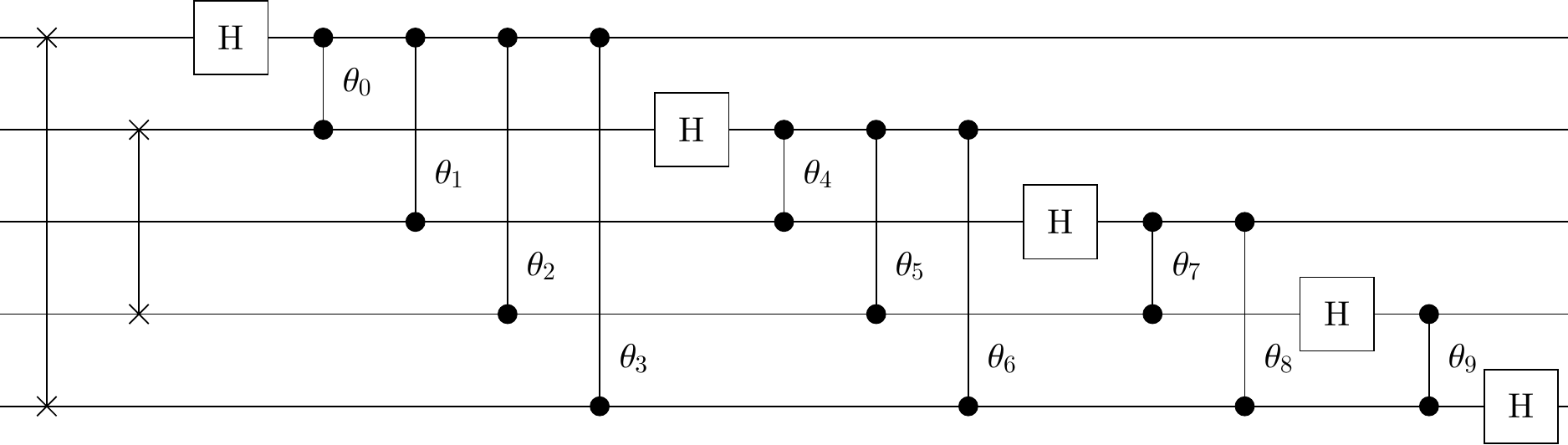}
    \caption{The QFT circuit used to transform from the $k$-space to the real-space representation. The circuit shown is for a 5-qubit sub-register. }
    \label{fig:qft_circuit_diagram}
\end{figure}

The sub-register as a whole will be transformed as
\begin{equation}
\ket{\psi}_{x}^\text{RS}=U_\text{QFT}\ket{\psi}_{x}^\text{KS}=\sum_{n=-\rho}^{\rho-1} b_n \ket{n},
\label{app:transformingToR}
\end{equation}
where 
\begin{equation}
b_n=\frac{1}{\sqrt{2\rho}}\sum_{k=-\rho}^{\rho-1} 
\exp(i \frac{n \pi}{\rho}k)\,a_k.
\end{equation}
The superscript RS indicates that we call this the real-space mode of representation; we will justify this label momentarily. When we wish to return to the original, $k$-space representation we employ the inverse QFT: 
\begin{equation}
\ket{\psi}_{x}^\text{KS}=U_\text{QFT}^{-1}\ket{\psi}_{x}^\text{RS}=\sum_{k=-\rho}^{\rho-1} a_k \ket{k},
\label{app:transformingToK}
\end{equation}
where of course
\begin{equation}
a_k=\frac{1}{\sqrt{2\rho}}\sum_{n=-\rho}^{\rho-1} 
\exp(-i \frac{k \pi}{\rho}n)\,b_n.
\label{app:ak_from_bn}
\end{equation}

We can now ask, what wavefunction must be represented by each computational basis state $\ket{n}$ appearing in Eqn.~(\ref{app:transformingToR})? We can answer by setting all $b$ coefficients to zero except for $b_n$ (i.e. $b_m=\delta_{m,n}$), transforming to $k$-space using Eqns.~(\ref{app:transformingToK}) and (\ref{app:ak_from_bn}), and finally referring back to our original declared mapping in Eqn.~(\ref{appeqn:k-element-map}). We find that each $\ket{n}$ maps to a wavefunction peaked at (but not strictly localised around) the spatial point $x_n=\frac{n}{\rho}\frac{L}{2}$. Specifically, the inferred mapping is
\begin{equation}
\phi_n(x)=P_{n_r}^n(x)
\ \ \ \leftrightarrow\ \ \
\ket{n}
\label{appeqn:1Dpixel}
\end{equation}
with
\begin{equation}
P^n_{n_r}(x)=
\exp({-\frac{i\pi x^\prime}{L}})\sqrt{\frac{2}{\rho L}}
\ \sum_{j=1}^{\rho}\cos\frac{\pi(2j-1)x^\prime}{L}
\label{appeqn:Pfunction}
\end{equation}
where $x^\prime=x-x_n$ with $x_n= \frac{nL}{2\rho}$, and $\rho=2^{n_r}-1$. 

Within our simulation box $-\frac{L}{2}<x<\frac{L}{2}$, the function $P^n_{n_r}(x)$ serves the role of an approximation or ``smear" of the Dirac delta function $\Delta(x-x_n)$, the sharpness increasing as $2^{n_r}$ tends to infinity (strictly speaking it tends to a Dirac comb since the function is periodic beyond the simulation box; the function  can also be recognised as a difference of two Dirichlet kernels e.g. for $L=\pi$, $P^0_{n_r}(x)\propto D_{2\rho}(x)-D_{\rho}(2 x)$). Shortly we will motivate the informal term `pixel functions' for our $P_{n_r}^n(x)$. The separation between the peaks of adjacent pixel functions, e.g. $x_2-x_1$, is 
$\delta r=L/2^{n_r}$, and
we now define the model's {\it spatial resolution} as the reciprocal of this quantity, i.e. as the number of pixel functions per unit distance:
\begin{equation}
\delta r^{-1}=2^{n_r}/L.
\label{appeqn_resolution}
\end{equation}

For the present paper, we note that there are four key observations regarding the set of spatial wavefunctions $P_{n_r}^n(x)$ which are true for any given qubit count $n_r$. They are mutually orthogonal, each is peaked at $x_n$, and importantly each is zero at all points $x_{m\neq n}$. That is to say, where a given spatial wavefunction has its primary peak, all other wavefunctions are strictly zero. Finally, we note that at its peak each function has the value 
\[
P_{n_r}^n(x=x_n)=\frac{1}{\sqrt{\delta r}}.
\]

These properties mean that a suitable decomposition to represent any 1D wavefunction $\Psi(x)$ in our quantum register is very intuitive:
\begin{equation}
\Psi(x)\approx
\frac{C}{\sqrt{\delta r}}\sum_{n=-\rho}^{\rho-1} \Psi(x_n)P^n_{n_r}(x)
 \leftrightarrow\ 
\frac{C}{\sqrt{\delta r}}
\sum_{n=-\rho}^{\rho-1}\Psi(x_n)\ket{n}\ 
\label{appeqn:1DpixelatedWF}
\end{equation}
i.e. the required amplitude of $\ket{n}$, the state representing the wavefunction peaked at point $x_n$, is found simply by sampling the target wavefunction at that point. Here $C$ is a normalisation constant that will be close to unity providing that (a) the target wavefunction has negligible amplitude outside of the simulation box and (b) the target wavefunction varies slowly with respect to $\delta r$. 
If $C$ differs significantly from unity then the model does not properly capture $\Psi(x)$, however $C$ can be above or below unity and $C=1$ does not imply optimality.

So far we have considered the role of one sub-register, which corresponds to one dimension of a given particle. The generalisation to 2D or 3D is the natural one: the sub-registers tensored together to form the complete representation of a given particle. The 3D analogue is
\begin{eqnarray}
\Psi(\mathbf{r}) &\approx& C\delta r^{-\frac{3}{2}}\sum_{\mathclap{n,m,l=-\rho}}^{\mathclap{\rho-1}} \Psi(x_n,y_m,z_l) P^n_{n_r}(x)P^m_{n_r}(y)P^l_{n_r}(z) \nonumber \\  
&\leftrightarrow& \ 
C\delta r^{-\frac{3}{2}}\sum_{\mathclap{n,m,l=-\rho}}^{\mathclap{\rho-1}} \Psi(x_n,y_m,z_l) \ket{n} \ket{m}\ket{l}.\label{appeqn:3DrealSpace}
\end{eqnarray}
(Our 2D numerical results of course omit the third sub-register.) When we generalise to represent a $P$-particle wavefunction $\Psi(\mathbf{r}_1,...,\mathbf{r}_P)$ we need only extend in the natural fashion,
\begin{equation}
C\delta r^{-\frac{3P}{2}}\sum_{\mathclap{\{n_1..l_P\}=-\rho}}^{\rho-1} \Psi(x_{n_1},y_{m_1},...,z_{l_P}) \ket{n_1} \ket{m_1}...\ket{l_P}.\label{appeqn:3DrealSpacePparticles}
\end{equation}

The first figure in the Introduction section of the main text shows the ground state of 2D hydrogen plotted from the analytic solution (lower left side, cyan) and constructed as a approximation using the 2D variant of Eqn.\,\ref{appeqn:3DrealSpace} for the case of $n_r=6$ qubits per sub-register (right side, orange). This wavefunction will be discussed in more detail presently, and is one of the initial states we will use in our numerical simulations. It is apparent by inspection that the approximation is good over the majority of the simulation box. The analytic state of course extends beyond the box, where our modelled state is undefined; taking the modelled state to be zero there, the fidelity of the modelled state with respect to the analytic state (integrating over all space) is $0.99946$. If instead we project the analytic state into the box and renormalize it, then the infidelity between the model and the analytic solution falls below $10^{-4}$. The discrepancy is primarily localised at the Coulomb singularity, where the analytic expression has a gradient discontinuity. Of course, the reconstruction must have a continuous gradient (since all constituent functions have this property). The sharpness of the approximation is characterised by $\delta r^{-1}$ and so increases exponentially with $n_r$. 

We also show examples of $P_{n_r}^n(x)$ for the same case of $n_r=6$ (but with the phase component removed so that the plots can be seen without an imaginary axis) in the same Figure.  Recall that in our two's complement notation, integer indices can denote negative or positive coordinates; the register state $\ket{00..0}=\ket{n=0}$ corresponds to the wavefunction peaked at the origin (the blue lines in the left and middle panels). The orthogonality property is clear from the middle panel, showing $\ket{n=0}$ in blue and $\ket{n=1}$ in orange.

Intuitively, one can think of these spatial wavefunctions analogously to pixels as used in conventional digital photographs: A pixel has a single colour corresponding to the (real, continuous) scene sampled at that point. Similarly each spatial basis state $\ket{n,m,l}$ in Eqn.\,\ref{appeqn:3DrealSpace} is associated with a single complex number corresponding to the continuous wavefunction sampled at that point. A photo should have sufficient pixels that all features of interest to the viewer are fully captured, and our simulation should have sufficiently many spatial basis states that all features of the wavefunction are adequately captured. Because of this close analogy, we sometimes refer to the $P^n_{n_r}$ wavefunctions as `pixel functions'. As we will see presently, a moderate resolution $\delta r^{-1}$ can prove to be sufficient and thus the number of qubits required can be remarkably low.

It is obvious that pixel wave functions resemble standard $\sinc$ functions; indeed if we increase the number of qubits $n_r$ while commensurately increasing $L$ so that the resolution $\delta r^{-1}$ is held constant, then the pixel function $P^0_{n_r}$ tends to a sinc function in the interval $[-\frac{L}{2},\frac{L}{2}]$. This is reminiscent of Whittaker-Shannon interpolation and indeed there is a close association between the grid-based quantum model and the theory of signal processing.

The representation described above involves simple Cartesian grids in momentum-space and position-space, with the conversion effected by a QFT; in principle, other coordinate systems could be used (cylindrical, spherical etc) or alternative transformations~\cite{boyd2001chebyshev}. In the present paper we will consider the hybrid coordinates needed to evaluate particle-particle interactions. 

A spin degree of freedom is trivial to add into the simulation: for spin one-half particles we would simply require a single additional qubit per complete register, while a spin-$S$ particle would require $\ln(S)$ qubits. However in our results, we do not consider any Hamiltonians with a spin-dependent term and therefore we do not model the particle's spin. Importantly, particle symmetry is properly preserved by the SO-QFT; symmetry can be confirmed at any stage using e.g. a SWAP test~\cite{PhysRevLett.87.167902} on the registers representing the relevant pair of particles. 

\subsection{Split-Operator Time Propagation}
We now make some remarks regarding the SO-QFT method.
We first note that Both the first-order ($O(\delta t^2)$) and second-order ($O(\delta t^3)$) Trotterisation split-operator have well-documented numerical advantages \cite{doi:10.1063/1.5092611, doi:10.1063/1.5094046, tannor2007introduction, Kosloff1988}, and higher order Lie-Trotter-Suzuki sequences can also be relevant~\cite{suzuki1991,childsTrotter2021}. The real and $k$-space Cartesian grid dual representations are natural options for state representation when computing the approximate time evolution operator. In the $k$-space representation, the kinetic part of the Hamiltonian $\hat{H}_\text{kin}$ is separable and exactly local (diagonal). In the real-space representation the interaction part of the Hamiltonian $\hat{H}_\text{int}$, is approximately diagonal.

On a quantum computer it is relatively efficient to switch between the two representations using the QFT; the number of gates required per sub-register is quadratic in its number of qubits $n_r$, and all sub-registers can be transformed independently. Thus we can implement the two parts of the SO cycle each in their preferred, diagonal basis. We perform the following to approximate a time step $\delta t$ using our quantum computer:
\begin{equation}
\ket{\psi(t+\delta t)}^\text{RS}=U_\text{SO}(\delta t)\,\ket{\psi(t)}^\text{RS}
\nonumber
\end{equation}
where we recall that the superscript RS denotes the real space representation, and
\begin{equation}
U_\text{SO}(\delta t)=e^{-iD_\text{int}\delta t}\left(U_\text{QFT}^\dagger\ e^{-iD_\text{kin}\delta t}\  U_\text{QFT}\right).
\label{appeqn:usingD}
\end{equation}
Here $D_\text{kin}$ and $D_\text{int}$ are diagonal real matrices as explained above, and $U_\text{QFT}$ is the quantum Fourier transform applied to all sub-registers. 

We now discuss the evaluation of these operators on quantum computers in detail. 
We observe that (regardless of basis) any period of evolution under purely the kinetic part $\hat{H}_\text{kin}$, i.e. any operator
\[
U_\text{kin}(\delta t)=e^{-i \hat{H}_\text{kin} \delta t}
\]
will separate exactly into a product of operators acting independently on each particle and in each dimension:
\[
U_\text{kin}(\delta t)=\prod_{p=1}^P\prod_{q\in \{x,y,z\}}\exp(i\frac{\hbar^2\delta t}{2m_p} \frac{\partial^2}{\partial q^2})
\]
because these components commute. Now suppose that a given quantum sub-register is currently in the k-space representation defined by  Eqn\,(\ref{appeqn:kSpaceMapping}), and recall Eqn.\,(\ref{appeqn:k-element-map}) for the meaning of each computational basis state, viz.
\[
\phi_k(x)= L^{-\frac{1}{2}}\,\exp(\frac{i 2\pi k x}{L})\ \ \leftrightarrow\ \ 
 \ket{k}.
\]
We see that the proper way to account for the action of $U_\text{kin}(\delta t)$ on the molecule is to introduce phases onto our computational basis states according to
\begin{equation}
\ket{k}\,\Rightarrow\,e^{i C\delta t\, k^2 }\ket{k}
\label{appeqn:littleKineticTime}
\end{equation}
with constant $
C=2\hbar^2\pi^2/(L^2m_p)$.
Thus our task is simply to apply phase operations to our quantum computer, independently for each sub-register (and in parallel if our hardware has that capability). Each element of the matrix $D_\text{kin}$ appearing in Eqn.\,(\ref{appeqn:usingD}) is just the sum of $C(k_x^2+k_y^2+k_z^2)$ over all particles. It is trivial to implement the required phases as, for example, a sequence of single- and two-qubit phase gates (see for example \cite{Ollitrault2020, Cody_Jones_2012}).
The number of gates required goes as the square of the number of qubits in the register.

We now consider the challenge of modelling evolution under the potential parts of $\hat{H}_\text{tot}$. Consider first $\hat{H}_U$, which represents the energy of each particle in classical fields. We are primarily interested in an attractive Coulomb potential representing a nucleus (although we discuss a variation including a static electric field presently). For the Coulombic case we write 
\begin{equation}
\hat{H}_\text{U}=\sum_{p=1}^P\frac{Q_p}{|\mathbf{r}|_p}=Q\sum_{p=1}^P \frac{1}{\sqrt{x_p^2+y_p^2+z_p^2}}.
\label{eqn:HnuclearOnepart}
\end{equation}
Here we took $Q$ outside of the sum since all $Q_p$ are the same in our atomic and molecular systems of interest; however there would be no difficulty in retaining distinct values. As before, we can write an time evolution operator 
\[
U_\text{U}(\delta t)=\prod_{p=1}^P\exp(-i\frac{Q\delta t}{\sqrt{x^2+y^2+z^2}}).
\]
The operations are independent between the registers corresponding to different particles, but not independent between sub-registers assigned to a given particle. 

We will be in the real space representation when we implement the corresponding dynamics. Recall that the computational basis states of each sub-register will thus correspond to locally-peaked 1D single-particle wavefunctions,
\[
\phi_n(x)=P_{n_r}^n(x)
\ \ \ \leftrightarrow\ \ \
\ket{n}
\]
with each such `pixel function' having its primary peak at $x_n=nL/2^{n_r}=n\delta r$. Similar expressions describe the roles of the $y$ sub-register states $\ket{m}$, peaked at $y_m$, and the $z$ sub-register states $\ket{l}$ peaked at $z_l$.

We proceed to apply a period of time evolution under $\hat{H}_U$ using an approximation that {\it would} be exact in the limit of infinite spatial resolution. In that case we could apply a series of phase changes to our quantum registers to properly describe any time step $\delta t$.
\begin{eqnarray}
\ket{n}\ket{m}\ket{l}\,&\Rightarrow&\,
\exp(\frac{-iQ\,\delta t}{\sqrt{x_n^2+y_m^2+z_l^2}} )\ket{n}\ket{m}\ket{l}\nonumber\\
&=&\, \exp(\frac{-iQ\,\delta t}{\delta r\sqrt{n^2+m^2+l^2}} )\ket{n}\ket{m}\ket{l}\ \ 
\label{appeqn:littleNuclearTime}
\end{eqnarray}

We see that the task of implementing the latter is only modestly more complex than the former: Now, we need to compute on an entire register of all three sub-registers and the phases required are inverse-square-root instead of merely square. Efficient evaluation of functions such as the inverse-square-root on quantum computers is an active area of development. In particular, Kassal\,\textit{et al.}~\cite{Kassal2008}, Jones\,\textit{et al.}~\cite{Cody_Jones_2012} and H\"aner \textit{et al.}~\cite{haner2018optimizing} proposed procedures for reversible computation of the inverse-square-root using fixed-point arithmetic and Newton-Raphson iteration. 
Polynomial function interpolation using quantum read-only memory (QROM) type circuits is another promising approach~\cite{PRXQuantum.1.020312}.
Recent work from Poirier\,\textit{et al.}~\cite{poirier2021efficient, poirier2021fulldimensional} combined with other formulations of the Coulomb potential may also ease the efforts of computing the inverse-square-root. 
We refer to \cite{haner2018optimizing, Cody_Jones_2012} for resource estimates, but here it suffices to note that the number of gates required can potentially scale merely quadratically with the number of qubits $n_r$.
We discuss this further in Supplementary\,\ref{appendix:numerical}.

We motivated the expression above by noting that this would be the exactly correct process if our spatial states were Dirac delta functions. Those states in fact have finite spatial extent but we can expect that, for any sufficiently short time $\delta t$ and given a smoothly varying potential, an adequate spatial resolution $\delta r^{-1}$ will result in dynamics that converge to the exact behaviour. However the Coulomb potential is singular at $\mathbf{r}=\mathbf{0}$ and consequently we will need to investigate the behaviour in this region carefully. As we presently explain, it is possible to augment the basic SO-QFT cycle of kinetic and potential evolution with a third phase, whose purpose is to stabilise behaviour at the core singularity.

Finally we must consider the part of the Hamiltonian corresponding to particle-particle interactions $\hat{H}_\text{V}$. This is straightforward to handle using the same approximation above, i.e. by updating our registers {\it as if} the states they represent are Dirac delta functions. Then the operator,
\begin{equation}
U_\text{V}(\delta t)=\exp(i\delta t\sum_{p,q=1; p\neq q}^P \frac{q_{p,q}}{|{\mathbf r}_p-{\mathbf r}_q |})
\end{equation}
will be approximated as having diagonal form (and will therefore commute with the $U_U(\delta t)$ operator). Consider the two registers, each composed of three sub-registers, which represent a given pair of particles $p=1$ and $p=2$. Then basis states will be updated as 
\begin{equation}
\ket{n_1\,m_1\,l_1}\ket{n_2\,m_2\,l_2}
\,\Rightarrow\,
e^{-i \Theta}\ket{n_1\,m_1\,l_1}\ket{n_2\,m_2\,l_2}
\nonumber
\end{equation}
where
\begin{eqnarray}
\Theta&=& \frac{Q\,\delta t}{\sqrt{(x_{n_1}-x_{n_2})^2+(y_{m_1}-y_{m_2})^2+(z_{l_1}-z_{l_2})^2}} \nonumber\\
&=& \frac{Q\,\delta t}{\delta r\sqrt{(n_1-n_2)^2+(m_1-m_2)^2+(l_1-l_2)^2}}
\label{appeqn:interactingForm}
\end{eqnarray}
One means of computing the required phases is discussed in the main text. We would pair every particle with a partner and perform the following process (which can occur simultaneously over each pair of registers), before repeating with a different pairing, and so on until all pairings are considered. For each pair, we perform the computation
\begin{eqnarray}
&&\ket{n_1,m_1,l_1}\ket{n_2,m_2,l_2}
\,\Rightarrow\,\nonumber\\
&&\ \ \ \ket{n_1-n_2,m_1-m_2,l_1-l_2}\ket{n_2,m_2,l_2}
\label{eqn:subtraction}
\end{eqnarray}
which is a straightforward instance of quantum arithmetic on pairs of sub-registers. One suitable method uses the QFT which of course our machine will already have been optimised for~\cite{draper2000addition}. This approach is frugal in terms of the qubit count: we need only a single additional qubit to fully represent each difference $n_1-n_2$, etc (as the maximum magnitude is within a factor of two of the maximum magnitudes of $n_1$ and $n_2$). With the registers in this form, we can simply use the same procedure employed for the single-particle phases. Once the desired phases are applied, we perform the reverse of the computation Eqn.\,(\ref{eqn:subtraction}) and then repeat the whole process with another set of particle pairings. For a system of $P$ particles, the total number of particle pairings is $\frac{1}{2}P(P-1)$. However as we discuss in the Results section of the main text, there is a natural parallelism so that the time cost should scale only linearly with $P$. Moreover, there are methods to trade greater hardware resources to reduce the time cost; Jones~\textit{et al.} introduced a parallelised scheme for speeding up the computation of the particle pairings using only $O(\log P)$, or even $O(1)$, circuit depths~\cite{Cody_Jones_2012}. Regardless of the implementation method used, the elements of the diagonal matrix $D_\text{int}$ are thus simply sums of phases.

In the approach just described, we are finding the relative coordinates of each particle pairing. There was no need to compute the complete transformation by mapping the second particle's register to $\ket{n_1+n_2,m_1+m_2,l_1+l_2}$. However, this would have been possible. A motivation for completing this full transformation is that general, non-diagonal operations could be performed if we wished; this observation relates to the augmented split-operator concept developed presently.

\section{Numerical modelling}\label{appendix:numerical}
Here we describe further details of the hardware, software, and configuration details relating to the numerical simulations presented in the main paper.

\subsection{Emulation software and hardware}
\label{appendix:classicalResources}
All numerical results in this manuscript were ultimately obtained through the Quantum Exact Simulation Toolkit ({QuEST})~\cite{QuESTandHPC}, though interfaced through the \texttt{Python} {pyQuEST}~\cite{pyquest} and \texttt{Mathematica} {QuESTlink}~\cite{QuESTlink} software packages. The core QuEST simulator is written in \texttt{C} behind a hardware agnostic interface, allowing redeployment of the simulations in this manuscript between laptops, GPUs and distributed supercomputing facilities. This enabled multithreaded and GPU simulation of the modestly sized systems, such as the $25$-qubit 2D Hydrogen scattering presented, directly within QuESTlink. While open-source, we note that the use of QuESTlink requires a Wolfram Engine environment, obtainable either through the commercial Mathematica product, or the recently released Wolfram Engine standalone~\cite{wolf_eng}.

Results for the largest scale systems we considered were generated with pyQuEST, which enables relatively low-level access to QuEST simulation primitives through a high-level \texttt{Python} interface. Importantly, pyQuEST inherits the capacity to run distributed tasks, where the numerical representation of a quantum state is partitioned between compute nodes cooperating over a network. This allows both the representation of states too large to fit into the memory of any single compute node, and their concurrent simulation -- each multicore node is further able to parallelise its local simulation tasks through multithreading. In this manuscript, distributed pyQuEST was used to emulate 36-qubit quantum computers in our study of 3D helium, employing up to 32 compute nodes of the Oxford Advanced Research Computing (ARC) facility~\cite{oxfordARC}. Each node contains 48 CPU cores, and took roughly 52 hours ($\approx 50\,000$ core hours) to process its 64\,GiB partition of the full 1\,TiB quantum state-vector.

\subsection{Emulation of SO-QFT}

The QuEST family of emulators perform numerical simulation at the level of individual gates. Supported gates include all commonly used operations, and general unitaries of any number of control and target qubits, and therefore all operations required of the quantum algorithms presented in this manuscript.
However, completing this work required the repeated simulation of circuits within which a series of contiguous gates could be more efficiently effected by a single invocation of a bespoke function. We implemented this optimisation to accelerate the simulations in this manuscript, which has since been integrated into the QuEST emulators. The interested reader may view these documented facilities \href{https://quest-kit.github.io/QuEST/group__type.html#gaa7d869b117ba5024d6b84938e8cdfc65}{here}. We now describe these optimisations, and where they were invoked, in detail.

In the Methods section in the main text, we described how our grid-based description of 3D multi-electron systems meant that particle-field and particle-particle interactions
admit a unitary time evolution operator which can be split into step operators
\begin{gather}
    U_{\text{U}} \ket{n\,m\,l} = \exp(-i\frac{Q\,\delta t}{\sqrt{n^2 + m^2 + l^2}}) \ket{n\,m\,l}, \label{eq:appendix_splitop_nml} \\
    \tag*{}
    U_{\text{V}} \ket{n_1 m_1 l_1}\ket{n_2 m_2 l_2} = \exp(- i \Theta)
    \ket{n_1 m_1 l_1}\ket{n_2 m_2 l_2}, \\
    \tag*{}
    \Theta = 
    \frac{2^{1-n_r} Q \delta t}{\sqrt{ 
        (n_1-n_2)^2 + 
        (m_1-m_2)^2 + 
        (l_1-l_2)^2
    }},
\end{gather}
where sub-register $\ket{n\,m\,l}$ encode $(z,y,x)$ coordinates of a single particles, which may be negative (encoded with two's complement binary).  An experimentalist must effect these ($d \,n_r$)- and ($2 \,d \,n_r$)-qubit operators through $\mathcal{O}({n_r}^2)$ single-qubit and double-qubit phase gates. However, we can leverage that $U_{\text{U}}$ and $U_{\text{V}}$ are diagonal in the real space representation to numerically simulate them cheaper than even a \textit{single} general unitary gate, in an \textit{embarrassingly parallel} manner. 
We formally present such a scheme to effect $U_{\text{U}}$, the simplest of the split-operators, upon a distributed statevector in Algorithm~\ref{alg:phase_func_bespoke_sim}. For clarity, we have excluded the pseudocode for additional functionality critical to our actual implementation, such as the ability to override the phases at particular $l,m,n$ values to avoid phase divergences.

    \begin{algorithm}[tb]
    \caption{
        Embarassingly parallel distributed simulation of the 3D $U_{\text{U}}$ split-operator upon an $N$-qubit statevector $\ket{\psi}$, which is uniformly distributed between $2^k$ nodes. $\vec{\psi}$ is the $2^{N-k}$ vector of complex amplitudes stored in each node, with $i$-th element $\vec{\psi}[i]$, indexed from $0$. Each node has a unique rank $0 \le r < 2^k$. Integers $q_l$, $q_m$, $q_n$ are the starting qubit indices of the contiguous $n_r$-qubit registers which together form substate $\ket{n}\ket{m}\ket{l}$ of Equation~\ref{eq:appendix_splitop_nml}, via a two's complement signed binary encoding. 
        Symbols $\&$ and $\gg$ notate bit-wise AND and bit right-shift operators respectively.
        The outer \textbf{for} loop of our algorithm is trivially parallelised using multithreading or GPU acceleration.
    }
    \label{alg:phase_func_bespoke_sim}
    
    \hspace{1pt} \textbf{apply}$\bm{U}_{\mathbf{U}}$($\vec{\psi}$, $Q$, $\delta t$, $q_l$, $q_m$, $q_n$, $n_r$)
    
    \hspace{10pt}
    \codecomment{iterate every local basis state}
    
    \hspace{10pt}
    \textbf{for} $i$ \textbf{in} $\{0, \; \dots, \; 2^{N-k}-1\}$
    
    \hspace{10pt}\hspace{10pt}
    \codecomment{determine global index of basis state}
    
    \hspace{10pt}\hspace{10pt}
    $j = r \, 2^{N-k} + i$
    
    \hspace{10pt}\hspace{10pt}
    \codecomment{determine sub-register values of basis state}
    
    \hspace{10pt}\hspace{10pt}
    $l$ = \textbf{getRegVal}($j$, $q_l$, $n_r$)
    
    \hspace{10pt}\hspace{10pt}
    $m$ = \textbf{getRegVal}($j$, $q_m$, $n_r$)
    
    \hspace{10pt}\hspace{10pt}
    $n$ = \textbf{getRegVal}($j$, $q_n$, $n_r$)
    
    \hspace{10pt}\hspace{10pt}
    \codecomment{evaluate the phase}
    
    \hspace{10pt}\hspace{10pt}
    $\theta = Q \; \delta t \; / \sqrt{l^2 + m^2 + n^2}$
    
    \hspace{10pt}\hspace{10pt}
    \codecomment{modify the amplitude}
    
    \hspace{10pt}\hspace{10pt}
    $\vec{\psi}[i] = \exp(-i \, \theta) \; \vec{\psi}[i]$
    
    \vspace{5pt}
    
    \hspace{1pt}
    \codecomment{returns $(b\ge0)$-th bit of integer $j$}
    
    \hspace{1pt}
    \textbf{getBit}($j$, $b$)
    
    \hspace{10pt}
    \textbf{return} $(j \gg b) \; \& \; 1$
    
    \vspace{5pt}
    
    \hspace{1pt}
    \codecomment{returns the signed value of the $n_r$ contiguous qubits, starting at index $q$, in the $j$-th basis state}
    
    \hspace{1pt}
    \textbf{getRegVal}($j$, $q$, $n_r$)
    
    \hspace{10pt}
    $v = 0$
    
    \hspace{10pt}
    \textbf{for} $k$ \textbf{in} $\{0, \; \dots, \; n_r-2\}$
    
    \hspace{10pt}\hspace{10pt}
    $v = v + 2^k$ \textbf{getBit}($j$, $q+k$)
    
    \hspace{10pt}
    \textbf{if} \textbf{getBit}($j$, $q+n_r-1$) \textbf{is} 1 
    
    \hspace{10pt}\hspace{10pt}
    $v = v - 2^{\,n_r-1}$
    
    \hspace{10pt}
    \textbf{return} $v$
    
    \end{algorithm}

Use of Algorithm~\ref{alg:phase_func_bespoke_sim} and several similar strategies for the other split-operators was crucial in order to numerically study the various simulation scenarios presented in the main paper without incurring impractical time costs.



%
We stress that the use of these high level functions, which do not involve specifying a circuit-level description, does not compromise the exact nature of our numerical emulation of a quantum processor. It is simply an expediency allowing us to arrive at that the exact behaviour with a more modest use of classical resources that would be required if we were to explicitly use gate-level operations throughout. 
In the real device, such circuit-level prescriptions must of course be used. 


For the $U_\text{kin}$ operator, which simply applies a phase dependent on the square of the binary number $k$ in each of the system's sub-registers, this is straightforward: we require only a series of single- and two-qubit phase gates applied directly to the sub-registers~\cite{somma2016quantum, Ollitrault2020}. Of $O(n_r^2)$ such gates would be required for each sub-register of $n_r$ qubits.

The implementation of $U_\text{U}$ and $U_\text{V}$ is more complex as it involves the Coulomb potential $1/|\mathbf{r}|$. In the main paper, we describe the arithmetic that occurs in the case that we are applying $U_\text{V}$ and therefore require sub-registers to represent $x_i-x_j$ etc. One could then apply a circuit $C$ to compute the desired phase $~1/r$ into an ancilla register (i.e. a binary representation of the desired phase to any given accuracy), followed by actually implementing that phase by a series of single-qubit phase gates applied to the ancilla qubits. Finally one would apply circuit $C^\dagger$ to un-compute the ancilla's state, disentangling it from the main registers.

Any classical algorithm that computes $1/\sqrt{x^2+y^2+z^2}$ from inputs $x$, $y$, $z$ will in principle suffice; such an algorithm might be expressed in terms of irreversible operations (AND, OR, XOR etc) but any such circuit be be re-expressed as classical reversible circuit and thus provide a suitable $C$. It is of course interesting to optimise this, especially since quantum gate sets are can potentially implement classically-reversible circuits more efficiently and we will wish to be as frugal as possible with the use of ancilla qubits. However even an inefficient implementation would suffice in the sense that it would not alter the overall scaling of the SO-QFT method: the computation of $1/r$ happens at the register level, and register size scales only very weakly with the complexity of the simulated system as argued in the main text.


\section{Multi-qubit Phase Estimation and Fourier Analysis}\label{appendix:fancyPhase}
In order that the present paper can be a useful resource for introducing the grid-based method, we now summarise the standard multi-qubit phase estimation method using the present paper's notation. This method is of course more powerful than the frugal single-ancilla technique described in the main paper.

The logarithmically-scaling phase estimation of Kitaev~\cite{kitaev1995quantum},  trades higher qubit overhead for measuring the phase with far less sampling; in suitable cases, even in a single shot. Consider an ancillary register with $S$ qubits and a working register containing the state of interest. We first conditionally apply $2^{S-1}N$ split-operator steps to the state, controlled by the most significant ancillary qubit. We then repeat by conditionally applying $2^{S-2}N$ split-operator cycles, now controlled by the next most significant ancillary qubit, and repeat again until every ancillary qubit has been used and the least significant qubit controls only $N$ split-operator steps. At this point the quantum computer is in the state
\begin{equation}
    \frac{1}{\sqrt{2^S}}\sum^{2^S-1}_{m=0} \ket{m} U^{mN}(\delta t)  \ket{\Psi}.
\end{equation}
If $\ket{\Psi}$ is an eigenstate $\ket{\Psi_n}$ with energy $E_n$, the state is thus
\begin{align}
    & \frac{1}{\sqrt{2^S}}\ket{\Psi_n}\sum^{2^S-1}_{m=0} e^{-iE_nmN\delta t} \ket{m} \nonumber\\
    =& \frac{1}{\sqrt{2^S}}\ket{\Psi_n}\sum^{2^S-1}_{m=0} e^{-i\phi m} \ket{m}
\end{align}
where $\phi=E_nN\delta t$. The phase information $\phi$ is now encoded in the ancillary qubit register, and can be efficiently extracted using an inverse QFT on the ancillary register. Indeed, if $\phi=2^k$ where $k\in \mathbb{N}$, then a QFT of the ancillary register should yield the binary representation of $k$ upon measurement.

\begin{figure}[!tbp]
    \centering
    \includegraphics[scale=0.6]{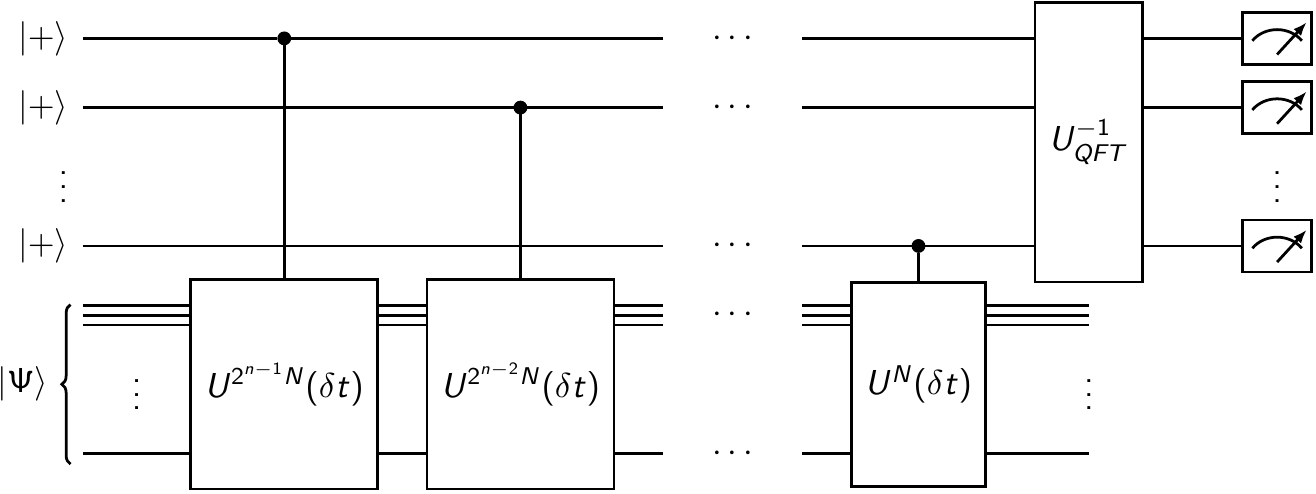}
    \caption{Generalisation of the single-ancilla IPE to the `canonical' phase estimation using an $n$-qubit ancilla register.  The $i^\text{th}$ most significant ancilla qubit controls the propagation of $2^{n-i}N$ split-operator steps in the main register.}
\end{figure}

We also mention that classical Fourier analysis of the autocorrelation signal can also reveal the power spectrum, with resonance peaks that correspond to the stationary states of the system. This is in many ways the classical analogue of the canonical phase estimation method. Specifically, the power spectrum is peaked at the eigenvalues of $\hat{H}_\text{tot}$ with heights corresponding to $|c_n|^2$
\begin{equation}
    \frac{1}{\sqrt{2\pi}}\int \bra{\Psi(t)}\ket{\Psi(0)} e^{i\omega t}dt = \sqrt{2\pi}\sum_n|c_n|^2\delta(\omega-E_n)
\end{equation}
In practice, the discrete and finite time integration limits the resolution of the power spectrum, given by
\begin{equation}
    \delta \omega = \frac{2\pi}{M_\text{tot}N\delta t}
\end{equation}
where $M_\text{tot}$ is the total number of time samples.


\section{State Preparation}\label{appendix:statePrep}
\subsection{State loading}
As discussed in the main paper, our use of emulated quantum computers means that we can side-step the challenge of preparing a known, analytic initial state on our device -- we simply load it into the emulator's memory. For completeness, we briefly review some of the literature that describes methods and costs for real quantum processors.

In 2009, Ward \textit{et al.}~\cite{doi:10.1063/1.3115177} extended the work of Zalka \cite{doi:10.1098/rspa.1998.0162} to develop such a method for generating physically relevant states in grid representations, which we summarise here. We begin with the task of loading suitable initial states into the particle quantum register.
The objective is to distribute a single particle state $\phi$ centred about $(x_0, y_0, z_0)$ within a simulation box of width $L$ across the amplitudes of an $N$-qubit register
\begin{equation}
    \ket{0}\ket{0}\ket{0} \Rightarrow \phi\left(\mathbf{r}\right) \ket{n}\ket{l}\ket{m}
\end{equation}
where $\mathbf{r} = (n\delta r-x_0, l\delta r-y_0, m\delta r-z_0)$ and $\delta r = L/2^{n_r}$. $L$ must be sufficiently large to capture the significant parts of the wavefuntion and the sum of all the amplitudes squared should be unity. In 1998, Zalka proposed a constructive algorithm for loading single particle distributions onto a register \cite{doi:10.1098/rspa.1998.0162}, which was subsequently either rediscovered or generalised, most notably by Grover and Rudolph in 2002 \cite{grover2002creating}, in a number of contexts \cite{kaye2001quantum, 10.5555/2011670.2011675, marinsanchez2021quantum, kitaev2008wavefunction}. The method splits the norm of the single particle function $N$ times across the $N$ qubits by performing qubit rotations, the angles for which are determined by computing exponentially many integrals. Complex-valued functions will require additional procedures for including the phase in each amplitude. There are conflicting opinions on the efficiency of the method; some argue that the method has an exponential $O(2^N)$ overhead \cite{9259933, marinsanchez2021quantum}, while others claim that it can be efficient \cite{PhysRevApplied.15.034027}, perhaps using quantum computers to solve the integrals \cite{doi:10.1063/1.3115177}. In recent years, more novel approaches for initial state loading techniques have been proposed. 
Ollitrault \textit{et al.} prepared Gaussian amplitudes using hybrid variational algorithms~\cite{Ollitrault2020}.
Holmes and Matsura used matrix product states to represent functions, which maps to state preparation circuits with linear-depth~\cite{9259933}. 
In particular, a noise-resistant technique for efficient preparation of distributions from Rattew \textit{et al.} \cite{rattew2021efficient, Rattew2022} shows promise for generalisation to the single particle functions required.

\subsection{Alternative antisymmetrisation method}
\label{appendix:secondAntisymmMethod}
In the main paper we described, and numerically tested, a means for antisymmetrising a state; the input to the procedure is simply as a tensor product of $P$ unique single particle states represented by our $P$ registers. 

We now outline the second approach, also exploiting $H_\text{synth}$. Instead of fully representing all the tag registers, we can consider what happens if we compute and uncompute the energies of the states $\ket{\psi}$ `on the fly'. We assume that we have available a sorting network formed of swap operations and `controller qubits' that determine whether each swap operation will occur. Moreover we take it that these controller qubits are already properly primed in the appropriate superposition to create an antisymmetric permutation (as per the method of Berry {\it et al.}~\cite{berry2018improved}). As we push our initially canonically-ordered state `backwards' through this network, as per Berry, we must erase each controller qubit as we go: wherever two inputs $\ket{A}\ket{B}$ have indeed been swapped to $\ket{B}\ket{A}$, we must now flip the controller qubit from $\ket{1}$ to $\ket{0}$. 

Ref.~\cite{berry2018improved} describes this erasure for the case of integer labels, noting that if $\ket{A}$ and $\ket{B}$ are any two superpositions of integers, but all integers in $\ket{A}$ are smaller than all those in $\ket{B}$, then the reordering implied by the swap can be used to erase the controller. We must instead erase the controller qubit with an analogous condition based on the energies of our grid-based states. This can be achieved by introducing a set of $t$ ancilla qubits in the state $\ket{+}^{\otimes t}$ and performing a phase estimation process, where the two registers involved in the swap operation are subject to controlled-$e^{i\,H_{synth}C\pi}$ and controlled-$e^{-i\,H_{synth}C\pi}$, respectively. Here, as earlier, $C$ denotes the appropriate power of two for each ancilla qubit. As a result, the ancilla now encodes the QFT of a number that is either positive or negative; by performing an inverse-QFT, that sign is made manifest as the most-significant bit, and can be used to unset the swap controller qubit before the entire computation is reversed.

This latter approach has the merit that not all tag registers need to coexist simultaneously; the process can occur sequentially using only $t$ rather than $P\,t$ qubits (albeit incurring a time cost). However the method is more computationally costly, requiring more phase-estimation processes: multiple uses to compute and uncompute at {\it each} swap in the sorting network. An interesting distinction is that there will be more validating measurements, one confirming each controller qubit is in state $\ket{0}$ after its swap process is complete. It would be interesting to evaluate the overall pros and cons of the two approaches in a further study.

\subsection{Driving loaded state to desired state}
Loading multiple uncorrelated single particle states onto separate quantum registers is not sufficient to represent many-particle eigenstates of real molecular Hamiltonians. As noted in the main text, a popular method is to drive an initial state that can be efficiently loaded onto a quantum computer towards the correlated molecular ground state. An advantage of this approach is that we need not have the ability to analytically describe the target state. Techniques include adiabatic evolution~\cite{Farhi2000QuantumCB}, projective eigenstate filtering~\cite{Lin2020nearoptimalground}, phase estimation~\cite{kitaev1995quantum, PhysRevLett.102.130503, berry2018improved}, and imaginary time evolution~\cite{liu2021probabilistic, kosugi2021probabilistic}.
In this work we demonstrate a small subset of these methods that only require a single ancilla qubit, targetting early fault-tolerant quantum computers; these included a probabilistic variant of imaginary time evolution, and a protective IPE method. Recent advances in eigenstate filtering and quantum signal processing using only a single ancilla~\cite{https://doi.org/10.48550/arxiv.2204.05955} can potentially be applicable to preparing ground states of grid-based states.

An alternative proposed by Ward \textit{et al.}~\cite{doi:10.1063/1.3115177} is to load occupied Hartree-Fock orbitals~\cite{szabo1996modern} into real-space particle registers using the aforementioned method by Zalka, followed by (anti)symmetrisation, to generate Slater-determinant type states. They further generalise this to preparing superpositions of Slater-determinant type states in the real-space particle registers, yielding exact molecular ground states through the creation of Full Configuration Interaction (FCI) states. Their method however demands the classical or quantum precomputation of the electronic structure in second-quantisation, loading the result in ancilla qubits, and achieving a transformation which loads the information from the second-quantised ancilla to the real-space particle registers. The authors did not detail how such a transformation can be achieved, and loading a factorial number of Slater determinants from the FCI state may pose challenges.

For completeness, we also discuss the general approach for using the IPE for state preparation. Assuming we know the energy $E$ of the state we wish to prepare, we apply controlled evolution as before, but choose a total time $T=N\delta t=2n\pi/E$ for some integer $n$, to measure the ancilla. If the initial state had been exactly the desired state then the ancilla would now certainly be found in the $\ket{+}$. An initial state with a different energy $E_k\neq E$ would in general not yield the $\ket{+}$ with certainty but rather with probability $\cos^2(T E_k)$. We therefore demand a $\ket{+}$ outcome, and restart if we do not see it. On success, the post-selected state on the main register has a boosted amplitude associated with the desired state because other states have been suppressed by the factor $\sin^2(T E_k)$, analogous to PITE. We repeat the process to probabilistically purify the desired state (or if degeneracy is present, a state in the degenerate subspace with energy $E$). Our probability of success is precisely the total probability associated with the target state's component of the initial state, therefore it is important to prepare an initial state having reasonable overlap with the target. The total time required will depend on the inverse of the energy gap $E_k-E$ between the desired state and the nearest unwanted component.

\section{The Coulomb potential: Demands on spatial and temporal resolution}
\label{appendix:resolutiondemands}
The standard split-operator and indeed the SO-QFT method involves three critical approximations: (a) a discretised representation of the state, (b) propagation of the potential phase assuming the interaction operator $\hat{H}_\text{int}$ is exactly diagonal, and (c) the Trotter error from discretisation of $e^{-i \hat{H}_\text{tot}\delta t}$ into the two non-commuting $e^{-i \hat{H}_\text{kin}\delta t}$ and $e^{-i \hat{H}_\text{int}\delta t}$ phases; see Ref.~\cite{Childs2022}. Both (a) and (b) become exact in the limit of infinite spatial resolution, while (c) becomes exact in the limit of infinitesimally small time steps. 

It is worth stressing that approximation (a) has consequences for both the wavefunction representation and moreover for the operators forming the SO-QFT cycle: the representation of these continuous operators as a discrete matrix is an approximation, and moreover the nature of the off-diagonal terms in the potential part will depend on the resolution. 
The severity of the impact depends on the modelled interactions; if we were using  quadratic potentials $V(\mathbf{r})\propto r^2$, then a modest resolution could ensure that the potential is almost constant over interval between one spatial pixel function and the next, $x_{n+1}-x_n=\delta r$. We remark that the preliminary testing and debugging of our emulation code was performed on `easy' scenarios of this kind.

However, the results presented in this paper concern the far more challenging Coulomb potentials which are singular at $\mathbf{r}=\mathbf{0}$, where the potential changes rapidly between neighbouring pixel functions. In this case, a relatively modest resolution can still suffice if the wavefunctions involved have near-zero amplitude at the origin; for example in 2D hydrogen, states with quantum number $m\neq 0$ have this property, and this motivates our use of such states in several of the numerical studies presented in the main paper. Nevertheless, many important states (for example the ground state of atomic electrons) have significant amplitudes at the origin. 
It is well-established in the literature that any reasonable observable property of our system will converge properly as the spatial resolution $\delta r^{-1}$ is increased (provided that other parameters are suitably updated; notably $\delta t$ must decrease as explained below).
This may not be intuitively obvious since increasing the resolution introduces new pixel functions that are closer to the origin and for whom the diagonal approximation is `worse'.
The most direct way to see this is by considering the unitary nature of the simulation; the SO-QFT method, while approximate, is nevertheless strictly unitary. As the resolution increases, the pixels that are closest to the Coulomb singularity may indeed be more imperfectly updated by our simulation steps, but they constitute an exponentially vanishing component of the entire simulation, both in terms of the Hilbert space and (crucially) the total amplitude associated with them. Thus it ultimately becomes impossible for any observable to distinguish between a state evolved under the ideal $e^{-i \hat{H}_\text{tot} \delta t}$ and the same state evolved using SO-QFT at high resolution.

We can therefore be confident that there is {\em some} finite resolution $\delta r$ that will suffice to adequately approximate the Coulomb operator, addressing the first two sources of error.
But this comes with an important caveat that relates to the Trotter error. The problematic cost is not the qubit count but rather the time required for our simulation; as $\delta r$ decreases we must reduce the time step $\delta t$ proportionately and thus more time steps will be needed to simulate a given period of dynamics. 
The intuitive reason is clear: the effect of the kinetic operator is to change the shape of the wavepacket in real space, but if $\delta t$ is too large, the change is significant on the scale of multiple pixel widths $\delta r$. Amplitude can then pass through otherwise impermeable features such as the Coulomb singularity. Instead, $\delta t$ should be small enough that the kinetic part of the SO-QFT step only modestly alters the amplitudes associated with each pixel function.

One means of optimising the approach would be selecting an appropriate Trotter sequence~\cite{tannor2007introduction,suzuki1991,childsTrotter2021}. In this paper we use only the simplest 1$^\text{st}$ order sequence where the error $\epsilon = \frac{1}{2}[\hat{H}_\text{kin},\hat{H}_\text{int}]\delta t^2$. Any order Trotter sequence will ultimately have an error involving on the same commutator (in nested form), which in turn depends on the higher order derivatives of the interaction potentials. In our finite discretisation of the Coulomb potential, it will again be elements very close to the singularity, where the derivatives are the largest, that are the most problematic, and the more so as the spatial resolution increases. Thus we anticipate that $\delta t$ should still commensurately decrease, albeit its severity will vary with the Trotter order.
Ultimately this cost will not prevent successful simulation of interesting molecules on a quantum computer; as discussed in the main text, the proper choice of $\delta r^{-1}$ does not scale with system size but only with the highest nuclear charge. Nevertheless it is an impactful cost.

We conclude this Section by noting that methods for handling the Coulomb singularity in real-space have been explored in the literature. One could cap or truncate the potential~\cite{Fernandez_1991, Hall2009, Chen2021, Childs2022}, or describe it with a Fourier expansion so that it cannot be more highly curved than the highest Fourier component (which might be chosen to match the highest $k$ plane waves in the state's representation) \cite{Spencer2008, PhysRevX.8.011044}.
The nuclear potential could also be replaced with an effective potential representing which accounts for the inner shell electrons \cite{Rogers1970}. Methods of this kind may ultimately prove to be the optimal approach for quantum computer enabled simulations too. For the present paper we attempt to deal with the Coulomb potential in its `natural' form by effectively capping it; we always define the singular origin of the potential between two grid points throughout this work.

\section{Placement of nuclear origin}
\label{appendix:origin}
We briefly report that placement of the nuclear origin between grid points does not significantly affect the dynamics (summarised in Table~\ref{tb:originplacement}). When we time propagate the $\Psi_{1,1}$ state with the Coulomb potential centred at different fractions between two neighbouring grid points, the state visibly oscillates where grid points lie asymmetrically around the Coulomb origin. However, the magnitude of the loss in fidelity is limited.
Most importantly, the energy from the phase estimation is not influenced at all by the skew. We conclude that discrepancies in the placement of grid points relative to the system can, if needed, be alleviated with a brute force increase in the spatial resolution and correspondingly the time resolution, but its influence on observables related to state fidelity can be negligible.
\begin{table}[htb]
    \centering
    \begin{tabular}{l|ccccc}
                          & 0.5 & 0.4 & 0.3 & 0.2 & 0.1  \\ 
    \hline
    Max loss & $7\times 10^{-7}$   & 0.00085   & 0.00021 & 0.0042 &  0.0022  \\
    $\epsilon$ ($10^{-5} E_h$) & -1.84  & -1.11   &  0.29 & -1.66  & 5.98 \\ 
    \hline
    \end{tabular}
    \caption{Placement of the Coulomb nuclear potential between grid points and its influence on the autocorrelation as well as the energy predicted through phase estimation.}
    \label{tb:originplacement}
\end{table}

\section{Resource scaling}
\label{appendix:scaling}

In this work we considered the way in which the grid-based method's resources can be expected to scale. We noted that there are detailed studies in the literature. The contribution to this topic made by the present paper is that its emulated algorithms may elucidate some of the constants that arise in any scaling analysis. 

\subsection{Space Cost}
We argued that the qubit count should scale with particle count $P$ according to $3n_rP$ with
\begin{equation} \label{suppeqn:scaling}
    n_r\approx
C_3+\log_2(Z_\text{max})+\frac{1}{3}\log_2(P)    
\end{equation}
(reiterating from main text) where $Z_\text{max}$ accounts for the highest nuclear charge in the modelled system. We noted that this crude expression does not account for the fact that a given atom's radius has a sub-linear dependence on the number of electrons; accounting for this would lower the outcome.

Since $n_r$ really depends on more properties of the problem besides $Z_\mathrm{max}$ and $P$, $C_3$ will not be an absolute constant, but rather fluctuate with other system properties, such as the geometry and electron configuration. We can proceed to estimate the required resources for the two molecules we identified as interesting in the Introduction to arrive at two example values for $C_3$. These will then give us a rough idea of its range.

We begin by noting that in the numerical modelling reported here, remarkably accurate and stable simulations can be achieved with as few as 6 qubits per $x$, $y$ or $z$ sub-register (thus, 18 qubits per 3D particle). Methods such as iterative phase estimation, requiring only one additional qubit, can then obtain eigenenergies with accuracy up to 6 decimal places. Our results using the ASO method suggest that even core-peaked electronic states can be modelled with only a modest increase in resolution. For present purposes we therefore take an optimistic stance and assume that $n_r=7$ can suffice for systems with $P = Z = 1$. For larger values of $P$ and $Z$, more data points per dimension are necessary to compensate for differently sized simulation boxes and higher curvature of core states. We discuss momentarily how we estimate these required changes.

For molecules that are already well-understood, the ionization potential can be used to estimate their long-range behaviour~\cite{PhysRevA.23.1030}, however one may wish to use a quantum simulator to explore molecular systems that have not been experimentally charatcterised. Therefore we will adopt a first-principles argument based only on the constituent atoms and their presumed locations.  We place the centre of the highest occupied hydrogen-like wave function of each atom at the coordinates of each nucleus, and calculate its radial charge density according to Ref.~\cite{ghosh2002theoretical}, which also semi-empirically accounts for nucleus shielding. Summing up the contributions from all atoms gives us an approximation to the total charge density if the electrons of different atoms were not interacting with each other. We then find the maximum of this total density on the surface of the simulation box, which informs us about how strongly the electrons will interact with its boundary. For molecules of interest here, the atomic locations in their equilibrium geometry are presented in  Ref.~\cite{cccbdb}.

To get an idea of what an acceptable charge density at the box surface could be, we utilise the calculation of 3D helium without electron-electron interaction discussed in the main text. From the electron configuration and box size of that simulation, we derive its maximum surface electron density $\rho_0$ using the approximate method described above. Consequently, the simulation box for any other molecule might be considered sufficiently large whenever the method above yields a maximum surface electron density that does not exceed $\rho_0$. Taking a relatively optimistic line in our resource estimation, we will refer to this rule; but note that because it is only a rough approximation to the extent of the electron cloud, and specific scenarios might necessitate an increase. Interesting dynamics with substantial numbers of moving particles may require significantly larger boxes, as to not let significant amplitude collide with the simulation boundary (or to allow an adequate attenuation regions as described in the main paper).

Besides varying simulation box sizes, we should also account for changes in the required spatial resolution. For $Z > 1$, the features of hydrogenic wave functions shrink by a factor of exactly $Z^{-1}$, which is usually also a good approximation for low-lying core states. To accurately resolve these electronic states, we must therefore increase the number of grid points per unit length by a factor of $Z_\mathrm{max}$.

For Ammonia (NH$_3$), the more modest of the two molecules we identified as interesting in the Introduction, an optimistic resource audit can proceed as follows:
The highest charge in the molecule is $Z_\mathrm{max} = 7$. Therefore, the number of grid points must increase to 7 times that of our estimate for $Z = 1$. At the same time, the above mentioned method to determine the box size yields a side length of $\approx 1.1$ times the length required for a single hydrogen atom, giving a total factor of $\approx 1.1 \times 7 = 7.7 < 8 = 2^3$, which means we must increase $n_r$ by $3$ from its reference value of 7, giving $n_r = 10$ qubits per particle and dimension. Substituting these values in Eq.~\ref{suppeqn:scaling} and rounding where appropriate leads to a value of $C_3^{\mathrm{NH}_3} \approx 6$. For 14 particles (10 electrons and 4 nuclei) in 3 dimensions, we arrive at a total of $3 \times 10 \times 14 = 420$ qubits. We might round this to $450$, recognising that multiple ancillas may be required even in a very frugal implementation.

The more challenging case we mentioned was C$_2$F$_6$. The highest-charge nuclei are those of fluorine, giving us $Z_\text{max}=9$, thus increasing the required spatial resolution 9-fold. Surprisingly, as the electron densities of C and F drop off quite rapidly with increasing distance from the atom, the above described method suggests that the required side length of the simulation box could be as small as 0.85 times that of a single hydrogen atom. This gives a total factor of $\approx 0.85 \times 9 = 7.65 < 8 = 2^3$, again meaning that 3 qubits must be added to the reference value, giving a total of $n_r = 10$. Using Eq.~\ref{suppeqn:scaling} again, we arrive at $C_3^{\mathrm{C}_2\mathrm{F}_6} \approx 5$. The total number of required qubits is $3 \times 10 \times 74 = 2220$, which we might round up to $2250$ to allow for a frugal ancilla overhead.

From these two data points, we cautiously estimate $C_3$ to be on the order of $\sim 10$, but note that it might be lower for advantageous circumstances as in the examples given, or may also increase in unfavourable situations.

\subsection{Time cost}
It is difficult to provide meaningful estimate of total `wall clock' time for a significant quantum simulation, even to within an order of magnitude. We will nonetheless use some broad assumptions to provide generally indicative numbers for the time cost.

The timescale of the simulation depends heavily on the energy scales across which the events probed occur; the models in the paper (electron scattering, ionisation) correspond to high energy `fast events' that occur on a sub-femtosecond timescale, whereas resulting dynamics (e.g. bond breaking, molecular fragmentation, energy transfer) are `slow events' that happen at $O(1)$ even up to $O(10)$ picosecond timescales. We postulate this to be independent of system size. The resolution by which we discretise this total simulation length by $\delta t$ depends only on the $\delta r$ sufficient to describe the most tightly bound electrons of the element with highest nuclear charge, which also does not change with system size. Moreover, the methods explored here, and indeed proposals of e.g. Ref.\,\cite{Lin2020nearoptimalground}, indicate that energy estimation and state preparation could also be achieved using a number of time steps that does not scale significantly with system size. 

The time to complete each SO-QFT cycle does however scale with system size. There is a modest scaling due to the computation associated with introducing proper phases (equivalent to binary addition/subtraction, squaring, and inversion) which scales with the number of qubits in each register -- although not, of course, with the number of qubits in total. Since the $n_r$ scales only logarithmically with simulation box size $L$, this time factor should be poly-logarithmic in e.g. the number of particles $O(\log^k P)$. A more significant factor comes from the need to compute the electron-electron interaction for all possible pairings of electrons. The task can be divided into a series stages within each of which all electrons are paired-off and  we can assume that computations relevant to electron-electron dynamics occur in parallel (see~\cite{Cody_Jones_2012}). The number of such stages is $P-1$ for $P$ electrons, so that the total time for a complete SO-QFT implementation is of the order $O(P\,\log^k P)$. Considerations such as the time required to move from one pairing to another should also be borne in mind, but one finds that this should not scale with $n$ even in devices with restricted connectivity. 

Our QFT implementation used an explicit quantum circuit, while the conditional phases were applied directly using a bespoke algorithm (see Section~\ref{appendix:numerical}); if we focus on the QFT component and use the same implementation employed in our emulator (Fig\,\ref{fig:qft_circuit_diagram}) then $O(n_r^2)$ gates are required. A full SO-QFT cycle may involve operations equivalent to several QFTs, since they constitute an approach to the addition and squaring operations that we employ. Moreover other operations may require $n_r^2$ gates, such as the application of the $\propto k^2$ phases for the kinetic operator. If we therefore take $10n_r^2$ as a plausible suggestion for the gate cost of each SO-QFT sub-step in a parallelised implementation, then for the C$_2$F$_6$ molecule for each SO-QFT cycle we have a total gate depth of $260\times 10\times 10^2$ which we conservatively call $O(10^6)$ (and $260/2=130$ such processes occurring in parallel throughout that cycle). 

Many of the simulations presented in the main paper used $O(10^3)$ SO-QFT cycles to generate dynamics with good precision. This count would diminish with the adoption of more sophisticated Trotter sequences than the lowest-order pattern employed in the paper; one might expect an order of magnitude reduction is possible here. Thus, given the conclusion of the above paragraph that $O(10^6)$ gates are required for each SO-QFT cycle, one could assume that understanding a `fast event' such as electron ionisation requires an algorithmic gate depth of hundreds of millions. In the main paper we indicate how such numbers may translate to clock time, finding one execution of the algorithm requires only minutes.

However, we might ask whether it is possible to encapsulate a `slow event' like the subsequent dynamics of a molecule immediately after an interesting `fast event'. Simplistically, this might seem to imply the need for a corresponding multiplicative factor in the number of SO-QFT cycles required for the complete simulation. Fortunately however, as we explored in our two-particle scattering simulation, it might be natural to use a coarser time stepping according to the speed of the physics that is unfolding. Accounting for these various considerations, we assume that $O(10^5)$ SO-QFT cycles should suffice to reveal desired `slow', longer time quantum molecular dynamics. One concludes that the gate depth of the parallelised quantum algorithm for evolving such a system in the picosecond timescale is about a hundred billion. As noted in the main paper, this leads to a total clock time of the order of a day for this more challenging simulation. 

\end{document}